\documentclass[12pt,preprint]{aastex}

\slugcomment{Submitted to ApJ?}

\shorttitle{Star Formation in M51}
\shortauthors{Calzetti et al.}

\begin{document}

\title{Star Formation in NGC~5194 (M51a): The Panchromatic View from GALEX to 
Spitzer\altaffilmark{1}}

\author{D. Calzetti\altaffilmark{2}, R.C. Kennicutt\altaffilmark{3},
L. Bianchi\altaffilmark{4}, D.A. Thilker\altaffilmark{4},
D.A. Dale\altaffilmark{5}, C.W. Engelbracht\altaffilmark{3},
C. Leitherer\altaffilmark{2}, M.J. Meyer\altaffilmark{2},
M.L. Sosey\altaffilmark{2}, M. Mutchler\altaffilmark{2},
M.W. Regan\altaffilmark{2}, M.D. Thornley\altaffilmark{6},
L. Armus\altaffilmark{7}, G.J. Bendo\altaffilmark{3}, 
S. Boissier\altaffilmark{8}, A. Boselli\altaffilmark{9}, 
B.T. Draine\altaffilmark{10}, K.D. Gordon\altaffilmark{3}, 
G. Helou\altaffilmark{7}, D.J. Hollenbach\altaffilmark{11},
L. Kewley\altaffilmark{12}, B.F. Madore\altaffilmark{8}, 
D.C. Martin\altaffilmark{16}, E.J. Murphy\altaffilmark{13}, 
G.H. Rieke\altaffilmark{3}, M.J. Rieke\altaffilmark{3}, 
H. Roussel\altaffilmark{7}, 
K. Sheth\altaffilmark{7}, J.D. Smith\altaffilmark{3}, 
F. Walter\altaffilmark{14}, B.A. White\altaffilmark{3}, 
S. Yi\altaffilmark{15}, N.Z. Scoville\altaffilmark{16}, 
M. Polletta\altaffilmark{16}, \& D. Lindler\altaffilmark{17}}
%\affil{}
%\email{calzetti@stsci.edu}

%\author{ }
%\affil{}

%\author{}
%\affil{}

%\author{}
%\affil{}

%\author{}
%\affil{}

%\and

%\author{}
%\affil{}

\altaffiltext{1}{Based on observations obtained with the Spitzer Space
Telescope and with GALEX.}  \altaffiltext{2}{Space Telescope Science
Institute, 3700 San Martin Drive, Baltimore, MD 21218, USA;
calzetti@stsci.edu} \altaffiltext{3}{Steward Observatory, University
of Arizona} \altaffiltext{4}{Department of Physics and Astronomy, The
Johns Hopkins University} \altaffiltext{5} {Dept. of Physics,
University of Wyoming} \altaffiltext{6}{Department of Physics,
Bucknell University} \altaffiltext{7}{Spitzer Science Center, CalTech}
\altaffiltext{8}{Observatories of the Carnegie Institution of
Washington} \altaffiltext{9}{Laboratoire d'Astrophysique de Marseille,
Marseille} \altaffiltext{10}{Princeton University, Observatory}
\altaffiltext{11}{NASA/Ames Research Center}
\altaffiltext{12}{Institute for Astronomy, University of Hawaii}
\altaffiltext{13}{Dept. of Astronomy, Yale University}
\altaffiltext{14}{Max--Planck Institute for Astronomy, Heidelberg}
\altaffiltext{15}{Physics Department, Oxford}
\altaffiltext{16}{CalTech} \altaffiltext{17}{Sigma Computers
Corporation}

\begin{abstract}
Far ultraviolet to far infrared images of the nearby galaxy NGC~5194
(M51a), from a combination of space--based (Spitzer, GALEX, and Hubble
Space Telescope) and ground--based data, are used to investigate local
and global star formation, and the impact of dust extinction. The
Spitzer data provide unprecedented spatial detail in the infrared,
down to sizes $\sim$500~pc at the distance of NGC~5194. The
multiwavelength set is used to trace the relatively young stellar
populations, the ionized gas, and the dust absorption and emission in
HII--emitting knots, over 3 orders of magnitude in wavelength
range. As is common in spirals, dust extinction is high in the center
of the galaxy (A$_V\sim$3.5~mag), but its mean value decreases
steadily as a function of galactocentric distance, as derived from
both gas emission and stellar continuum properties. In the
IR/UV--UV~color plane, the NGC~5194 HII knots show the same trend
observed for normal star--forming galaxies, having a much larger
dispersion ($\sim$1~dex peak--to--peak) than starburst galaxies. We
identify the dispersion as due to the UV emission predominantly
tracing the evolved, non--ionizing stellar population, up to ages
$\sim$50--100~Myr. While in starbursts the UV light traces the current
SFR, in NGC~5194 it traces a combination of current and recent--past
SFR. Possibly, mechanical feedback from supernovae is less effective
at removing dust and gas from the star formation volume in normal star
forming galaxies than in starbursts, because of the typically lower
star formation rate (SFR) densities in the former.  The application of
the starburst opacity curve for recovering the intrinsic UV emission
(and deriving SFRs) in local and distant galaxies appears therefore
appropriate only for SFR densities
$\gtrsim$1~M$_{\sun}$~yr$^{-1}$~kpc$^{-2}$. Unlike the UV emission,
the monochromatic 24~$\mu$m luminosity is an accurate {\em local} SFR
tracer for the HII knots in NGC~5194, with a peak--to--peak dispersion
of less than a factor of 3 relative to hydrogen emission line tracers;
this suggests that the 24~$\mu$ emission carriers are mainly heated by
the young, ionizing stars. However, preliminary results show that the
ratio of the 24~$\mu$m emission to the SFR varies by a factor of a few
from galaxy to galaxy; this variation needs to be understood and
carefully quantified before the 24~$\mu$m luminosity can be used as a
SFR tracer for galaxy populations. While also correlated with star
formation, the 8~$\mu$m emission is not directly proportional to the
number of ionizing photons; it is overluminous, by up to a factor
$\sim$2, relative to the galaxy's average in weakly ionized regions
and is underluminous, by up to a factor $\sim$3, in strongly ionized
regions. This confirms earlier suggestions that the carriers of the
8~$\mu$m emission are heated by more than one mechanism.
\end{abstract}


\section{Introduction}

Over the past decade, discoveries of galaxy populations at earlier and
earlier cosmic times has rekindled interest in star formation rate
(SFR) indicators, estimated from a variety of monochromatic and
non--monochromatic emission measurements across the full spectrum. Of
particular interest for cosmological studies are indicators exploiting
measurements at restframe ultraviolet (UV), optical, and
mid/far--infrared (MIR/FIR) wavelengths; the interest has, however,
accompanied a renewed awareness that potential limitations are not
fully quantified yet. Presence of even small amounts of dust
extinction in the early galaxies hampers significantly SFR
measurements from the rest-frame UV emission of the high--redshift
galaxies (e.g., the Lyman-break galaxies,
\citet{stei99,erb03,giav04,reddy04}). At the other end of the
spectrum, our still limited understanding of the infrared spectral
energy distribution of galaxies may decrease the SFR prediction power
of the sub--mm emission of the IR-bright SCUBA sources
\citep[e.g.,][]{barg00,smail00,chap03,chap04}. Even in the local
Universe, the widely different angular scales that have characterized
until recently UV, optical, and infrared observations of galaxies,
ranging from arcsec/subarcsec resolution for UV/optical to
multiple~arcsecs/arcminutes for FIR (IRAS, ISO), have so far limited
our ability to undestand in detail the applicability of each indicator
within the realm of physically complex systems
\citep{kenn98,kew02,rosgon02}. This in turn has inhibited attempts at
cross--correlating calibrations of SFR indicators.

The issue lays in how well tracers at each wavelength can measure the
actual SFR. The main problem afflicting UV and optical SFR indicators
is dust obscuration. There are two aspects to this problem. One is
that regions with moderate amounts of dust will be {\em dimmed} in a
way that depends not only on the amount of dust, but also on the
distribution of the emitters relative to the absorbers. This problem
is exacerbated by the fact that populations of different ages suffer
different amounts of dust extinction
\citep{calz94,charl00,zari04}. Recently it has been shown that
quiescently star forming galaxies follow a different dust
opacity--reddening relation than starburst galaxies
\citep{bua02,bell02,gord04,bua05,seib05,laird05}. In particular,
their IR/UV ratio, a measure of dust opacity, is on average lower than
that of starbursts for the same UV color, a measure of dust reddening,
and shows a larger spread; differences in the `{\em b} parameter' (the
ratio of current--to--lifetime SFR) between star--forming and
starburst galaxies has been invoked as an explanation for the observed
difference \citep{kong04}.

A second problem is the unknown fraction of star formation that is
{\em completely obscured} by dust at UV and optical wavelengths. The
UV and FIR may, indeed, probe different regions/stages of star
formation. Heavy obscuration is generally tied to the first temporal
phases of star formation, $\sim$a few~Myr; as the stars age, they tend
to drift off the parental cloud and diffuse in regions of lower
gas/dust density, or to disperse the natal gas/dust cloud
\citep{leisa88}. Estimates indicate the fraction of completely
obscured star formation to be relatively small in the local Universe,
$\sim$20\%--30\% \citep{calz95,heck99,calz01}, but uncertainties are
large and their impact on the calibration of SFR indicators mostly
unprobed.

A comprehensive attack to these problems is a core goal of the Spitzer
Infrared Nearby Galaxies Survey project \citep[SINGS,][]{kenn03}. This
paper presents the first case study based on the well--known
grand--design spiral galaxy NGC~5194 (M51a, Whirpool Galaxy). We use a
multi--wavelength dataset of the galaxy by combining UV images from
GALEX, ground--based optical images, infrared emission line images
from HST/NICMOS \citep{scov01}, and Spitzer 3.5--160~$\mu$m
images. These data provide a panchromatic view of the star formation
in this galaxy, both locally (on the scales of star--formation
complexes) and globally. Spitzer and GALEX observations of nearby
galaxies (closer than $\sim$10~Mpc) are uniquely suited for
investigating issues of dust obscuration and star formation, thanks to
a combination of comparatively high angular resolution (a few arcsec)
and the large field-of-views (many arcminutes). We use the
multi--wavelength data to investigate the opacity--reddening
properties of this quiescently star--forming galaxy on a detailed
spatial scale. The UV, MIR and FIR emission are then compared with the
optical (nebular lines) emission, both locally and globally, to test
the viability of each as a SFR indicator. For instance, the 8~$\mu$m
emission is a potentially attractive SFR indicator at high-redshifts,
as the restframe $\sim$8~$\mu$m PAH bands are redshifted to
$\lambda\gtrsim$24~$\mu$m for z$\gtrsim$2, thus still within the
regime probed by, e.g., Spitzer.  In addition, unlike tracers that
probe directly the stellar light, MIR/FIR SFR tracers are little
affected by dust extinction.

There are a number of reasons for why NGC~5194 is an optimal target for
this study. At a distance of about 8.2~Mpc (from \citet{tull88}'s
systemic velocity and H$_o$=70~km~s$^{-1}$~Mpc$^{-1}$), the typical
angular resolution of our mid--infrared data,
5$^{\prime\prime}$--13$^{\prime\prime}$, corresponds to
$\sim$200--520~pc, or the size of a large star-formation complex. The
relatively high level of spatial detail enables us to investigate the
nature of the difference in the opacity--reddening properties between
starbursts \citep{meur99} and quiescent star--forming galaxies
\citep{kong04}. The total SFR, $\sim$3.4~M$_{\sun}$~yr$^{-1}$, and the
SFR/area, $\sim$0.015~M$_{\sun}$~yr$^{-1}$~kpc$^{-2}$, of this galaxy
places it among the `quiescently' star--forming systems, despite its
interaction with the early-type galaxy NGC~5195 (M51b), the latter
located about 4$^{\prime}$.4 (10.5~kpc) to the North.

NGC~5194 is a nearly--face--on (i$\sim$20$^o$), grand-design spiral
(SAbc), with intense star formation in the center and along the spiral
arms. Its OB associations population, the gas they ionize, as well as
the diffuse ionized medium, have been extensively investigated at
optical and infrared wavelengths
\citep{kenn89,scov01,thilk02,hoope03}. The UV emission shows a strong
color gradient as a function of distance from the nucleus, becoming
bluer at larger galactocentric distances, based on the GALEX images
\citep{bian04a}. This is similar to what was previously found by
\citet{hill97} from UV$-$U radial color trends, and a comparison with
IRAS images suggest this color gradient to be induced by a gradient in
the dust extinction \citep{boiss04}.  NGC~5194 is a metal--rich galaxy
(12$+$log(O/H)$\sim$8.7--8.9, \citet{breso04}), with a weak
metallicity gradient as a function of distance from the nucleus out to
at least 10~kpc radius, or $\sim$75\% the B$_{25}$ radius
\citep{zari94}. When comparing properties of HII knots within the
galaxy, the shallow metallicity trend enables us to investigate
stellar population ageing effects unencumbered by the metallicity
variations that affect galaxy--to--galaxy comparisons.

The present paper is organized as follows: section~2 presents the
observations, relevant data reduction considerations, and the main
characteristics of the dataset; section~3 is a short overview of the
galaxy's morphology at different wavelenghts; section~4 presents the
measurements of HII-emitting regions performed on the images;
section~5 describes the observed properties of these HII--emitting
regions; section~6 presents dust extinction properties; section~7
analyzes the properties of popular star formation rate indicators,
while the results are discussed in section~8. A summary is given in
section~9.

\section{Observations and the Dataset}

\subsection{Spitzer Images}

The Spitzer images of M51 (NGC~5194/NGC~5195) were obtained with both
IRAC (3.6, 4.5, 5.8, and 8.0~$\mu$m) and MIPS (24, 70, and
160~$\mu$m), as part of the SINGS Legacy project. A description of
this project and the observing strategy can be found in
\citet{kenn03}.

Each of the four IRAC images is a combination of two mosaics, each
resulting from a 6$\times$9 grid covering a
18.5$^{\prime}\times$25$^{\prime}$ field. Observations of each mosaic
were obtained on 18 and 22 May 2004, allowing a separation of a few
days between the two to enable recognition and exclusion of asteroids
and detector artifacts. Total exposure times in each filter are 240~s
in the center of the field, and 120~s at the edges (outer
$\sim$2.5$^{\prime}$). The SINGS IRAC pipeline was used to create the
final mosaics, which exploits the sub-pixel dithering to better sample
the emission, and resamples each mosaic into 0.76$^{\prime\prime}$
pixels \citep{rega04}. The measured 8~$\mu$m PSF FWHM is
2.1$^{\prime\prime}$, and the 1~$\sigma$ sensitivity limit in the
central portion of the 8~$\mu$m mosaic is
1.2$\times$10$^{-6}$~Jy~arcsec$^{-2}$.

A `dust--emission' image at 8~$\mu$m is obtained by subtracting the
stellar contribution using the recipe of \citet{pahr04}. The
stellar--emission--dominated 3.6~$\mu$m and 4.5~$\mu$m images are
combined assuming colors appropriate for an MIII0 star
([3.6]-[4.5]=-0.15 in Vega mag, \citet{pahr04}), and then rescaled
under the same assumption to create a stellar--only image at 8~$\mu$m
([3.6]-[8.0]=0.0 in Vega mag). A few percent adjustement of the
rescaled `stellar' image is used to optimize the subtraction from the
8~$\mu$m image. 

Potentially, the 3.6~$\mu$m and 4.5~$\mu$m images can contain, in addition to
photospheric emission from stars, also a component of hot dust emission. The
impact of this component relative to the stellar contribution is different in
the two images, with flux ratios
[f(dust)/f(star)]$_{3.6}\sim$0.3--0.7~[f(dust)/f(star)]$_{4.5}$ (depending on
the adopted stellar population), for dust with temperature
T$\lesssim$1000~K. To test whether a hot dust contribution may affect the
derivation of the 8~$\mu$m `dust--emission' image, we have produced a second
stellar--continuum--subtracted 8~$\mu$m image, using only the rescaled
3.6~$\mu$m image as `stellar continuum'. The two dust images differ from each
other by less than 3\% across the entire region analyzed, suggesting that hot
dust is not significantly impacting the stellar continuum subtraction process.

MIPS observations of the galaxy were obtained on 22 and 23 June 2004.
The reduction steps for MIPS mosaics are described in
\citet{gord05}. The final mosaics have size
27$^{\prime}\times$60$^{\prime}$, fully covering M51 and the
surrounding background. At 24~$\mu$m, 70~$\mu$m, and 160~$\mu$m, the
PSF FWHM is $\sim$5.7$^{\prime\prime}$, $\sim$16$^{\prime\prime}$, and
$\sim$38$^{\prime\prime}$, respectively. The 1~$\sigma$ detection
limits are 1.1$\times$10$^{-6}$~Jy~arcsec$^{-2}$,
8.7$\times$10$^{-6}$~Jy~arcsec$^{-2}$, and
2.6$\times$10$^{-5}$~Jy~arcsec$^{-2}$, respectively, for the
24~$\mu$m, 70~$\mu$m, and 160~$\mu$m images.  The three MIPS images
are considered `dust' images for all purposes, as contributions from
the photospheric emission of stars is negligible at these wavelenghts.

Consistency between the MIPS and IRAS flux scales has been checked by
comparing the MIPS24 with the IRAS25 fluxes and, to a lesser extent,
the MIPS70 with the IRAS60 fluxes. We get f(24)$\sim$12.3~Jy for
NGC~5194, or about 20\% lower than the IRAS25 value of 14.8~Jy (as
measured from IRAS HiRes images); for the MIPS70 channel, we get
f(70)$\sim$105~Jy, in better agreement with the IRAS60 value of
110.3~Jy, despite the slight offset between the two wavebands. The
total far--infrared luminosity of NGC~5194, L(IR)=L(3--1100~$\mu$m), as
derived from the MIPS fluxes (equations~4 of \citet{dale02}), is
Log[L(IR)/(erg s$^{-1}$)]=44.1, about 7\% lower than the same quantity
obtained from the IRAS fluxes (and using equation~5 of
\citet{dale02}). The nominal MIPS calibration uncertainties,
$\sim$10\% at 24~$\mu$m and $\sim$20\% in the longer wavelength bands,
account for most of the discrepancies between the MIPS and IRAS fluxes
and luminosities, with the possible exception of the 24~$\mu$m
band. However, removal/editing of the companion NGC~5195 is
non--trivial in the low resolution IRAS images, and this may account
for some of the discrepancy. Indeed, the MIPS24 flux of the whole M51
pair (NGC~5194$+$NGC~5195), f(24)$\sim$13.5~Jy, is in agrement with the
25~$\mu$m flux, $\sim$13~Jy, obtained by COBE DIRBE (from the DIRBE
Point--Source Photometry Browser,
http://lambda.gsfc.nasa.gov/product/cobe/browser.cfm).

\subsection{GALEX Images}

GALEX \citep{marti04} imaging observations are centered at
1529~\AA~ for the far--UV (FUV, 1350--1750~\AA) and at 2312~\AA~ for
the near--UV (NUV, 1750--2750~\AA) bands. Data for M51 were obtained
on June 19--20, 2003, as part of the Nearby Galaxies Survey (NGS,
described by \citet{bian04b,bian04c}). The exposure time of 1414~s
yields a NUV(FUV) 1-$\sigma$ sensitivity limit of
1.4$\times$10$^{-19}$ (3.6$\times$10$^{-19}$)
erg~s$^{-1}$~cm$^{-2}$~\AA$^{-1}$~arcsec$^{-2}$ at the PSF scale
(FWHM=4.6$^{\prime\prime}$).  More details on the GALEX data, and a
comparison with previous UIT data, are given by \citet{bian04a}. The
latest photometric calibrations (IR1 release, November 2004)
were applied to the two GALEX images of M51.  The measured FWHM of the
GALEX PSF is only slightly smaller than the MIPS 24~$\mu$m PSF, making
the comparison between the two sets of images straightforward.

Distorsions present in the FUV image  were corrected by
application of non--linear geometric transformations to the image,
using the optical images as reference. Residual distorsions amount to
$\lesssim$1$^{\prime\prime}$.2, negligible for the purpose of this
analysis (that employs $\sim$10 times larger apertures to perform
photometry, see section~4).

\subsection{HST/NICMOS Images}

Observations with HST/NICMOS are available for the central region of NGC~5194
in the P$\alpha$ emission line (1.8756~$\mu$m, F187N narrow--band filter) and
the adjacent continuum (F190N narrow--band filter). The image is a 3$\times$3
NIC3 mosaic (GO-7237, P.I.: Scoville) that spans the central
144$^{\prime\prime}$, or the inner $\sim$6~kpc of the galaxy.  Details of the
observations, data reduction, and mosaicing are given in \citet{scov01}.

Because of the proximity in wavelength of the two narrow--band
filters, the line--only image is obtained by subtracting the
continuum--only image, previously rescaled by the ratio of the
filters' efficiencies, from the line$+$continuum image.  The NIC3
0$^{\prime\prime}$.2 pixels undersample the NICMOS PSF, although this
is not a concern for the diffuse ionized gas emission of interest
here.

The continuum--subtracted P$\alpha$ image shows a diagonal tilt in the
background, which is removed by fitting an inclined linear surface to the
image (using the task IMSURFIT in IRAF). The resulting image shows a
relatively flat background. The sensitivity is variable, being lower at
the seams of the 9 images that form the mosaic. The average 1~$\sigma$
sensitivity limit of the continuum--subtracted image is
1.8$\times$10$^{-16}$~erg~s$^{-1}$~cm$^{-2}$~arcsec$^{-2}$.

The region of the galaxy imaged in P$\alpha$ offers a unique
opportunity, in conjunction with the H$\alpha$ image (next section),
to directly probe the impact of dust obscuration on the ionized gas,
and to measure star formation using an indicator (P$\alpha$) weakly
affected by dust. An extinction of 1~mag at V produces an extinction
of 0.15 magnitudes at P$\alpha$, i.e., a small, $\sim$14\% change in
the line intensity. We adopt an intrinsic ratio
H$\alpha$/P$\alpha$=8.734 \citep{oste89}, and differential value
k(H$\alpha$)$-$k(P$\alpha$)=2.08 for the extinction curve. 
The central $\sim$6~kpc of NGC~5194 are characterized by
observed H$\alpha$/P$\alpha$ ratios that are smaller than the
theoretical unreddened ratio, implying attenuations at V in the range
A$_V\sim$1--3.4. In what follows, the central region imaged in
P$\alpha$ will be referred to as the Inner Region, while areas
external to this will be globally referred to as Outer Region.

\subsection{Ground--based Optical Images}

H$\alpha$--centered narrow--band, B--band, and R--band images were
obtained on 28 March 2001, with the Direct Camera at the 2.1--m KPNO
telescope, as part of the SINGS ancillary data program
\citep{kenn03}. Exposure times were 1800~s, 720~s, and 360~s for
H$\alpha$, B, and R, respectively. Standard reduction procedures were
applied to the images.  Both images are mosaics of 2 frames, displaced
along to N--S direction to include both NGC~5194 and NGC~5195. Because
of vignetting along one edge of the camera, correction procedures were
applied; the photometric integrity along the seam of the mosaic was
verified from comparing measurements of stars along the vignetted side
of the mosaic with the same stars on the non--vignetted side. Standard
stars observations were obtained during the observing run to derive
photometric calibrations.

A U--band image of the galaxy obtained on 20 June 2004 with the
Steward 90--inch Prime Focus Camera \citep{willi04} is also used in
this analysis to construct the stellar population age--sensitive color
U$-$B. The final combined U image is the result of 2 dithered images,
with a total exposure time of 1200~s. Photometric calibrations were
also obtained for these observations \citep{engel05}. A comparison of
the calibrated U band image of NGC5194 with the analogous image from
the Sloan Digital Sky Survey indicates a disagreement between the two
calibration scales of 28\%, with our image being bluer than the SDSS
one. We adopt our own calibration for this work, discussing the impact
of the different photometric calibration from SDSS in section~6.1.

The R--band image is rescaled and subtracted from the H$\alpha$ image,
which is then corrected for the contribution of the two
[NII](6548,6584~\AA) emission lines.  We adopt a fixed ratio
[NII](6584~\AA)/H$\alpha$=0.5, although it should be noted that this
ratio covers a wide range in NGC~5194, from $\sim$0.3 in individual
HII regions up to $\sim$1.9 in the diffuse H$\alpha$ component
\citep{hoope03}. The ratio [NII](6584~\AA)/H$\alpha$=0.5 is typical of
the spatially integrated line emission from a metal--rich galaxy
(e.g., compare with M83 in \citet{mcqua95}). For a ratio
[NII](6548~\AA)/[NII](6584~\AA)=0.335, the observed line emission is
1.617$\times$H$\alpha$. The shallow metallicity trend as a function of
galaxy radius \citep{zari94} justifies the use of a single
[NII]/H$\alpha$ ratio for all HII knots in M51.

The absolute photometry of the ground--based H$\alpha$ image is
checked against archival HST/WFPC2 H$\alpha$ images of M~51.  The WFPC2
images cover approximately the same region as the P$\alpha$ mosaic,
i.e., just the inner galaxy region; more details are given in
\citet{scov01}. The HST images are used to check for three effects:
(a) absolute photometry, since our ground--based H$\alpha$ frames were
obtained in marginally photometric conditions; (b) [NII]
contamination, since the ground--based images require a large
correction, while the HST/WFPC2 H$\alpha$ filter (F656N) is narrow
enough that only a few percent of the total flux is due to [NII]
\citep{scov01}; (c) potential oversubtractions in regions with large
H$\alpha$ equivalent widths from using the R--band image (which
includes H$\alpha$ within its bandpass) as underlying continuum, as the
HST continnum images exclude H$\alpha$.

Points (a) and (b) are not independent, and flux comparisons between
12 common, isolated HII regions, with H$\alpha$ fluxes between
5.5$\times$10$^{-15}$ and 1.8$\times$10$^{-13}$ and equivalent widths
between 25~\AA~ and 500~\AA, indicate a mean systematic offset of
about 20\% between the ground--based and the HST images, the former
having lower mean flux than the latter; we correct the ground--based
image for this offset. The dispersion around this value is about
20\%--25\%, and can originate from intrinsic variations of the
[NII]/H$\alpha$ ratio in the HII regions, as observed by
\citet{hoope03}. We do not correct our individual datapoints for this
case--by--case dependent offset, but carry the uncertainties
accordingly.

The 1~$\sigma$ sensitivity limit of our final H$\alpha$ image is
1.8$\times$10$^{-17}$~erg~s$^{-1}$~cm$^{-2}$~arcsec$^{-2}$. The
measured PSF in the optical images is 1.9$^{\prime\prime}$, smaller
than both the GALEX and Spitzer--MIPS data, and comparable to the PSF
of the Spitzer--IRAC data.

\section{The Morphology of the Star Formation Tracers}

The high--resolution (by infrared standards) maps obtained for this
study allow for the first time a comparison of the spatial location of
the emission at each wavelength, from the ultraviolet, through the
optical, to the infrared for this galaxy. A three--color composite of
M51 using three widely used star formation rate indicators
(Figure~\ref{fig1}, left) shows that the FUV, H$\alpha$ and 24~$\mu$m
emission do not always arise from the same regions. In particular, the
FUV radiation emerges predominantly along the outer edges of the
spiral arms, indicating relatively low dust extinction in these
regions, while the FIR dominates the inner edges. The H$\alpha$
emission appears to preferentially follow the infrared emission, down
to a very detailed level, both in knots and in areas of diffuse or
filamentary emission. Presence of filamentary dust emission in the
inter--arm regions is better appreciated in the higher angular
resolution image which combines continuum--subtracted H$\alpha$,
3.6~$\mu$m continuum emission from the aged, diffuse stellar
population, and 8~$\mu$m dust emission (Figure~\ref{fig1}, right). The
complex structure of the dust emission contrasts the relatively smooth
stellar emission from the 3.6~$\mu$m IRAC image, while, quite
expectedly, the H$\alpha$ emission clusters along the spiral arms as
do the brightest knots of dust emission. Along the outermost regions
of the spiral arms of NGC~5194, H$\alpha$ appears relatively
unextincted, while dust emission (and extinction) increases steadily
towards the center. The 8~$\mu$m and 24~$\mu$m images also differ in
the level of contrast between the luminosities of the arms and
inter--arm regions: the contrast is lower, by a factor of 3--4, for
the 8~$\mu$m dust emission than for the 24~$\mu$m emission within the
central 15~kpc of NGC~5194.

How can diagnostics, like the FUV and the FIR, derived from light
emerging at seemingly different locations effectively provide a good
calibration of the star formation?

\section{Multiwavelength Photometry of Star Forming Regions}

\subsection{Aperture Photometry}

For the multiwavelength comparisons that are the main goal of this
work, all images have been registered to the same coordinate system
and pixel scale, using the H$\alpha$ image as reference. The MIPS
24~$\mu$m (MIPS24) images have the lowest resolution, with a PSF
FWHM$\sim$6$^{\prime\prime}$\footnote[1]{Images of observed (IRAC) or
simulated (MIPS) PSFs were downloaded from the SSC Instruments' pages
at: http://sscspitzer.caltech.edu/obs/ ; FWHMs and aperture
corrections (see next section) are calculated from these images.}, and
thus will be driving the minimum spatial scale that can be
investigated. We choose apertures of 13$^{\prime\prime}$ diameter,
which correspond to about 520~pc at the distance of the galaxy.

We perform photometry of 166 circular, 13$^{\prime\prime}$--diameter,
regions in the FUV, NUV, U, B, H$\alpha$, dust--only 8~$\micron$, and
the 24~$\mu$m images, across NGC~5194 (Figures~\ref{fig2} and
\ref{fig3}). The regions are selected primarily as being emission
peaks in the MIPS24 image, though a second pass is made through the UV
images to ensure that bright regions in these are also included. The
apertures are selected to be as much as possible non--overlapping, but
for a few of them some overlap is unavoidable. In these cases, checks
were performed to verify that the flux contained in the overlap region
did not exceed 5\% of the total flux in either of the two apertures.
Of the 166 apertures, 54 are located within the Inner Region probed by
the HST P$\alpha$ image (Figure~\ref{fig3}), and for these, the
P$\alpha$ flux is also measured.

In about half of the regions the infrared emission peak and the UV
emission peak are visibly displaced relative to each other
(Figure~\ref{fig4}). The displacement of peaks is of the order of a
few arcsec, larger than any displacement expected from
mis-registration of the images or from the residual distortions in the
GALEX FUV image ($\approx$1$^{\prime\prime}$, see section~2.2
above). The large apertures still allow us to encompass both UV and
infrared emission, but the measured fluxes are clearly emerging from
slightly different regions. Conversely, there is a high degree of
coincidence, within the accuracy afforded by the images' resolution,
between the infrared, both 8~$\mu$m and 24~$\mu$m, and H$\alpha$
emission peaks.

Because of crowding, background annuli are generally difficult to
define around each aperture, without including a neighboring
region. We thus adopt a different approach for background removal from
the measured fluxes. The 166 apertures are divided into 12 `areas'
(one of these being the region probed by the HST P$\alpha$ image),
where the local background at each wavelength is fitted and globally
removed from each area\footnote[2]{The local background is fitted
interactively in each image and for each region, using the IRAF
routine MSKY written by M. Dickinson (1993). MSKY allows the user to
define the interval in the pixels distribution where the mode and the
variance are calculated. The interactivity of the procedure produces a
robust result even in the absence of source masking.}. Checks
performed on the few isolated apertures that can be identified in the
images indicate that this process of background removal is robust for
the relatively small galactic areas selected (e.g., the Inner Region
corresponds to about 20\% of D$_{25}$). The 12 regions used for local
background fitting are identified as rectangular areas in
Figure~\ref{fig2}. The local background has fairly different values
from region to region; for instance, it changes by a factor close to
10, from $\sim$10$^{-5}$~Jy~arcsec$^{-2}$ to
$\sim$10$^{-4}$~Jy~arcsec$^{-2}$ among the 12 regions in the MIPS24
image. As a reference, in the central region background levels at
8~$\mu$m and 24~$\mu$m represent 32\% and 21\%, respectively, of the
total flux in the region.

The definition of background regions is a compromise between selecting
small enough areas that a `local background' can be defined and, at
the same time, large enough to include enough pixels that a mode can
be robustly derived (see above). Hence, small--level background
variations can still be present within each region. To prevent such
local variations from significantly affecting our analysis, we adopt a
strict definition of `detection': detections are defined as
background--subtracted fluxes that are at least 100\% of the local
background. Below this level, we define them as `upper limits'. Within
this definition, out of 166 apertures, 33 contain upper limits in one
or more bands; 29 of them are in the FUV image. This leaves us with
133 apertures with reliable measurements at all wavelengths. In the
Inner Region, of the 54 regions, 43 have measurements in all six
bands; for 7 regions P$\alpha$ is below our detection threshold, and
for 4 other regions the FUV is also an upper limit. One of the 133
regions coincides with the Seyfert~2 nucleus
\citep{ford85,goad85,tera01,sturm02}, which is excluded from our
analysis. The UV, optical, and near--IR photometry is corrected for
the small foreground extinction from our own Galaxy, E(B$-$V)=0.037
(from the NASA Extragalactic Database). Table~\ref{tbl-1} lists of the
positions and luminosities of the 132 HII knots defined as
`detections'.

For the $U$ and $B$ bands a stricter approach is adopted for
background removal, because of the proportionally larger contribution
and inhomogeneity of the underlying galaxy at these wavelengths. In
this case, background removal is checked for each aperture, and in
case of undersubtraction with the default background regions, smaller,
more appropriate background regions are applied as needed.

The uncertainties assigned to the photometric values are a quadratic
combination of three contributions: variance of the local background,
photometric calibration uncertainties, and variations from potential
mis-registration of the multiwavelength images. The variance on the
local background is derived from the original--pixel--size images,
after projecting each rectangular region on the original images. The
effect of potential misregistrations are evaluated by shifting the
images by 1$^{\prime\prime}$.2 (the magnitude of the residual
distorsions in the GALEX images, section~2.2) relative to each
other. Because of the large apertures employed for the photometry,
this contribution is either small (a few \%~ of the total uncertainty)
or negligible in all cases. An additional contribution to the
uncertainties assigned to the wavelength--integrated infrared
luminosities is discussed in section~4.3.

\subsection{Additional Sources of Uncertainty}

The star forming regions that are being studied here can be
considered, for all purposes, point sources at the MIPS24 resolution;
the aperture corrections are fairly substantial despite the relatively
large apertures (factor 1.67). Aperture corrections are smaller at
shorter wavelengths: about 6\% and 10\% for point sources in the the
FUV and NUV GALEX images\footnote[3]{Aperture corrections for the GALEX
photometry were measured from point sources contained in the GALEX 
images of NGC5194.}, respectively, and negligible for the ground-based, HST,
and IRAC images.

The large MIPS24 PSF can also lead to contamination of the aperture
photometry by nearby sources. The photometry in an aperture centered
13$^{\prime\prime}$ {\em away} from a source will include on average
4\% of the flux from the contaminating source. Photometry of the
target source will be affected in proportion to the flux ratio between
the two sources, which can be a significant fraction of the target's
flux if the contaminating source is significantly brighter than the
target. There are about a dozen apertures in our sample of 132 that
contain sources at least twice as faint as the adjacent one, and for
which the impact from the neighboring source on the 24~$\mu$m flux is
8\% or larger. Tests run using samples with or without data from these
apertures have produced results that are nearly identical for the
trends described in the next sections. This suggests that the
influence of those `affected' apertures is negligible on the general
trends. Use of the PSF--fitting method to improve photometry of
sources in crowded regions for our lower resolution images is being
investigated for possible application to future multi--wavelength
analyses of the SINGS galaxies.

Our analysis is based on the assumption that in each single region,
the emission at any wavelength is due to the stellar population and
dust located in that region. Effects of radiation transfer could in
principle be important for the IR measurements, as dust in a region
could be heated by UV photons produced outside that region. However:
(1) in general, local IR peaks have a one--to--one correspondence to
local H$\alpha$ peaks in our images, across the entire galaxy's disk
and center; although heating from out--of--region UV photons cannot be
excluded, the observed correspondence suggests that most of the
heating in those peaks is produced locally. (2) The subtraction of
`local' backgrounds removes IR flux contributions from the heating by
the diffuse stellar population. We thus conclude that for our purposes
we can assume the multi--wavelength emission in each aperture to be
due mainly to local stellar populations and locally--heated dust.

\subsection{Derivation of the Infrared Luminosities}

The 13$^{\prime\prime}$ diameter apertures, although already large by
`HII-regions-size' standards, are too small to contain a significant
fraction of the 70~$\mu$m (MIPS70) or 160~$\mu$m (MIPS160)
PSFs. Therefore, these images cannot be used directly to measure the
fluxes of our HII knots at the longer wavelengths, raising the problem
of estimating total far--infrared luminosities. Conversely, choosing
apertures appropriate for photometry in the MIPS70 and MIPS160 images,
i.e, $>$50$^{\prime\prime}$, would hamper any attempt to investigate
the properties of the young stellar populations, the gas they ionize,
and the dust they heat. Such apertures correspond to physical sizes
2~kpc or larger at the distance of M51, thus probing significant
fractions of the galaxy's integrated population.

We thus determine total infrared luminosities for our HII knots,
L(IR)=L(3--1100$\mu$m), by exploiting the correlation between the
8~$\mu$m--to--24~$\mu$m flux ratio and the 24~$\mu$m--to--total
luminosity ratio \citep{dale02}. We derive the best fit relation for
our HII knots as follows. Photometry in 21 70$^{\prime\prime}$
diameter apertures is performed on the 8~$\mu$m dust--only image and
on the 3 MIPS images; the 21 regions are selected to target peaks of
70~$\mu$m emission in the center of the galaxy and along the spiral
arms. The diameter of the regions, corresponding to $\sim$2.8~kpc, are
selected to encompass 65\% of the light from the 160~$\mu$m PSF; the
same aperture contains 100\%, 94\%, and 88\% of the light from the
PSFs at 8~$\mu$m, 24~$\mu$m, and 70~$\mu$m, respectively. Local
background values are subtracted from each region to minimize
contamination from dust heated by the the diffuse stellar
population(s). Six different local background regions are selected to
`hug' as close as possible small groups of apertures, using an
approach similar to the one described in section~4 above. The shape of
such background regions is rectangular, with the longer side aligned
as much as possible along the direction of the spiral arm containing
the photometric apertures. L(IR) for these datapoints is integrated
from the MIPS band measurements using equation~(4) of \citet{dale02};
this relation between the MIPS bands and the total infrared emission
is still applicable to our case, as the range of colors of the 21
regions, $-0.5\lesssim Log[L(8)/L(24)]\lesssim 0.1$ and $0.2\lesssim
Log[L(70)/L(160)]\lesssim 0.7$, is within the range of spectral energy
distributions parametrized by those authors.

The resulting plot Log[L(24)/L(IR)] versus
Log[$L_{\nu}(8)/L_{\nu}(24)$] (where L(24)=$\nu
L_{\nu}$(24~$\mu$m)\footnote[4]{From now on, the convention
L(band)=$\nu L_{\nu}$(band) is adopted for broad--band flux
measurements, where the band can be any of the GALEX, optical, IRAC,
or MIPS bands. These `monochromatic' luminosities are in units of
erg~s$^{-1}$.}) for the 21 regions is shown in Figure~\ref{fig5},
together with the best fit line through the data points; the model of
\citet{dale02}, appropriate for whole galaxies, is also shown for
comparison. The datapoints of Figure~\ref{fig5} are systematically
higher than the whole--galaxies predictions of Dale et al., as can be
expected if the 24~$\mu$m luminosity is proportionally a larger
fraction of the total infrared luminosity in the HII knots than in
whole galaxies, and the IR spectral energy distributions are typical
of hotter dust. Since we have subtracted the diffuse disk emission
from the aperture measurements, we expect the longer wavelength
(`cirrus') emisssion to be depressed relative to the case of entire
galaxies emission, as the cirrus emission can be heated by the diffuse
stellar field \citep{helo86,boul88}. Indeed, the location of the
emission from the whole galaxy in Figure~\ref{fig5} is along the
`lower envelope' of the locus defined by the HII knots, in line with
the above reasoning. The full discussion of this aspect of the IR
spectral energy distribution of NGC~5194 and other local star--forming
galaxies will be undertaken elsewhere \citep{dale05}.

The best fit straight line through the datapoints of Figure~\ref{fig5}
provides the following relation:
\begin{equation}
Log[L(IR)] = Log[L(24)]+0.908+0.793 Log[L_{\nu}(8)/L_{\nu}(24)],
\end{equation}
which we adopt in the following as our baseline relation for deriving
total infrared luminosities for the 13$^{\prime\prime}$--diameter
apertures. The scatter around the best fit line is $\pm$40\%, and we
factor this scatter in the uncertainty budget of L(IR). According to
this relation, the 24~$\mu$m luminosity represents 7\%--21\% of the
total infrared luminosity. The datapoints in Figure~\ref{fig5} cover a
somewhat smaller 8/24~$\mu$m ratio range than that covered by the data
in the smaller apertures
($-$0.4$\lesssim$Log[L$_{\nu}$(8)/L$_{\nu}$(24)]$\lesssim$0.2); we
assume that our best fit line can be extrapolated 0.2~dex towards
smaller 8/24~$\mu$m ratios, to include all the smaller aperture
datapoints.

Our selection of local background values for the
70$^{\prime\prime}$--diameter apertures should still be considered a
`best attempt', as the typical background region has sizes
$\sim$4$\times$8~kpc$^2$, thus encompassing a significant fraction of
the galaxy's population. We, thus, cannot exclude that the data in
Figure~\ref{fig5} and equation~(1), and in particular the 160~$\mu$m
measurements, are partially contaminated by cirrus emission extraneous
to the HII blob infrared emission. In what follows, we assume that
this contamination represents a small fraction of the total infrared
emission.

\section{Observed Properties of the Star--Forming Regions}

The selected 166 star--forming regions cover a factor $\sim$300 in
24~$\mu$m luminosity (L$\sim$10$^{39}$--10$^{41.5}$~erg~s$^{-1}$), and
$\sim$100 in FUV luminosity (Figure~\ref{fig6}). The knots in the
Inner Region tend to be overluminous at 24~$\mu$m for constant FUV
luminosity relative to the knots in the Outer Region, and to have on
average redder UV colors (section~6); not surprisingly, this reflects
higher dust extinction in the Inner Region relative to the
Outer. There is no significant difference in the distribution of the
L$_{\nu}$(8)/L$_{\nu}$(24) flux ratios or in the mean H$\alpha$
luminosity between the Inner and Outer Regions.

None of the selected knots, not even the UV--selected ones, is
consistent with the UV colors expected for a ionizing, dust--free
stellar population (younger than $\sim$10--12~Myr,
Figure~\ref{fig7}). Nearly all of the knots contain at least a small
amount of infrared emission implying presence of dust. Stellar
population ageing is also a possibility, especially for those cases
where there is a displacement between infrared and UV peaks within the
photometric apertures.

For both the Inner and Outer Regions, there is a clear trend for more
luminous H$\alpha$ regions to have hotter FIR spectral energy
distributions (lower 8/24~$\mu$m flux ratios, Figure~\ref{fig8}). This
correlation persists when extinction--corrected H$\alpha$ luminosities
are used (see section~6.1). As will be seen later, the anticorrelation
between the IR colors and the H$\alpha$ luminosity is due to the
8~$\mu$m luminosity becoming underluminous for increasing H$\alpha$
luminosity.

\section{Dust Extinction Properties}

\subsection{The Impact of Dust and Age on the UV}

The ratio of the infrared to far--UV luminosities is a measure of the
total dust opacity experienced by the UV stellar continuum in a
region, and this quantity has been shown to correlate tightly with the
UV colors of starburst galaxies \citep{meur99}. In NGC~5194, the total
UV opacity L(IR)/L(FUV) is also related to the UV colors for the 132
HII knots, but with a significantly larger scatter than in the case of
starbursts (Figure~\ref{fig9}, left).  In particular, at fixed UV
color, the IR/FUV ratio of the HII knots spans about one order of
magnitude larger range than the IR/FUV ratio of starburst galaxies. In
addition, the locus identified by the starburst galaxies in the
IR/FUV--UV~color plane represents the {\em upper IR/FUV envelope} to
the HII knots; with few exceptions, the HII knots in NGC~5194 have
IR/FUV ratios that are lower than those of starbursts, at fixed UV
color \citep[see, also][]{bell02b}. However, even among the HII knots
there are no regions that can be at the same time red and
IR--faint. Indeed, despite the large scatter, there is still a good
correlation between the HII knot datapoints; a non--parametric rank
test indicates that the correlation is significant at the 7.2~$\sigma$
level. Such level of correlation implies that the UV reddening still
follows the total UV opacity, albeit with a different slope and
scatter than starbursts.

To remove any doubt on the location of the HII knots' datapoints
relative to those of the starburst galaxies, we investigate whether
there may be an impact on such location from the way L(IR) is
calculated for the HII knots in NGC~5194, i.e., using the
approximation of equation~(1). The right--hand side plot of
Figure~\ref{fig9} shows the same IR/FUV versus UV color plot, where
L(IR) is replaced by the directly measured L(24). Data for 29
starburst galaxies from the sample of Calzetti et al. (1994) are
reported on the same plot, using the IRAS~25~$\mu$m measurements in
lieu of MIPS24. Again, the relation for the starburst galaxies and the
NGC~5194 knots are offset, with the starbursts defining the upper
envelope of the correlation.

In Figures~\ref{fig9}--\ref{fig12} the data are compared with
stellar populations models convolved with dust extinction
models. Stellar population models are from Starburst99 \citep{leit99},
either instantaneous bursts or constant star formation, with solar
metallicity (to roughly match the metallicity of NGC~5194) and
Salpeter stellar Initial Mass Function in the range
1--100~M$_{\odot}$. Dust models employ both the Milky Way \citep[MW,
with R$_V$=3.1, ][]{card89} and Small Magellanic Cloud \citep[SMC,
][]{bouc85} extinction curves. The dust geometries investigated
include foreground, non--scattering dust screens, homogeneous mixtures
of dust, gas, and stars, and the starbursts' dust distribution
\citep{calz94,meur99,calz00}. The latter is equivalent to a clumpy
shell surrounding the starburst volume, where the ionized gas suffers
about twice the attenuation of the stellar continuum
\citep{calz01}. Colors and luminosities are obtained by convolving the
stellar plus dust models with the appropriate filter's passband; the
infrared luminosity is calculated assuming that the scattering
component of the extinction averages out, and all extincted stellar
light is re-emitted by dust in the infrared.

The UV reddening of the starbursts is originally expressed as their
spectral slope $\beta_{26}$, defined as the UV slope in the
0.13--0.26~$\mu$m range (f($\lambda$)$\propto\lambda^{\beta}$,
\citet{calz94}); $\beta_{26}$ has been converted to equivalent GALEX
UV colors using the formula:
\begin{equation}
Log[L_{\lambda}(FUV)/L_{\lambda}(NUV)] = -0.1688 \beta_{26} - 0.0177,
\end{equation}
where L$_{\lambda}$(FUV) and L$_{\lambda}$(NUV) are luminosity densities 
expressed in units of erg~s$^{-1}$~\AA$^{-1}$; negative values of
$\beta_{26}$ correspond to blue UV colors. The conversion between
$\beta_{26}$ and the GALEX UV colors is not unique, and depends on the
physical parameter driving the color variation. Equation~(2) is valid for
variations in color due to variations in dust reddening, as is the
case for the starburst sample of \citet{calz94}. Were the variation 
driven by the ageing of a dust--free stellar population, the
conversion formula would be:
\begin{equation}
Log[L_{\lambda}(FUV)/L_{\lambda}(NUV)] = -0.1435 \beta_{26} + 0.04102,
\end{equation}
where $\beta_{26}$ is measured in the age range 2--300~Myr.  

To assess the origin of the large scatter for the
IR/FUV--UV~colors of the HII knots in NGC~5194 and their
variance relative to starburst galaxies, we investigate two possible
origins: (1) presence of ageing, albeit UV--emitting, stellar
populations; (2) differences in the dust geometries of the starbursts
and the HII knots.

\subsubsection{Stellar Population Ageing}

Although each of the HII knots is a region $\sim$520~pc in size, and
may encompass more than one stellar cluster (single--age population),
we still attempt to model the {\em dominant source of UV emission} in
each aperture as an instantaneous burst population. Figure~\ref{fig10}
shows the effect of including stellar population ageing in the
IR/FUV--UV~colors plot. The model lines in Figure~\ref{fig10} are
obtained by combining instantaneous burst populations (Starburst99,
\citet{leit99}) with the starburst opacity curve
\citep{calz94,calz01}. About 92\% of the HII--emitting knots in
NGC~5194 have their IR/FUV ratios and UV~colors well described by
instantaneous burst populations in the age range 2--300~Myr, provided
their dust attenuation and reddening characteristics are similar to
those of the starburst galaxies. Within this model, the maximum
opacity experienced by the HII knots corresponds to
A$_V\lesssim$2.8~mag\footnote[5]{In the context of this work, A$_V$
will be used only to refer to the attenuation of the ionized gas. If
the dust geometry in the region follows the starburst attenuation
curve, the attenuation at $V$ appropriate for the {\em stellar
continuum} is A$_V^{star}$=0.44~A$_V^{gas}$, where in our convention
A$_V^{gas}\equiv$A$_V$. This convention will be maintained also in the
case of evolved stellar populations, i.e. populations older than
$\approx$10~Myr, where no ionized gas is expected to be present.}.

Incidentally, the UV--selected knots have generally low IR/FUV colors (low
dust opacity) and, at the same time, red UV colors, suggesting that even in
this specific case, the UV--bright knots are consistent with ageing stellar
populations.

For the Inner Region, the H$\alpha$/P$\alpha$ line ratio offers an
independent measure of dust reddening to further investigate the
properties of the knots. The UV colors of knots in this region show a
trend to be redder for higher extinction A$_V$ (Figure~\ref{fig11},
left), as expected if dust is present.  The points show a fairly large
spread at fixed A$_V$. For a starburst opacity curve, this spread
cannot be reproduced by dust reddening alone; rather, for each value
of A$_V$, a spread in age, from 2 to $\sim$300~Myr is present. In this
plot, the HII knots in the NGC~5194 Inner Region do not appear
qualitatively different from the starburst galaxies, in the sense that
the age spread at fixed A$_V$ appears similar in both samples. This
`similarity' is likely due to the small dynamical range of the UV
colors, and is broken once a longer wavelength baseline is
considered. The FUV/P$\alpha$ luminosity ratio spans indeed a narrower
range, at fixed dust opacity, for the starburst galaxies than for the
NGC~5194 Inner Region knots (Figure~\ref{fig11}, right). We then use the
U$-$B color to better constrain the age spread of the latter.

The Log[L$_{\lambda}$(U)/L$_{\lambda}$(B)] color is a sensitive age
indicator, because it straddles the 4000~\AA~ break of stellar
populations. This color is not as dust--insensitive as the D(4000) age
indicator, extensively used in SDSS studies \citep{kauff03}, but it is
far less sensitive than any of the colors we have used so far. In
particular, Log[L$_{\lambda}$(U)/L$_{\lambda}$(B)] has more than twice
the dynamical range in age, and less than half the sensitivity to dust
extinction than the GALEX UV colors (for a non--MW dust,
Figure~\ref{fig12}, left). The extinction--corrected
Log[L$_{\lambda}$(U)/L$_{\lambda}$(B)] colors of the Inner Region's
knots indicate an age range 3--100~Myr (Figure~\ref{fig12}). This age
range would change to $\sim$5--200~Myr, if the SDSS calibration for
the U--band image of NGC~5194 were adopted (section~2.4). Changing
assumptions on the dust extinction curve or geometry will only
minimally affect the corrected U$-$B colors, implying that the knots
span a considerable age range, from a few Myr to 100--200~Myr. The
observed U$-$B colors of the apertures in the Outer Region are
consistent with those of the Inner Region, showing a similar spread in
age. Thus, the different behavior of starburst galaxies and NGC~5194
HII knots in the IR/FUV--UV~color plane is driven by ageing of the
stellar populations in the knots.

The question arising at this point is: how are the UV emission and the
ionized gas emission related to each other? Any knot with UV or U$-$B
color ages older than $\sim$10--12~Myr should not be producing
detectable ionized gas emission \citep{leit99}. This conundrum can be
reconciled if each aperture, covering $\sim$520~pc in the galaxy,
contains multiple stellar populations; within each region, these
populations are likely to cover a range of ages, masses, and dust
extinctions. The stellar populations which dominate the ionized gas
emission are not always the same ones that dominate the UV (and longer
wavelengths) stellar continuum emission. As an example,
Figures~\ref{fig10}--\ref{fig12} show the colors and flux ratios
expected for a model obtained by combining two stellar populations: a
5~Myr~old burst and a 300~Myr~old burst, reddened by the starburst
opacity curve. In this specific example, the 5~Myr~old cluster has a
mass 300 times lower than the 300~Myr~old cluster (section~8.2), and
an excess extinction $\Delta$A$_V$=$+$0.25~mag relative to the older
cluster (section~8.1.1), except for A$_V$=0, where both populations
are extinction--free. The tracks and colors identified by this very
simple 2--populations model account for the intermediate--to--old
range of ages for the datapoints in the
Figures~\ref{fig10}--\ref{fig12}. Varying the difference in age,
mass, and dust extinction, and the number of separate
populations, will likely account for the full observed range of colors
of the HII knots. The findings of section~4, where it was noted that
in about half of the knots the UV and line emission peaks are
spatially displaced relative to each other further support the case
for the presence of multiple stellar populations.

\subsubsection{Variations in Dust Geometry}

Although age differences in the stellar populations dominating the
knots' UV emission appear to be likely responsible for the
differences between starburst galaxies and NGC~5194 HII knots in
Figure~\ref{fig9}, we still need to investigate whether differences in
dust geometry between starbursts and the quiescently star--forming
galaxy could also contribute to the effect. 

The Galactic (MW) extinction curve fails to reproduce either the UV
colors or the IR/FUV ratios of the HII knots (Figures~\ref{fig9} and
\ref{fig11}), for both foreground dust or mixed dust--stars--gas
geometries. We take here the foreground and mixed distributions as two
extremes of a continuum of possible dust geometries. A more extreme
geometry than the mixed one is the Cloudy model \citep{witt00},
where a dust--free stellar population is located in front of a mixed
dust--stars distribution. However, the purely mixed geometry is
sufficient for our current purposes.  The clear answer we can gain on
the MW extinction curve is due to the characteristics of the GALEX
filters. The GALEX NUV filter is centered on the 2200~\AA~ bump of the
MW extinction curve, and the differential reddening relative to the
FUV filter is small. Hence, UV colors tend to be insensitive to
effects of dust extinction for a MW curve, irrespective of
geometry. The fact that, however, the HII knots show a clear trend to
have redder colors for higher opacity values is indicative that the
extinction curve in NGC~5194 has a weaker 2200~\AA~ feature than the MW
dust.

A Small~Magellanic~Cloud--like extinction curve appears to be more
adequate at explaining the datapoints in the UV~colors--A$_V$ plane
(Figure~\ref{fig11}). The HII knots data are bracketed between the SMC
mixed dust and SMC foreground dust geometries (in particular the
foreground geometry with A$_V^{star}$=0.5~A$_V^{gas}$, indicated as
SMC(0.5*A$_V$) in Figure~\ref{fig11}). However, the same geometries
are inadequate to account for the observed range of IR/FUV
(Figure~\ref{fig9}).  In particular, a considerable number of HII
knots has systematically too low a IR/FUV ratio at any given UV color
even for the foreground SMC dust, and even after taking into account
photometric undertainties.

Thus, although variations in the dust geometry between starbursts and
the HII knots cannot be completely excluded \citep{bell02b}, they are
unlikely to be the main drivers of the observed scatter below the
starburst curve mean trend (Figure~\ref{fig9}). Age variations between
the UV--emitting populations are more likely to be the second
parameter to the relation (Figures~\ref{fig10} and \ref{fig11}).

\subsubsection{Dust Opacity and Star Formation}

The total IR$+$FUV luminosity and the dust opacity, expressed as
IR/FUV (Figure~\ref{fig13}), are correlated for the HII knots in the
Inner and Outer Regions, with a 5.1~$\sigma$ significance, using a
non--parametric rank test. For actively star--forming galaxies, the
sum of the infrared and far--UV luminosities is for all purposes the
total UV light from the region \citep[both direct and dust--absorbed,
][]{wang96, heck98}, which is proportional to the star formation rate
\citep{kenn98}.  The existence of a such a correlation for the
NGC~5194 HII knots testifies to the dominant role of dust extinction
over population ageing trends, the latter being a secondary
parameter. Incidentally, the best fit line determined by
\citet{heck98} for starburst galaxies is also a reasonable fit to the
knots, modulo a vertical rescaling to account for the intrinsic
faintness of the HII knots relative to the starbursts
(Figure~\ref{fig13}). The peak--to--peak spread, $\sim$1~dex, around the mean
behavior is, as seen in the previous section, the effect of the
multiple--age stellar populations contributing to the emission in each
aperture, together with possible contribution from spatially variable
dust geometries. Interestingly, this spread is still smaller than the
one observed for starburst galaxies \citep{heck98}, owing to the more
homogeneous nature of our HII knots.

A tantalizing characteristic of Figure~\ref{fig13} is that the
majority of the UV--selected datapoints is located to the left of the
best fit correlation, and offset from the main locus of the other
data. This is expected if the UV emission in those regions is from
evolved stellar populations (Figure~\ref{fig10}), unrelated to the
current star formation. In such a case, the IR/FUV ratio decreases and
the IR$+$UV luminosity increases because of the addition of unrelated
UV emission, in a way that tends to push the datapoints away from the
main locus of all the other data.

\subsection{Radial Trends of Dust and Age}

The {\it median} A$_V$ decreases as a function of the distance from
the nucleus, while the {\em median} UV color becomes bluer
(Figure~\ref{fig14}). The trend for A$_V$ is reported for both our
baseline 13$^{\prime\prime}$--diameter apertures and the
4$^{\prime\prime}$--diameter apertures. The scatter around the median
A$_V$ also increases for larger nuclear distances. Indeed, regions as
reddened as the very central ones are present at 3~kpc as well, but
for increasing nuclear distance the number of regions with very low
extinction increases as well. In particular, while the area about
500~pc from the nucleus is characterized by A$_V$ values in the range
2.8--3.6~mag, the region $\sim$3~kpc away has A$_V$ values in the
range 1--3.4~mag. The knots closer to the center also display on
average lower MIR color ratio L(8)/L(24) (hotter IR SEDs) than the
more distant knots, but as in the case of A$_V$ the scatter also
increases as a function of distance. The UV colors follow the general
trend of A$_V$, but their scatter around the median remains roughly
constant with galactocentric distance, contrary to the increase of
scatter in both A$_V$ and L(8)/L(24).

The tighter scatter displayed by the UV colors as a function of
galactocentric distance is due to the combined effects of decreasing
dust attenuation and decreasing mean age of the HII knots
(Figure~\ref{fig15}, left). The extinction--corrected
L$_{\lambda}$(U)/L$_{\lambda}$(B) color shows indeed that mean ages
decrease as a function of distance from the nucleus, from $<$100~Myr
close to the center down to $<$20~Myr at a distance of $\sim$3~kpc from
the center.  While the knots closer to the nucleus are older than the
more distant knots, they are still more luminous in P$\alpha$
(Figure~\ref{fig15} right), suggesting that: (1) multiple--age
populations co-exist within each knot; (2) the inner knots are more
massive (at fixed age) than the more distant knots.

The distribution of A$_V$ values has median value $\sim$2.6~mag for
the 13$^{\prime\prime}$ apertures and $\sim$2.8~mag for the
4$^{\prime\prime}$ apertures (Figure~\ref{fig16}). For comparison,
\citet{scov01} measure a median value A$_V\sim$2.9, in
1$^{\prime\prime}$ apertures, with extinction values as high as
A$_V\sim$6~mag, after rescaling for the slightly different intrinsic
value of H$\alpha$/P$\alpha$ used in this paper. The trend for larger
apertures to yield smaller values of A$_V$ and narrower distributions
is a well known effect, due to the higher level of blending between
HII regions of different A$_V$ values and increasing amounts of
diffuse gas included in the larger apertures. An analysis of the
P$\alpha$ image shows that the level of blending is from a few to many
tens of HII regions in each aperture. Our observed range of A$_V$ is
$\sim$1--3.6~mag for the 13$^{\prime\prime}$ apertures, consistent
with the range measured by \citet{vander88} using the H$\alpha$/radio
ratio in apertures of comparable size. In NGC~5194, the diffuse gas
suffers from a lower mean extinction than the HII regions,
A$_V\approx$2~mag, as shown in \citet{scov01}.

\section{The Infrared Emission as a Star Formation Rate Indicator}

The high angular resolution of the Spitzer images enables a detailed
comparison of the infrared emission with the ionized gas emission for
the HII knots. Both types of emission are used as SFR tracers in
star--forming/starburst galaxies: the infrared emission measures the
dust-reprocessed stellar continuum emission from massive stars, while
the ionized gas emission is proportional to the number of ionizing
photons \citep[e.g.,][]{kenn98,kew02}. In principle, for the infrared
emission to be an accurate tracer of SFR, all of the UV light from
massive stars needs to be absorbed by dust \citep{kenn98}.  The large
dust attenuation (A$_V>$1~mag) and IR/FUV ratio values of the Inner
Region indicate that the approximation L(IR)$\sim$L(FUV) is legitimate
in the center of NGC~5194. In this region, the P$\alpha$ emission
line, which is only modestly impacted by dust obscuration, is used to
measure the number of ionizing photons.

The infrared luminosity of the Inner Region's HII knots
correlates tightly with the extinction--corrected P$\alpha$ luminosity
(Figure~\ref{fig17}), with a linear best fit:
\begin{equation}
Log [L(IR)] = (0.90\pm 0.07) Log[L(P\alpha)] + (7.4 \pm 2.8). 
\end{equation}
This relation is about 1.5~$\sigma$ discrepant from a slope of unity;
the error bars on the individual datapoints are large
(Figure~\ref{fig17}) and mainly driven by the uncertainties in
equation~(1). The deviation of the slope of equation~(4) from unity is
small, and it is in the direction expected if the infrared emission
from fainter HII knots is proportionally more contaminated by
contributions from evolved populations present in the aperture than
brighter HII knots are. These evolved populations are unrelated to the
current SFR probed by the ionizing photons. This effect is likely
impacting also the integrated infrared emission of the Inner Region,
which is larger than what predicted by either the best fit line or the
unity--slope line (Figure~\ref{fig17}).

The sum of the infrared and FUV light has been used as a SFR tracer
for star--forming/starburst galaxies, as it measures all of the
available ultraviolet light from massive stars, both directly observed
and dust-reprocessed \citep{wang96,heck98}. This appears not to be the
case in the Inner Region of NGC~5194, where the FUV emission is
probing a range of stellar populations, including ageing,
non--ionizing ones (section~6.1.1). For the HII knots, although
L(IR$+$FUV) is correlated with L(P$\alpha$), the best fit line is:
\begin{equation}
Log [L(IR+FUV)] = (0.81\pm 0.07) Log[L(P\alpha)] + (10.1 \pm 2.8), 
\end{equation}
with a slope much shallower than unity ($\sim$3~$\sigma$) than the one
of equation~(4). The contribution of the UV light mainly affects the
data at the faint IR end, hence the shallow slope. L(IR$+$FUV) is
probing star formation on a longer timescale ($\approx$100~Myr) than
L(P$\alpha$) ($\approx$10~Myr).

The P$\alpha$ and 24~$\mu$m luminosities of the HII knots show a tight
correlation (Figure~\ref{fig18}), and a linear fit through the data of
the 42 regions gives:
\begin{equation}
Log [L(24)] = (1.03\pm 0.04) Log[L(P\alpha)] + (0.9\pm 1.3),
\end{equation}
after removing the faintest 24~$\mu$m datapoint. This fit is not
significantly discrepant from a slope of unity. In terms of SFRs,
equation~(4) implies that the 24~$\mu$m emission is as good an
indicator as the P$\alpha$, with a typical dispersion around the
median of $\sim$0.2~dex. The extrapolation of equation~(4), which is
derived for the HII knots, to the integrated P$\alpha$ luminosity of
the Inner Region reproduces the 24~$\mu$m luminosity of this region
(Figure~\ref{fig18}), as expected if the two luminosities closely
trace each other. Equation~(6) quantifies the close spatial correlation
between 24~$\mu$m emission and star forming regions found 
by \citet{helo04} in the disk of NGC~300. 

Although equation~(6) provides so far the best linear correlation
between infrared luminosities and L(P$\alpha$), some care should be
taken in concluding that the 24~$\mu$m luminosity provides a reliable
SFR tracer for galaxies in general. The ratio L(24)/SFR is constant
within the relatively uniform environment of the central region of
NGC~5194, and provides a locally accurate SFR tracer. However, it also
changes from galaxy to galaxy. In particular, the locus in L(24)/SFR
ratio identified by UV--selected starbursts and by ULIGs is
systematically higher than the ratio from the NGC~5194 HII knots
(Figure~\ref{fig19}). As will be discussed later, local conditions can
determine the strenght of the 24~$\mu$m luminosity relative to SFR and
possibly account for the observed variability from galaxy to galaxy
\citep{dale01}.

At mid--infrared wavelengths, the 8~$\mu$m dust luminosity also
correlates tightly with the P$\alpha$ luminosity, implying that
regions brighter at one wavelength are also brighter at the other, but
the slope is significantly discrepant from unity (Figure~\ref{fig20},
left). A linear fit gives:
\begin{equation}
Log[L(8)] = (0.79\pm 0.02) Log[L(P\alpha)] + (10.6\pm 0.7),
\end{equation}
after removing the faintest 8~$\mu$m point.  The fitted slope is about
10~$\sigma$ away from unity. The extrapolation of the fit to the
P$\alpha$ value of the entire Inner Region would underestimate the
8~$\mu$m luminosity for this region by a factor 3.2.

The fairly large, 13$^{\prime\prime}$, apertures used to derive equation~(7)
include, potentially, a significant fraction of diffuse emission in the 
8~$\mu$m photometry. This may artificially increase the 8~$\mu$m emission of
fainter regions, thus flattening the overall trend. However, the same fit
repeated for photometry in the 78 smaller, 4$^{\prime\prime}$--diameter,
apertures provides a fit with slope $(0.80 \pm 0.02)$, which is within the
uncertainties, identical to the slope of the larger apertures fit. The
datapoints for the two sets of apertures are in remarkable agreement
(Figure~\ref{fig20}), despite the large aperture correction, a factor 1.75,
required by the 8~$\mu$m measurements in the smaller apertures.

For the small aperture photometry, one concern is of a potential
artificial increase of the 8~$\mu$m flux of those faint regions
adjacent to bright regions, because of the flux contribution from the
PSF wings of the bright region (section~4.1). In particular,
photometry in an aperture centered 4$^{\prime\prime}$ away from a
bright source would include on average 5.5\% of the bright neighbor's
flux, thus impacting measurements of any faint source in the
aperture. We have repeated the fit on the 4$^{\prime\prime}$--diameter
datapoints after removing all sources located less than 2~diameters
away from sources at least twice as bright. The remaining datapoints
(72) give a best fit slope $(0.83 \pm 0.03)$, very similar to the
slope of the entire sample of 78 sources.

The 8~$\mu$m luminosity still shows a very significant deviation from a slope
of unity in the combined Inner$+$Outer regions, when plotted as a function of
the 24~$\mu$m luminosity, used as a proxy for the P$\alpha$ luminosity
(Figure~\ref{fig20}, right). Again, the linear fit on all 132 HII knots is:
\begin{equation}
Log[L(8)] = (0.78\pm 0.02) Log[L(24)] + (9.4\pm 0.7).
\end{equation}
The extrapolation of this relation to the 24~$\mu$m luminosity of the
whole NGC~5194 would lead to an underestimate of its measured 8~$\mu$m
dust luminosity by a factor $\sim$5.1.  

In all cases (Figures~\ref{fig17}--\ref{fig20}), the integrated
luminosity for the Inner Region is measured on the local
background--subtracted images. The integrated 24~$\mu$m and 8~$\mu$m
emission is 14\% and 57\%, respectively, larger than the sum of the
individual HII knots. The 24~$\mu$m emission is thus almost entirely
concentrated in the HII knots, with a small excess due to the faint
knots discarded from our sample. Conversely, about a third of the
8~$\mu$ emission is outside the knots. This `diffuse' emission is
highly concentrated along filaments and is not the result of
insufficient background subtraction.

The use of the extinction--corrected (and intrinsically mildly
extinction--impacted) P$\alpha$ has a large effect on the slopes of
equations~(4)--(7). For instance, use of the uncorrected H$\alpha$
luminosity instead of the P$\alpha$ would change the slope of
equation~(7) to $(0.95 \pm 0.05)$, very close to unity \citep[see,
also,][]{dale05}. Even in the absence of extinction corrections, a
correlation between any of the above infrared luminosities and
L(H$\alpha$) is preserved, owing to the fact that more actively star
forming regions are also more extincted (Figure~\ref{fig13}). However,
absent or insufficient extinction corrections will have the effect of
artificially increasing the slope between the infrared luminosity and
the recombination line luminosity.

\section{Discussion}

The large wavelength range and the detailed spatial scale covered by
the present body of observations of NGC~5194 has enabled a detailed
investigation of the strenghts and limitations of various SFR indicators,
and of the impact of dust obscuration on measurements of such indicators.

\subsection{The Impact of Dust Obscuration}

\subsubsection{The Observed Trends}

Quiescently star--forming galaxies do not follow the IR/UV--UV~colors
of starburst galaxies
\citep{bua02,bell02,gord04,kong04,bua05,seib05}: they tend to cover a
much broader range in IR/UV at fixed UV color, spreading towards lower
IR/UV values than starbursts.

We recover a similar trend in the IR/UV--UV~colors plane for
individual HII knots ($\sim$500~pc in diameter) within a single
star--forming galaxy. The knots show a much broader trend in the
IR/FUV--UV~colors plane than starbursts, and the latter form an `upper
envelope' to the knots \citep[Figure~\ref{fig9}, see also
][]{bell02b}. Analysis of various photometric and color
characteristics of the knots shows that the broad spread ($\sim$1~dex
peak--to--peak) is likely due to a spread in age of the UV--emitting
population(s), between 2 and $\lesssim$100~Myr. None of the
investigated knots, each encompassing $\sim$500~pc, can be thought of
as containing a single age population. Most do contain multiple age
populations, with the young population component responsible for
ionizing the gas being, in many cases, spatially and temporally
separated from the dominant UV--emitting population.

The general trends observed in Figure~\ref{fig9} and
Figure~\ref{fig13} are still driven by effects of dust reddening and
opacity: more opaque objects are in general redder and more actively
star forming. This had already been established for starburst galaxies
\citep{heck98} and for quiescently star--forming galaxies
\citep{wang96}, and still holds true for individual star--forming
knots within a single galaxy. It is a straightforward consequence of
the Kennicutt--Schmidt law, plus (for galaxy samples) the
mass--metallicity relation. 

\citet{vander88} had already established for NGC~5194 that the dust
extinction affecting the clumped ionized gas is consistent with
foreground dust, by comparing the H$\alpha$/radio ratio with the
Balmer decrement. We confirm these earlier results, by finding a range
of A$_V$ from H$\alpha$/P$\alpha$ which is similar to the range found
by \citet{vander88} from H$\alpha$/radio. Regions that are completely
dust--buried are rare in this galaxy; of the 166 regions selected at
24~$\mu$m, only 2 were not detected in H$\alpha$. The vast majority of
HII knots has detectable H$\alpha$, suggesting that: (1) presence of
high extinctions does not imply large fractions of completely obscured
(optical) emission \citep{vander88,jame04}, and (2) timescales for
newly formed cluster to separate from the parental cloud
\citep[$\sim$1--3~Myr,][]{garm82,leisa88} are shorter than the
evolution timescale of the cluster itself.

The opacity of the stellar continuum appears to be
reasonably well described by a starburst--like opacity curve with
A$_V^{star}\lesssim$0.44~A$_V^{gas}$ \citep[Figures~\ref{fig10} and
\ref{fig11}, e.g.,][]{calz01}, modulo the age--dependence of the
observed UV luminosities and colors. The most extincted regions have
UV attenuations corresponding to A$_V^{star}\sim$1.2~mag, implying
A$_V^{gas}\sim$2.8~mag (Figure~\ref{fig10}). This is consistent with
the maximum value A$_V^{gas}\sim$3.5~mag derived for the ionized gas
(Figure~\ref{fig11}).

Local peaks in the 8~$\mu$m and 24~$\mu$m emission usually correspond to local
peaks in H$\alpha$, and for 87\% of the cases also to local peaks in the UV
emission, albeit often displaced from the IR/H$\alpha$ peaks. This spatial
correspondence has led us to treat dust heating by UV photons as a local
effect, for the most part circumscribed within the size of our apertures. 

The decrease of the median A$_V$ with galactocentric distance is quite
typical of spiral galaxies \citep[][e.g.,]{pele92, giova95, jone96,
jame04}. The increasing scatter as a function of distance from the
nucleus shows that there are large local variations in the gas
density, but the median gas extinction decreases from about 3.5~mag in
the center to 2~mag at a distance of 3~kpc. We hypothize that the
trend towards bluer UV colors for larger galactocentric distances
(Figure~\ref{fig14}) is mainly driven by dust reddening, although a
trend towards younger ages for more distant UV--emitting knots is also
contributing (Figure~\ref{fig15} and next section). 

However, the characteristics of the knots' UV emission
(Figures~\ref{fig9} and \ref{fig10}) imply that, unlike starburst
galaxies, there is not a 1--to--1 correlation between the IR/FUV ratio
and dust opacity or between the UV colors and dust reddening. In each
case, an assumption on the mean age of the population dominanting the
UV emission needs to be made. For instance, the opacity in the FUV,
A(FUV) ranges between 4~mag and 6~mag in the center and between 0~mag
and 1~mag at 13~kpc distance, for stellar populations in the age range
100-5~Myr. Our range of UV opacity values is larger, especially in the
central regions, than what found by \citet{boiss04} from a radial
analysis of NGC~5194 using FOCA data. FOCA's 2000~\AA~ mean wavelenght
is reasonably close to the GALEX NUV's mean wavelenght, allowing
direct comparisons. Those authors find that the central NUV opacity in
NGC~5194 is closer to 2~mag, about half of what we infer. Two reasons
could explain the discrepancy: (1) we use HII knots, rather than
azimuthally averaged information, thus excluding potentially less
extincted diffuse emission; (2) \citet{boiss04} use the IR/NUV ratio
as a tracer of UV dust opacity; this assumption will lead to
underestimates of the UV opacity if, as we have seen in this work, the
IR--dominating and UV--dominating populations do not coincide and show
significant age differences.

The separation between IR--emitting and UV--emitting populations is
particularly evident along the spiral arms, where about 1/2 of the HII
knots show a displacement, at our resolution, between the IR (and
H$\alpha$) peak and the UV peak within our apertures
(Figure~\ref{fig4}), and the IR (and H$\alpha$) emission is generally
comparatively brighter than the UV emission along the inner edge, and
fainter than the UV along the outer edge. This can be interpreted as
the youngest stellar populations to be preferentially located along
the inner edge of the spiral arms.

\subsubsection{A `Dust--Star Geometry' Scenario for Star--Forming Galaxies}

Why are the opacity properties of NGC~5194, and probably of other
star--forming galaxies, different from those of starbursts? A
qualitative scenario involves the much higher SFR density (SFR per
unit area), and the consequent presence of stronger mechanical
feedback, in starbursts than in star--forming galaxies.

In starbursts, the feedback action of massive star winds and
supernovae explosions is likely to eject large fractions of the gas
and dust from the starburst volume into the surrounding interstellar
medium. This will `expose' the starburst population, that, in addition
to ionize the gas and heat the dust, will also be responsible for the
observed UV emission. In more quiescently star forming galaxies, the
various HII regions/complexes are generally unconnected, implying a
broader range of, and typically less strong, impact from feedback
mechanisms. The most active HII complexes will still shed their dust
cocoons via the action of gas outflows, thus resembling
mini-starbursts, but increasingly less active HII regions will come
out of the parental dust cloud only through secular motions. This last
process takes time \citep[$\sim$1--3~Myr,][]{garm82,leisa88} and
delays the emergence of the young population, which in the meantime
will have aged. As a result, the UV--emitting populations in
quiescently star--forming galaxies are not only `old' ($\gg$12-15~Myr)
and mostly non--ionizing, but also show a wide range of ages, which
may depend on the local gas, dust, and star--formation conditions
\citep{parra03}. The ionizing stellar populations, conversely, remain
for the most part sufficiently obscured by dust as to not provide the
dominant UV contribution.

Our definition of `old', however, is relative: the UV emission is
still emerging from populations that are $<$100~Myr old in NGC~5194,
and thus trace the `recent', albeit not the `current', star
formation. This characteristic may question the use of the {\em b}
parameter for measuring the degree of deviation of a star forming
galaxy from the locus of starbursts in the IR/UV--UV~colors plane
\citep{kong04}. The observed parameters of NGC~5194,
log[L(IR)/L(FUV)]=0.76 and $\beta_{GLX}$=$-$0.84 (where $\beta_{GLX}$
is defined in \citet{kong04}), would place this galaxy among those
with $b\lesssim$0.3 according to the models of \citet[ their
Figure~4]{kong04}. However, the current to past star formation rate
ratio in NGC~5194 is $b\gtrsim$2, based on a 24~$\mu$m--derived
SFR=3.4~M$_{\sun}$~yr$^{-1}$ and a 2MASS H--band derived stellar
mass. The disagreement can be reconciled if a modified {\em b}
parameter is used, namely the ratio of the current to recent
($<$100~Myr) star formation rate; in this case
$b_{recent}\lesssim$0.6--0.8, for UV--emitting populations of average
age 50--100~Myr and conservative assumptions on the dust
obscuration. Thus, the deviation of star--forming galaxies from the
locus of starbursts in the IR/UV--UV~colors plane could be driven by
star formation over the last few hundred Myr, or a fraction of the
Hubble time, rather than the Hubble--time averaged star formation.
 
What are the conditions of applicability of the starburst opacity
curve \citep{calz94,meur99,calz01} in those cases, like distant galaxies,
where only integrated light information is often available? NGC~5194
classifies as a star--forming galaxy, rather than a starburst, despite
is relatively high SFR (3.4~M$_{\sun}$~yr$^{-1}$, about 10 times
higher than the local starburst galaxy NGC5253, \citet{calz04}). Its
SFR density is, however, 0.015~M$_{\sun}$~yr$^{-1}$~kpc$^{-2}$, or
about 10--100 times lower than some of the `weakest' starbursts
\citep{kenn98b,heck05}. Although it is unclear at this stage where the
exact transition for the applicability of the starburst opacity curve
is, a SFR density $>$1M$_{\sun}$~yr$^{-1}$~kpc$^{-2}$ is safely within
the regime where the curve was derived.

\subsubsection{The 2200~\AA~ Feature}

The UV color properties of the NGC~5194 HII knots suggest that the
2200~\AA~ extinction curve feature is much weaker in this galaxy than
what is observed in our own Galaxy (section~6.1.2). This is a
tantalizing result, as the 2200~\AA~ feature is widespread in the
Milky Way, and there is evidence that this feature is mostly due to
absorption \citep{calz95}. NGC~5194 has a metallicity at least as high
as, and possibly higher than, our Galaxy, and a SFR only a factor
$\lesssim$2 higher. A number of studies have pointed out that the
intensity of star formation may play a larger role than metallicity in
determining the presence or absence of the 2200~\AA~ feature in
galaxies \citep{gord03}: starburst galaxies lack the feature,
independently of metallicity \citep{calz94}, and both in the Milky Way
and SMC there are sighlines with extinction curves that deviate from
the `canonical' ones \citep{lequ82,gord98,vale03}. The exact mechanism
that may take place in the HII knots of NGC~5194 is not clear at this
point, although there could be mechanisms other than star formation
intensity able to modify the strenght of the 2200~\AA~ feature in the
extinction curve, some of which are at work in dense clouds in the
Milky Way itself \citep{whitt04}.

\subsection{Intrinsic Properties of HII `Complexes' in NGC~5194}

The intrinsic properties of the H$\alpha$--emitting regions in the
center of NGC~5194 have been extensively discussed by \citet{scov01},
and will not be repeated here. We only note that our brightest
H$\alpha$ knot has extinction--corrected luminosity
$\sim$10$^{40}$erg~s$^{-1}$, not dissimilar from the brightest HII
regions measured (at much higher resolution) in \citet{scov01}. The
range of intrinsic H$\alpha$ luminosities correspond to SFR densities
between 3$\times$10$^{-3}$~M$_{\sun}$~yr$^{-1}$~kpc$^{-2}$ and
0.07~M$_{\sun}$~yr$^{-1}$~kpc$^{-2}$, when averaged over our $\sim$500~pc
apertures. Some regions are easily above the threshold for starbursts
(see previous section), as their intrinsic sizes are smaller than our
resolution--driven apertures.

The brightest H$\alpha$ knot in our sample corresponds to a stellar
mass $\approx$2$\times$10$^5$~M$_{\sun}$ (for a \citet{krou01} IMF in
the range 0.01--100~M$_{\sun}$) within the $\sim$500~pc enclosed by
each knot. For comparison, the oldest ($\lesssim$100~Myr),
UV--emitting populations correspond to masses of
$\sim$6-8$\times$10$^7$~M$_{\sun}$, as inferred from their
extinction--corrected U and B emission, implying stellar masses
$\approx$300--400~times larger than in the brightest ionizing
population, but still consistent with the typical stellar mass
densities of spiral galaxies
($\approx$200~M$_{\odot}$~pc$^{-2}$). Because of their large masses,
these UV--bright knots are likely blends of multiple recently formed
star clusters. The intrinsic masses and sizes of HII regions in
NGC~5194 are in the range 10$^3$--10$^4$~M$_{\sun}$ and a few tens of
pc, respectively \citep{scov01}. Tidal forces are very effective at
disrupting such small clusters, and cluster lifetimes in NGC~5194 have
been found to be $<$10$^8$~yr \citep{lame05}, well matched to the
lifetimes we derive from the U$-$B colors (Figure~\ref{fig15},
left). Cluster stars will hence drift out of our apertures and diffuse
in the galaxy over timescales of $\approx$30~Myr, for an assumed space
velocity of 10 km~s$^{-1}$, which typical of OB stars in the Milky
Way. Thus, after $\approx$100~Myr, the clusters will no longer
contribute to the concentrated UV emission identified in our HII
knots.

We observe a radial trend for the UV--emitting regions to be younger
at larger galactocentric distances, even after correction for dust
extinction; the oldest regions ($\lesssim$100~Myr) are located closest
to the nucleus, and ages decrease down to $\sim$5--20~Myr at about
3~kpc distance (Figure~\ref{fig15}). A trend of decreasing mean age
with distance was already observed by \citet{bian04a} for NGC~5194;
they find, however, older ages ($>$300~Myr) than ours for the same
range of galactocentric distances, as they do not correct their colors
for the effects of dust extinction. We argue that the radial age trend
we observe in Figure~\ref{fig15} is real, and not an effect of
insufficient extinction correction. The U$-$B colors we use to
estimate ages are only moderately sensitive to dust reddening. Larger
extinction values, or steeper extinction curves would produce
unphysically blue UV colors and high UV luminosities much before
changing significantly U$-$B--estimated ages (Figure~\ref{fig12}
left). One possible interpretation is that, as dust extinction and
local pressure increase towards the center, later and later ages
become the dominant contributors to the observed UV emission, as
younger populations remain deeply embedded in dust for comparatively
longer times relative to their outer region's counterparts.

\subsection{Star Formation Rate Indicators}

\subsubsection{The Infrared Luminosities}

The analysis of the HII knots within the central 6~kpc of NGC~5194
shows tight correlations between the P$\alpha$ luminosity and a
variety of integrated and monochromatic infrared luminosities,
confirming that in the dusty environment of this galaxy more strongly
star--forming regions have larger infrared luminosities. Since the
extinction--corrected P$\alpha$ is a close tracer of current SFR, we
can discuss our results in light of their relevance for tracing SFR.

As already stressed in many previous papers \citep[for a few recent
ones, see][]{kenn98,helo00,kew02}, the integrated 3--1100~$\mu$m
luminosity, L(IR), is a reasonable tracer of SFR in galaxies. We
confirm this result for individual HII knots in the dusty environment
of NGC~5194, although we observe a mildly significant deviation from
linearity (L(IR)$\propto$L(P$\alpha$)$^{(0.90\pm0.07)}$). The infrared
emission longward of $\sim$50~$\mu$m can receive substantial
contribution from large grains heated by a variety of stellar
populations field \citep{helo86,boul88}, including those that are no
longer young and ionizing. This may be the case for the less strongly
star--forming HII knots in our sample, whose infrared emission may
receive comparatively higher contamination from the non--ionizing
populations present in our apertures, thus driving the exponent
towards values smaller than one.

The closest tracer of SFR in NGC~5194 is the monochromatic luminosity
L(24), on the local scales of the HII knots. We find a linear 
correspondence between L(24) and the P$\alpha$ luminosity,
L(24)$\propto$L(P$\alpha$)$^{(1.03\pm0.04)}$. This implies that the
very small grains responsible for the IR emission closely trace the
young ionizing stars \citep{cesa96,helo00,haas02,helo04}. We have
quantified this correlation to have a dispersion of a factor 2.5--3
peak--to--peak in the center of NGC~5194, where metallicity variations
from region to region can be expected to be small \citep{zari94}, and
where HII regions are located in a relatively uniform environment.

However, the use of L(24) as a general SFR indicator for galaxies
should be regarded with caution at this stage. We have seen that the
ratio L(24)/SFR changes by a factor of a few from galaxy to galaxy
(Figure~\ref{fig19}), at least when considering
star--formation--dominated galaxies. Ionizing stars may heat the dust
to different average `effective temperatures', that may depend on the
local galactic conditions. The resulting variations in the emerging
infrared SEDs will imply variable fractions of L(24)/L(IR), hence
L(24)/SFR, from galaxy to galaxy \citep{dale01}. Further, L(24) can be
strong in galaxies dominated by nuclear non--thermal sources. In
NGC~5194, the relatively faint nuclear non--thermal source represents
only 2.2\% of the 24~$\mu$m luminosity of the whole galaxy, and 6\% of
the 24~$\mu$m luminosity within the central 6~kpc. For comparison, the
extinction--corrected P$\alpha$ luminosity of the nuclear source is
about 4\% of the integrated luminosity in the inner 6~kpc, thus
L(24)$_{AGN}$/L(24)$_{Inner Region}\approx$1.5--2
L(P$\alpha$)$_{AGN}$/L(P$\alpha$)$_{Inner Region}$; this implies that
L(24) is moderately overluminous relative to L(P$\alpha$) in dusty
and gas--rich non--thermal sources. In light of all the above,
analysis of a large sample of galaxies is needed to constrain the
galaxy--to--galaxy variation of L(24)/SFR, and establish whether this
luminosity can be effectively used as a SFR tracer.

The monochromatic 8~$\mu$m luminosity is more diffuse than either the
hydrogen recombination line emission and the 24~$\mu$m
luminosity. L(8) correlates tightly with L(P$\alpha$), but with a
significant non--linearity,
L(8)$\propto$L(P$\alpha$)$^{(0.79\pm0.02)}$. The emission in the
8~$\mu$m and other MIR bands is attributed to Polycyclic Aromatic
Hydrocarbons (PAH, \citet{lege84}), large molecules transiently heated
by single UV and optical photons \citep{sell90}, and which can be
destroyed, fragmented, or ionized by harsh UV photon fields
\citep{boul88,boul90,helo91,pety05}. Spitzer data of the nearby galaxy
NGC300 show, indeed, that the 8~$\mu$m emission highlights the rims of
HII regions and is depressed inside the regions \citep{helo04}. The
analysis of M33 by \citet{hinz05} shows that high intensity radiation
fields destroy the PAH \citep[see, also, e.g.][]{helo91,contu00}. In
addition, dust--poor environments are ineffective at shielding the
carriers from destruction by the UV emission \citep{bose04}, and the
PAH emission nearly disappears in galaxies with metallicities below
$\sim$25\% solar \citep{enge05b}.

The non--linear correlation between L(8) and the hydrogen
recombination lines, also found by \citet{peete04} in the Milky Way,
suggests that a second mechanism in addition to star formation is
responsible for the heating of the 8~$\mu$m carriers. The `second
mechanism' can indeed be a combination of multiple mechanisms,
including dissociation or ionization of the PAH molecules in
correspondence of regions of intense star formation \citep{tacco04},
and heating of the 8~$\mu$m dust by the UV photons in the general
radiation field \citep{li02,haas02,bose04}, possibly from B stars
\citep{peete04}.

It is perhaps premature to generalize the above result, obtained
locally within galaxies, to galaxies as a whole. For instance,
\citet{rous01} and \citet{forst04} find a linear relation between the
PAH emission and the hydrogen recombination line emission in a sample
of local galaxies \citep[see, however,][]{bose04}. More analysis is
clearly needed to ascertain the conditions under which the MIR PAH
emission might be used as a SFR indicator.

\subsubsection{The UV Luminosity}

Only $\sim$40\% of the detected UV light from the Inner Region's
HII knots is from `young' systems, where `young' is defined as any
region with intrinsic UV and U$-$B colors typical of a $\le$30~Myr old
cluster.  Less than half of the observed UV emission comes from
currently star forming regions, and the rest is associated with recent
past (last $\approx$50--100~Myr) star formation. This questions the
use of the UV emission for measuring {\em current} SFRs in `normal'
star--forming galaxies, in addition to complicate attempts to remove
effects of dust opacity from the observed UV light. 

The application of attenuation--correction techniques like, e.g., the
starburst opacity curve, to the observed UV emission of star--forming
galaxies may lead to overcorrections of the UV emission, due to the
extraneous contribution from the evolved populations, and
overestimates of the SFR(UV) \citep{bua02}. For instance, if the `red'
UV color of NGC~5194 were interpreted as due to dust reddening only,
and the starburst opacity curve applied `as is', the resulting
A(FUV)=2.8~mag would lead to SFR(FUV)$\sim$13~M$_{\sun}$~yr$^{-1}$,
using \citet{kenn98}'s formula and a \citet{krou01} IMF. This is about
a factor of 4 higher than what derived from the 24~$\mu$m
luminosity. When the presence of populations as old as
$\sim$50--100~Myr is accounted for, A(FUV)$\sim$1.6~mag, corresponding
to a mean SFR(FUV)$\sim$4.3~M$_{\sun}$~yr$^{-1}$, only about 30\%~
higher than SFR(24~$\mu$m). This discrepancy stresses the difference
between normal star--forming and starburst galaxies. For starbursts,
mechanical feedback is likely to be strong enough that the observed UV
light traces the same population responsible for the gas ionization
and dust heating; hence, one can expect
SFR(FUV)$\sim$SFR(line)$\sim$SFR(IR). Conversely, in normal
star--forming galaxies the observed UV light traces evolved,
non--ionizing populations, and the relation between SFR(FUV) and
SFR(line) or SFR(IR) will depend on the recent star formation history
\citep[see, also,][]{kong04}.

\section{Summary and Conclusions}
 
The multiwavelength analysis of NGC~5194 using a combination of
Spitzer, GALEX, HST, and ground--based data has yielded new
information on the properties of dust opacity and star formation in
this galaxy.

The impact of dust is large on the observed properties of the HII
knots, ranging from A$_V\sim$3.5~mag in the center to A$_V\sim$0--1~mag
in the outskirts of this galaxy, as derived from both the ionized gas
emission and the stellar continuum UV emission. The trend of
decreasing extinction for increasing galactocentric distance is common
among spiral galaxies. Somewhat unexpectedly, though, we don't find
evidence for a strong 2200~\AA~ feature in the extinction curve of
NGC~5194, a feature that is instead ubiquitous in our own Milky Way.

We reproduce, for individual HII knots, the broadening in the
IR/UV--UV~color plane observed for normal star--forming galaxies. The
deviation from the starburst opacity curve is due to age effects: the
UV--emission of the knots traces evolved stellar populations, with the
oldest ones ($\sim$50--100~Myr) being further away from the locus of
the starburst opacity curve. In terms of spatial location, some of the
oldest UV--emitting regions are located near the center of the galaxy,
within $\sim$1~kpc of the nucleus. We also find evidence for HII knots
along the outer edge of spiral arms to be more evolved (less
ionizing) than the HII knots along the inner edges.

The UV emission in normal star forming galaxies does not trace {\em
current star formation}, but the {\em recent} one. In this respect,
the parametrization of the deviation from the starburst curve with the
`b' parameter {\citep{kong04} may be inadequate, as the relevant
quantity for measuring such deviation is not the ratio of the current
to the Hubble~time--averaged star formation but the current to the
recent ($<$100~Myr in NGC~5194) star formation. In addition, dust
extinction corrections developed for starburst galaxies will fail, by
factors of a few, to recover the intrinsic UV emission from normal
star--forming galaxies, due to the contribution to the UV emission
from non--star--forming populations in the latter galaxies. Although
it is unclear how to separate starbursts from star--forming galaxies
at high redshifts, where only spatially integrated quantities are
often accessible, a method involving the SFR density may offer a
useful discriminant. In particular, the validity of the starburst
opacity curve has been tested on starbursts in the SFR density range
$\sim$1--50~~M$_{\sun}$~yr$^{-1}$~kpc$^{-2}$.

The most accurate {\em local} tracer of current SFR in NGC~5194 is the
24~$\mu$m emission, which shows a linear correlation with nebular line
emission, with a peak--to--peak dispersion of a factor
2.5--3. However, the L(24)/SFR ratio, although constant within the
central region of NGC~5194, varies by a factor of a few from galaxy to
galaxy. Thus, the use of L(24) as a SFR tracer for galaxies in general
is premature, until further investigation with larger samples of
galaxies.

Conversely, the monochromatic 8~$\mu$m luminosity of the HII knots
does not show a linear correlation with the nebular gas emission. The
8~$\mu$m emission is overluminous relative to the galaxy's average for
weakly ionized regions, and underluminous for strongly ionized
regions. A combination of two effects, heating of the carriers by the
general radiation field and ionization/destruction/fragmentation in
hard UV radiation field, may explain the observed trend
\citep{haas02,bose04,peete04,tacco04}.

The conclusions reached so far are based on a single galaxy. In the
case of L(24) and L(8), these luminosities need to undergo further
scrutiny to gauge galaxy--to--galaxy variations, and test their
applicability as SFR tracers in galaxy samples. The SINGS sample of
local star--forming galaxies is optimally designed to address this
issue in the near future.

\acknowledgments

This work is part of SINGS, The Spitzer Infrared Nearby Galaxies
Survey, one of the Spitzer Space Telescope Legacy Science Programs,
and was supported by the JPL, Caltech, Contract Number 1224667. The
Spitzer Space Telescope is operated by JPL, CalTech, under NASA
Contract 1407. The SINGS team acknowledges the hard work by the IRAC,
MIPS, and IRS instrument teams and the Spitzer Science Center for
making possible the work presented here.

GALEX (Galaxy Evolution Explorer) is a NASA Small Explorer, launched
in April 2003. Bianchi, Madore, Martin (PI), and Thilker gratefully
acknowledge NASA's support for construction, operation, and science
analysis for the GALEX mission, developed in cooperation with the
Centre National d'Etudes Spatiales of France and the Korean Ministry
of Science and Technology.

DC and RCK gratefully acknowledge the hospitality of the Aspen Center
for Physics (Aspen, Colorado), where some parts of this work were
developed by the authors during the Summer 2004 Workshop on Star
Formation.

%\appendix

%\section{Appendicial material}

\clearpage


\begin{figure}
\figurenum{1}
%\plottwo{f1a.eps}{f1b.eps}
\caption{Two three--color composites of M51. {\bf (Left)} The far--UV
(blue), continuum--subtracted H$\alpha$ (green), and 24~$\mu$m dust
(red) emission of the galaxy pair.  The FUV and FIR images, from GALEX
and Spitzer, respectively, have closely--matched resolution
($\approx$6$^{\prime\prime}$, while the resolution of the
ground--based H$\alpha$ image has been degraded to match that of the
two space--borne images. {\bf (Right)} The continuum--subtracted H$\alpha$
(blue), 3.6~$\mu$m stellar continuum (green), and 8~$\mu$m dust (red)
emission of the galaxy pair. This second image exploits the higher
angular resolution of the IRAC images (about 2$^{\prime\prime}$ FWHM)
to provide higher level of detail. The stellar continuum emission
traces evolved (old) stellar populations. In this figure, a foreground
star appears pure green.  North is up, East is left.  The size of the
pictures is $\sim$8$^{\prime}$.6$\times$11$^{\prime}$.8.
\label{fig1}}
\end{figure}
\clearpage

\begin{figure}
\figurenum{2}
%\plotone{f2.eps}
\caption{The 24~$\mu$m MIPS image of NGC~5194 with 112 of the 166
apertures in which photometry has been performed overlayed as red
circles. The 112 apertures are located in the Outer Region (see
text). North is left, East down. The flux scale is logarithmic, with
larger fluxes in darker grey. The areas where the local background has
been measured for removal from the aperture fluxes are shown as
rectangular outlines on the image; they are sequentially numbered 1 to
11, and the one corresponding to the entire Inner Region is marked IR.
The linear sizes of the image are similar to those of
Figure~\ref{fig1}.
\label{fig2}}
\end{figure}

\clearpage 

\begin{figure}
\figurenum{3}
%\plottwo{f3a.eps}{f3b.eps}
\caption{Images of the Inner Region (the central area imaged in
P$\alpha$) are shown in the light of: continuum--subtracted
H$\alpha$ line emission (top--left), continuum--subtracted P$\alpha$
line emission (top--right), dust--only IRAC 8~$\mu$m
emission (bottom--left), and MIPS 24~$\mu$m emission
(bottom--right). The North--East direction is indicated by vectors on
the IRAC and MIPS images. 54 of the 166 apertures used for photometry
are located in the Inner Region and are overlayed on the four images;
they are sequentially numbered 1 through 54 on the MIPS24 image, with
the Seyfert~2 nucleus indicated as aperture~1.  The flux scale in each
panel is logarithmic, with larger fluxes in darker grey. The linear
size of each Inner Region image is 2$^{\prime}$.46, or $\sim$5.9~kpc.
\label{fig3}}
\end{figure}

%\clearpage 

%\begin{figure}
%\figurenum{3b}
%\plottwo{f3c.eps}{f3d.eps}
%\caption{Bottom--left and bottom--right panels of
%Figure~\ref{fig3}. See Figure~\ref{fig3} caption.
%\label{fig3b}}
%\end{figure}

\clearpage 

\begin{figure}
\figurenum{4}
%\plottwo{f4a.eps}{f4b.eps}
\caption{A comparison of aperture locations on a small section of the the
24~$\mu$m {\bf (left)} and FUV {\bf (right)} images, showing the displacement
between the IR and FUV peaks. Particularly obvious are the cases of apertures
07 and 08. The orientation of the images is the same as Figures~\ref{fig3}.
\label{fig4}}
\end{figure}

\clearpage 

\begin{figure}
\figurenum{5}
\plotone{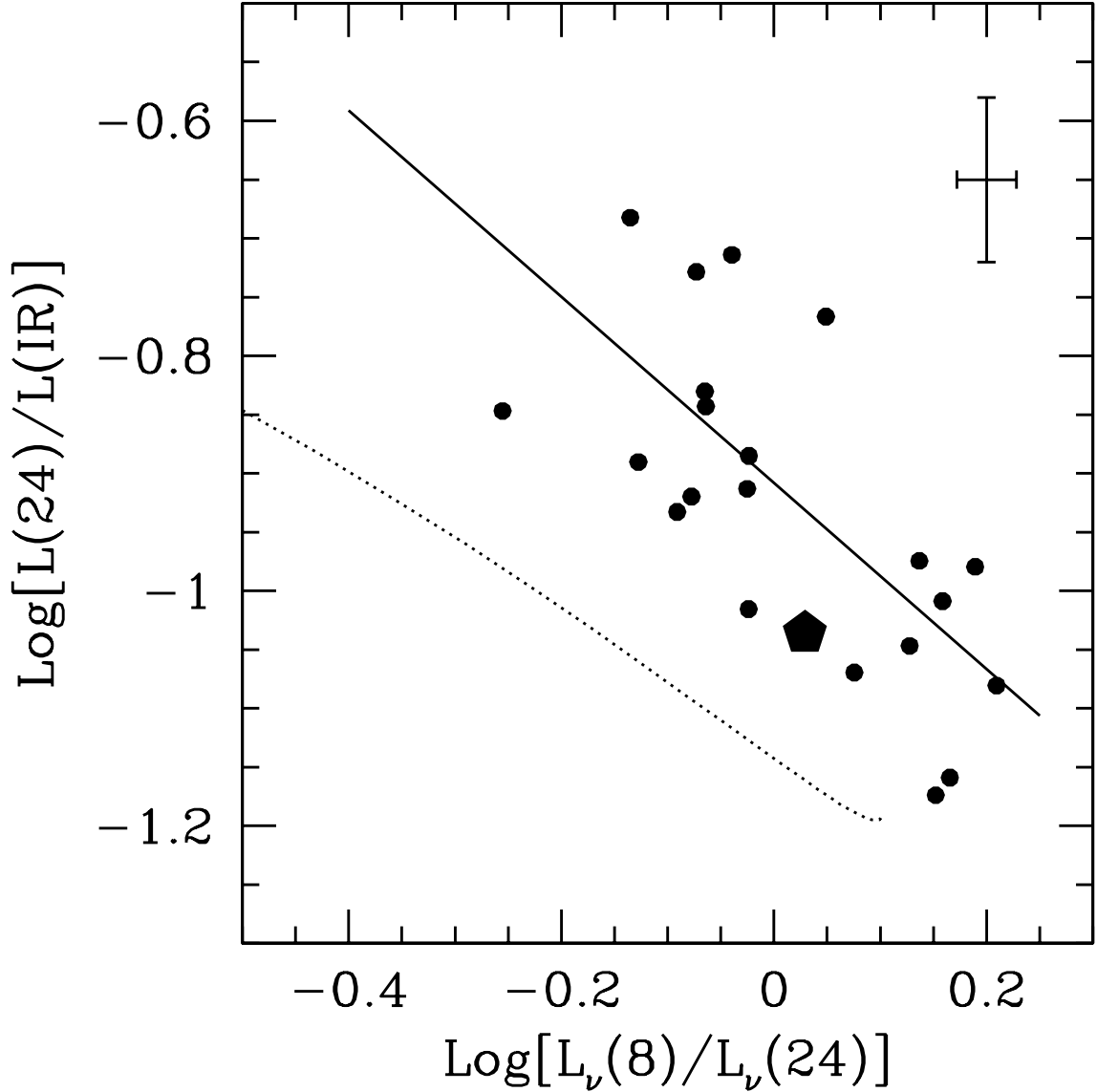}
\caption{The 24~$\mu$m--to--total infrared luminosity as a function of
the 8/24~$\mu$m flux ratio for 21 regions in the center and along the
spiral arms of NGC~5194, where L(IR)=L(3--1100~$\mu$m) and
L(24)=$\nu$L$_{\nu}$(24). The 21 regions have 70$^{\prime\prime}$
diameter, selected to properly sample the lowest resolution MIPS
channel (see text). A representative error bar appropriate for the
calibration uncertainties is shown at the top--right of the
figure. The datapoints show a mild correlation in the plane defined by
the two luminosity ratios, that we fit with a straight line
(continuous line). For comparison, the location of the whole galaxy on
this plane is shown as a large filled pentagon. Not surprisingly, the
whole galaxy location is along the lower envelope of the locus defined
by the HII knots. The prediction from the model of \citet{dale02},
appropriate for whole galaxies, is shown as a dotted line.
\label{fig5}}
\end{figure}

\clearpage 

\begin{figure}
\figurenum{6}
\plotone{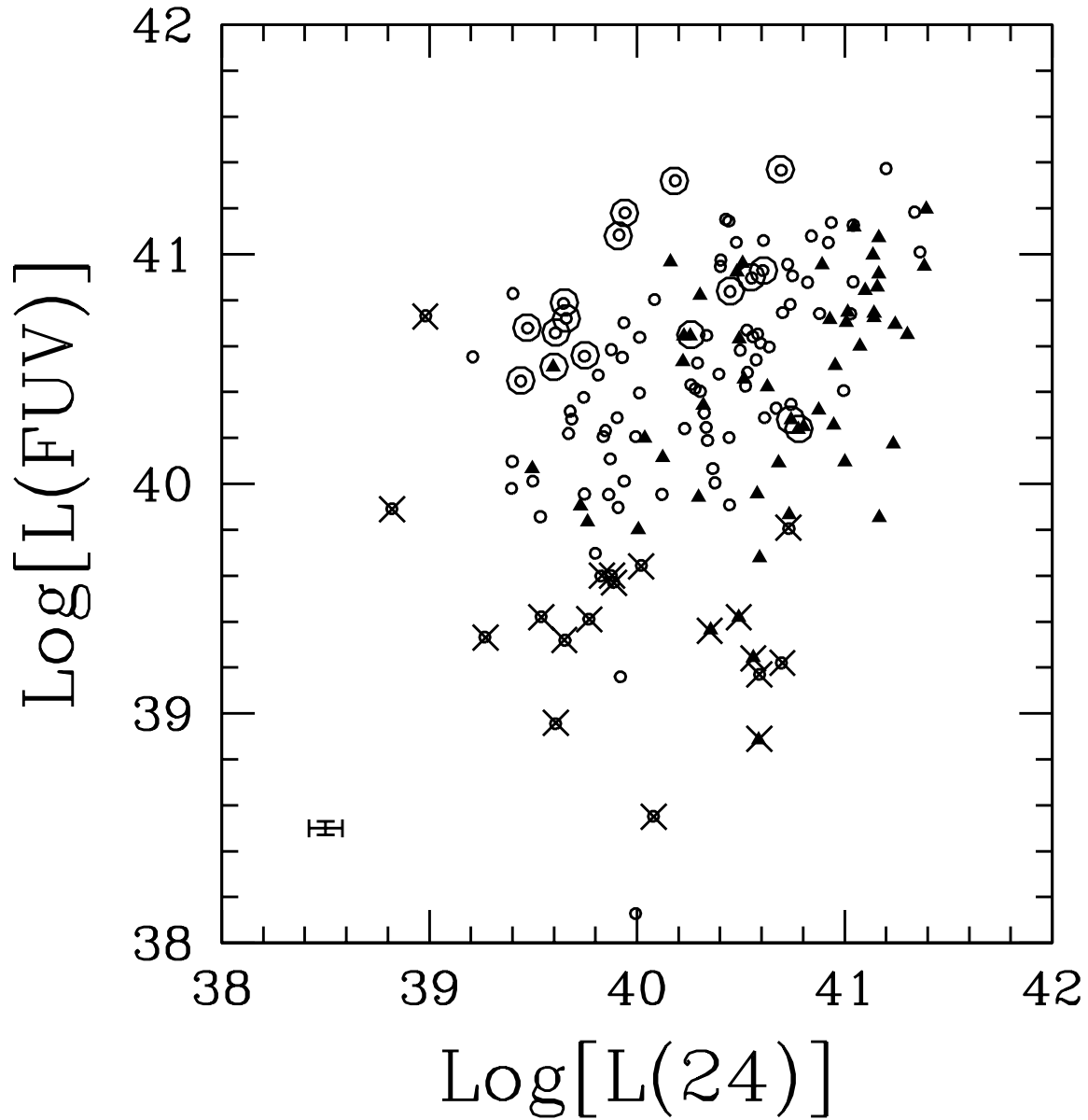}
\caption{Observed IR and FUV luminosities for the 166 regions with aperture
photometry. The IR is at 24~$\mu$m and the FUV is at 0.153~$\mu$m.  Different
symbols are used for apertures in the Inner Region (filled triangles) and the
Outer Region (small empty circles). Most regions are IR--selected; the
UV--selected regions are marked with large empty circles.  Upper limits (see
text) are indicated by crosses. The median error bar is shown at the
left--bottom of the plot. HII knots in the Inner Region tend to have, on
average, higher 24~$\mu$m luminosity than the Outer Region's knots, at fixed
UV luminosity. Also, the UV luminosities span close to the full range at each
fixed 24~$\mu$m luminosity, implying that there is no correlation between the
two quantities. 
\label{fig6}}
\end{figure}

\clearpage 

\begin{figure}
\figurenum{7}
\plotone{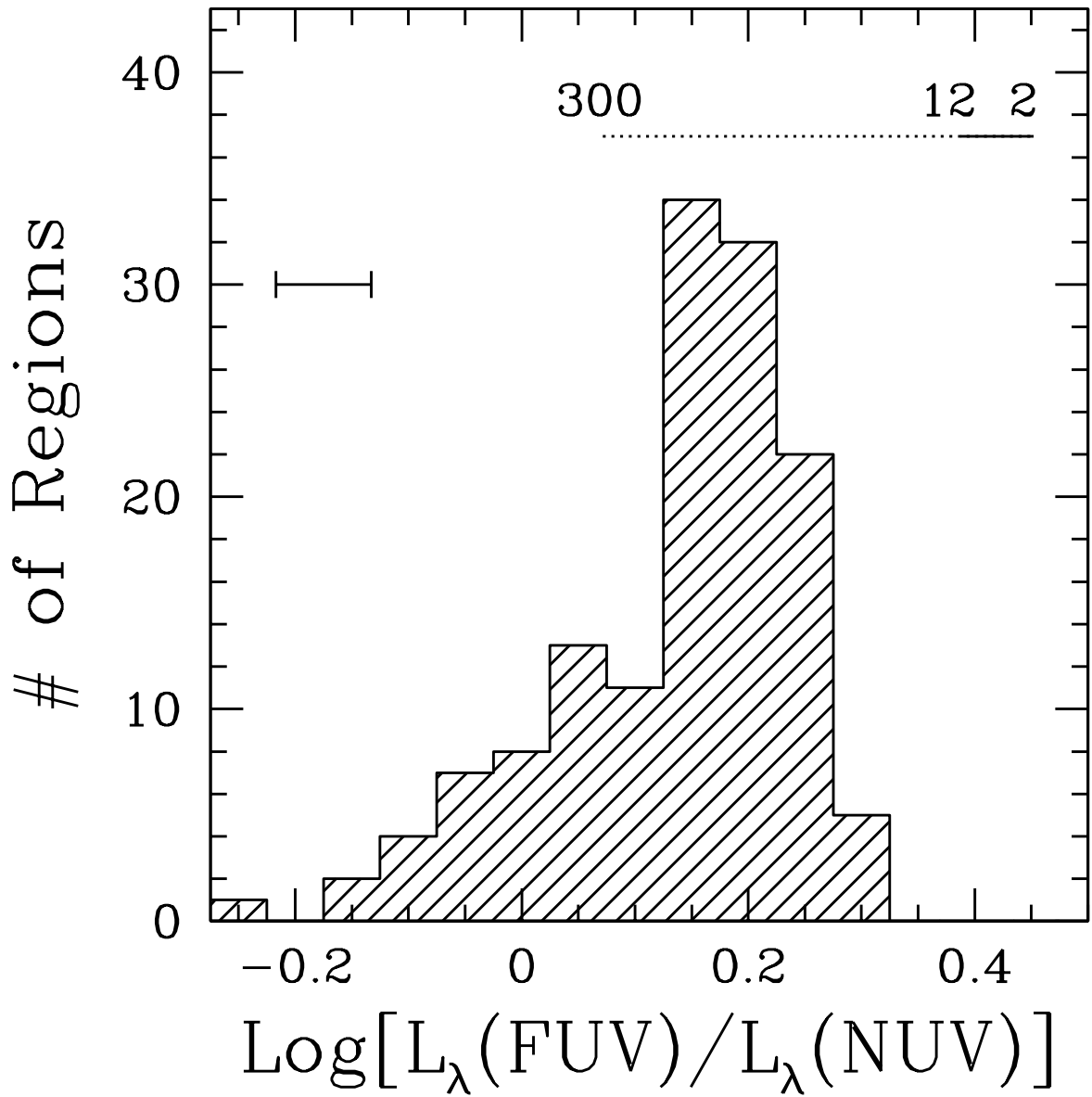}
\caption{Histogram of the observed UV colors for the 132 regions with
aperture photometry (upper limits are omitted). The UV color is the
0.153/0.231~$\mu$m flux ratio (GALEX FUV/NUV). The top--right
horizontal bar indicates the intrinsic UV colors of dust--free,
instantaneous--burst stellar populations for a range of ages
\citep[from Starburst99, ][]{leit99}: 2--12~Myr (continuous line,
ionizing population), and 12-300~Myr (dotted line, non--ionizing
population). A median error bar on the colors is shown at the
top--left of the plot. Noticebly, none of the observed HII knots has
UV colors compatible with a dust--free, ionizing stellar population.
\label{fig7}}
\end{figure}

\clearpage 

\begin{figure}
\figurenum{8}
\plottwo{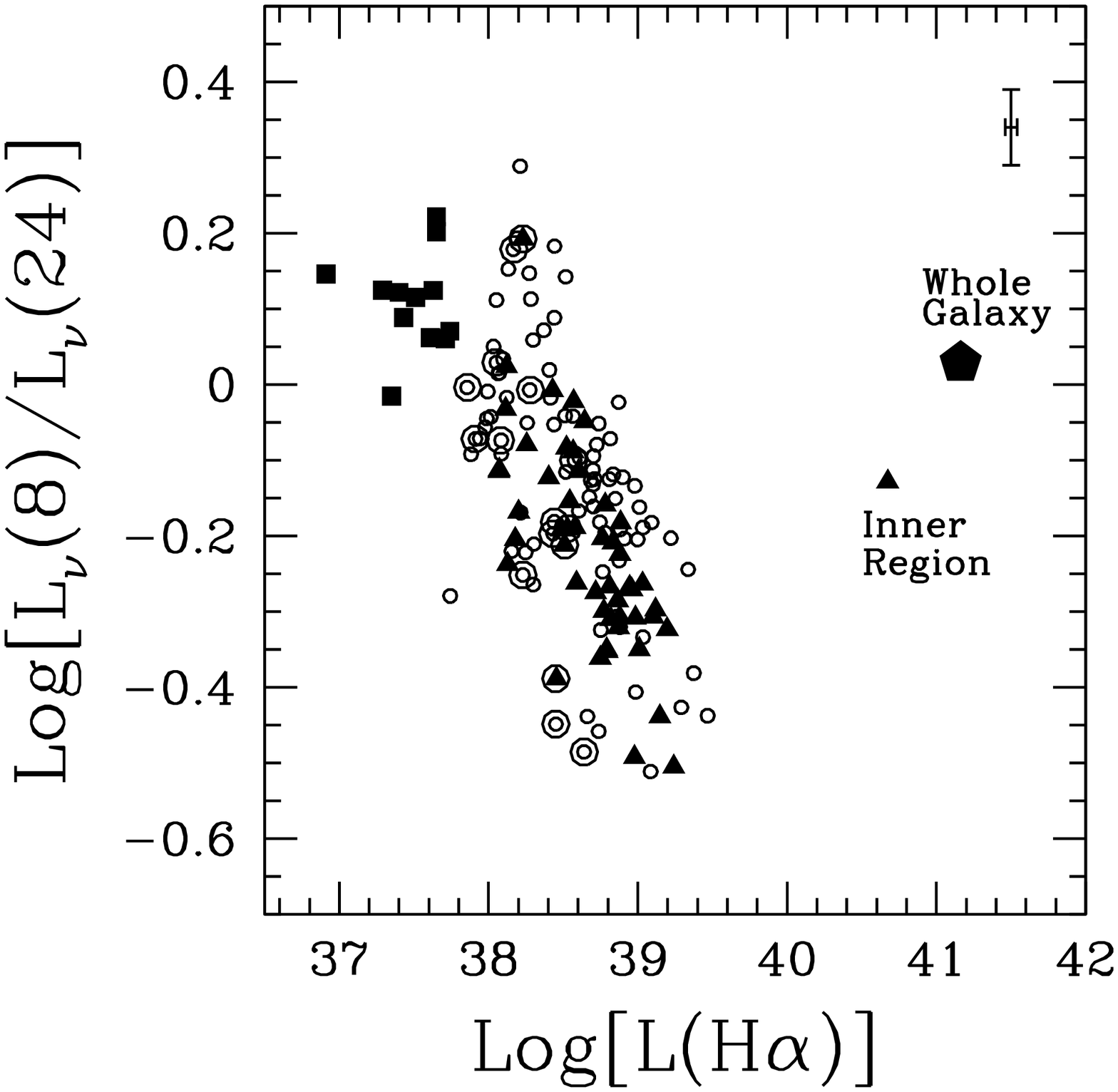}{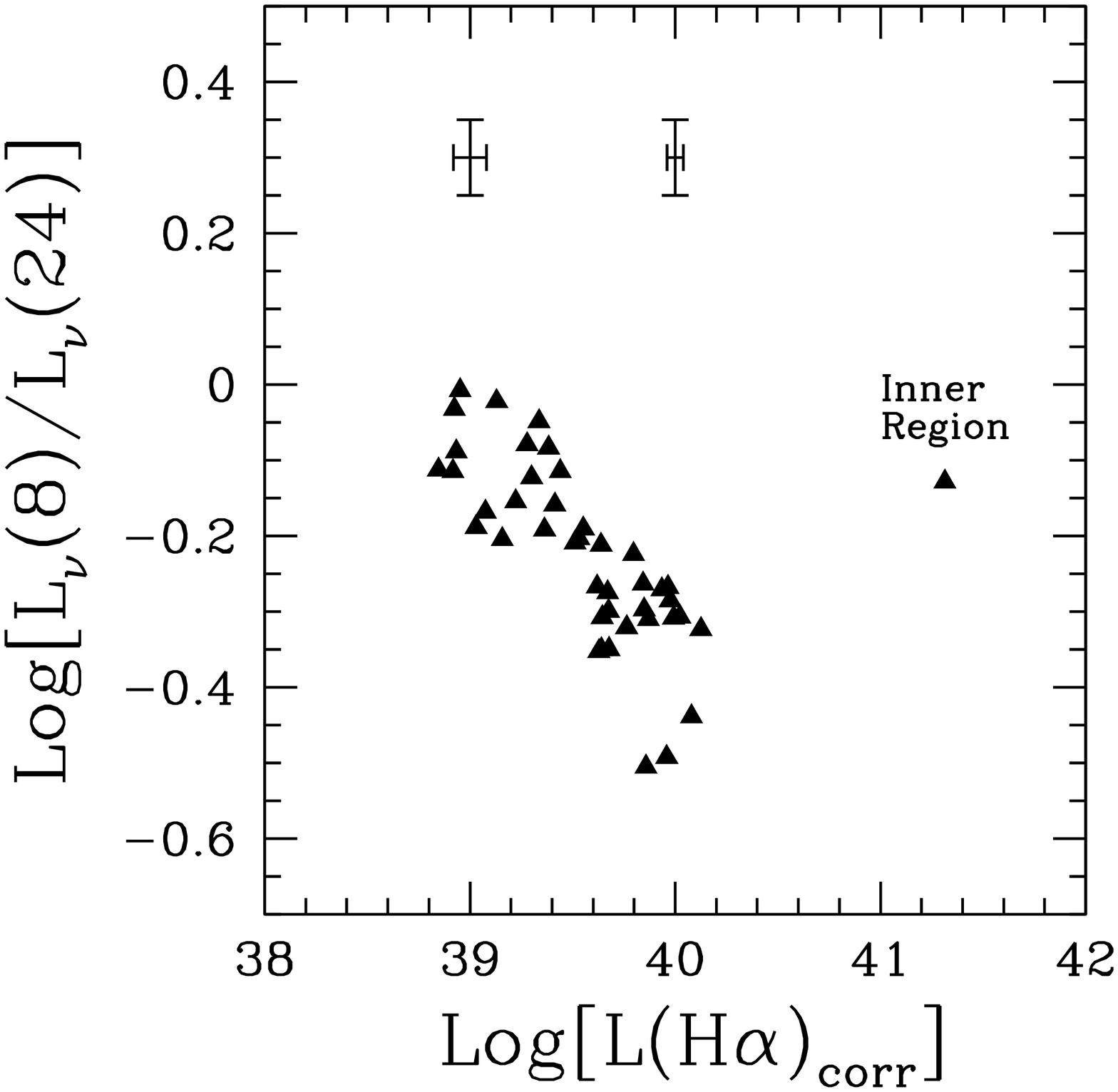}
\caption{{\bf (Left)} Observed IR colors as a function of the observed
H$\alpha$ luminosity, for both the HII knots in both the Inner
(triangles) and Outer Regions (circles); the integrated colors for the
Inner Region and the whole galaxy are shown as marked. For comparison,
the infrared colors of the 12 background regions are shown as filled
squares; the H$\alpha$ luminosity of the backgorund regions is
calculated over the 13$^{\prime\prime}$ diameter apertures also used
for the HII knots. {\bf (Right)} Observed IR colors as a function of
the {\it extinction--corrected} H$\alpha$ luminosity for the Inner
Region. The symbols are the same as Figure~\ref{fig6}. Upper limits
have been omitted. Median error bars are shown in both plots.
\label{fig8}}
\end{figure}

\clearpage 

\begin{figure}
\figurenum{9}
\plottwo{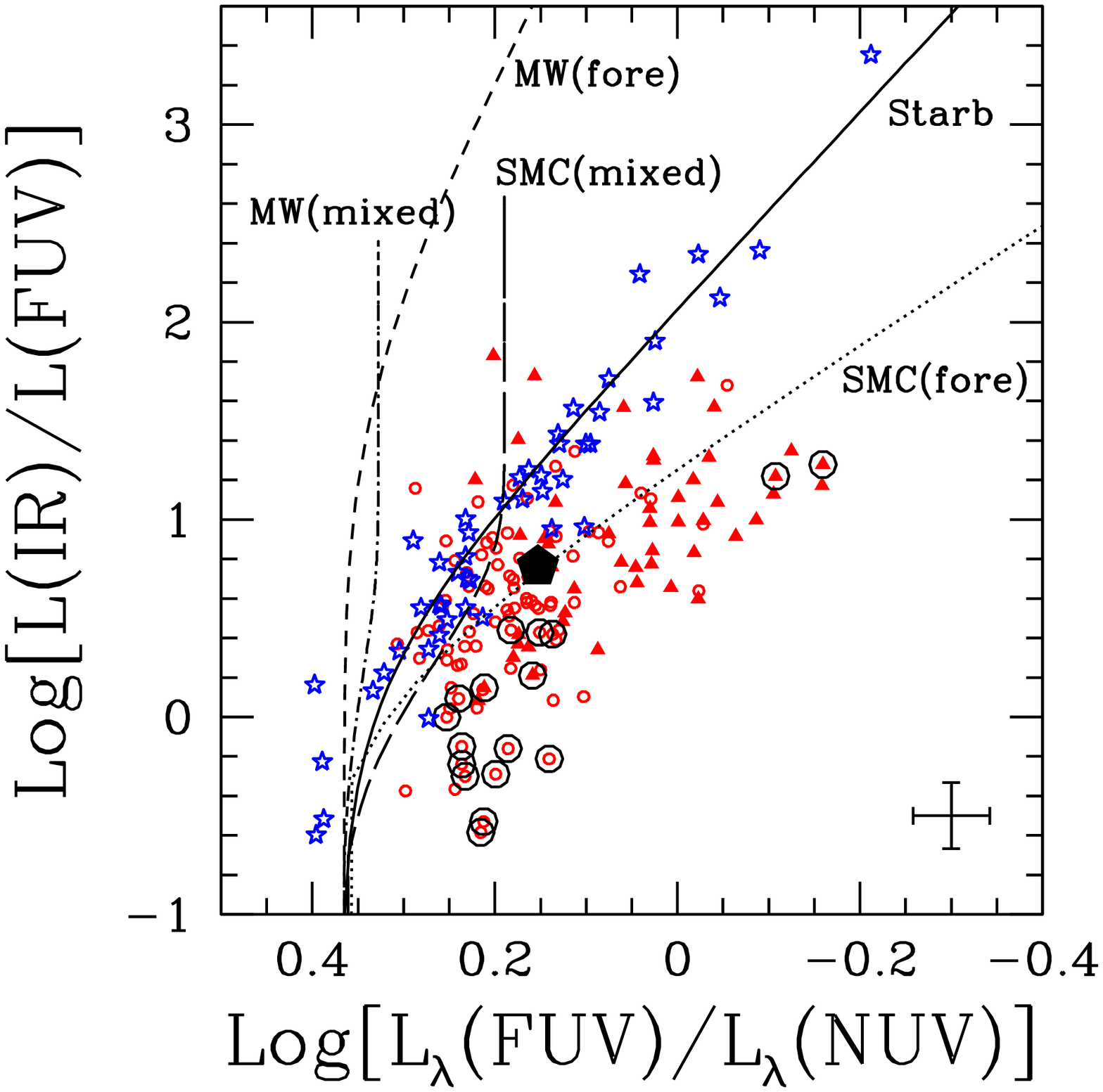}{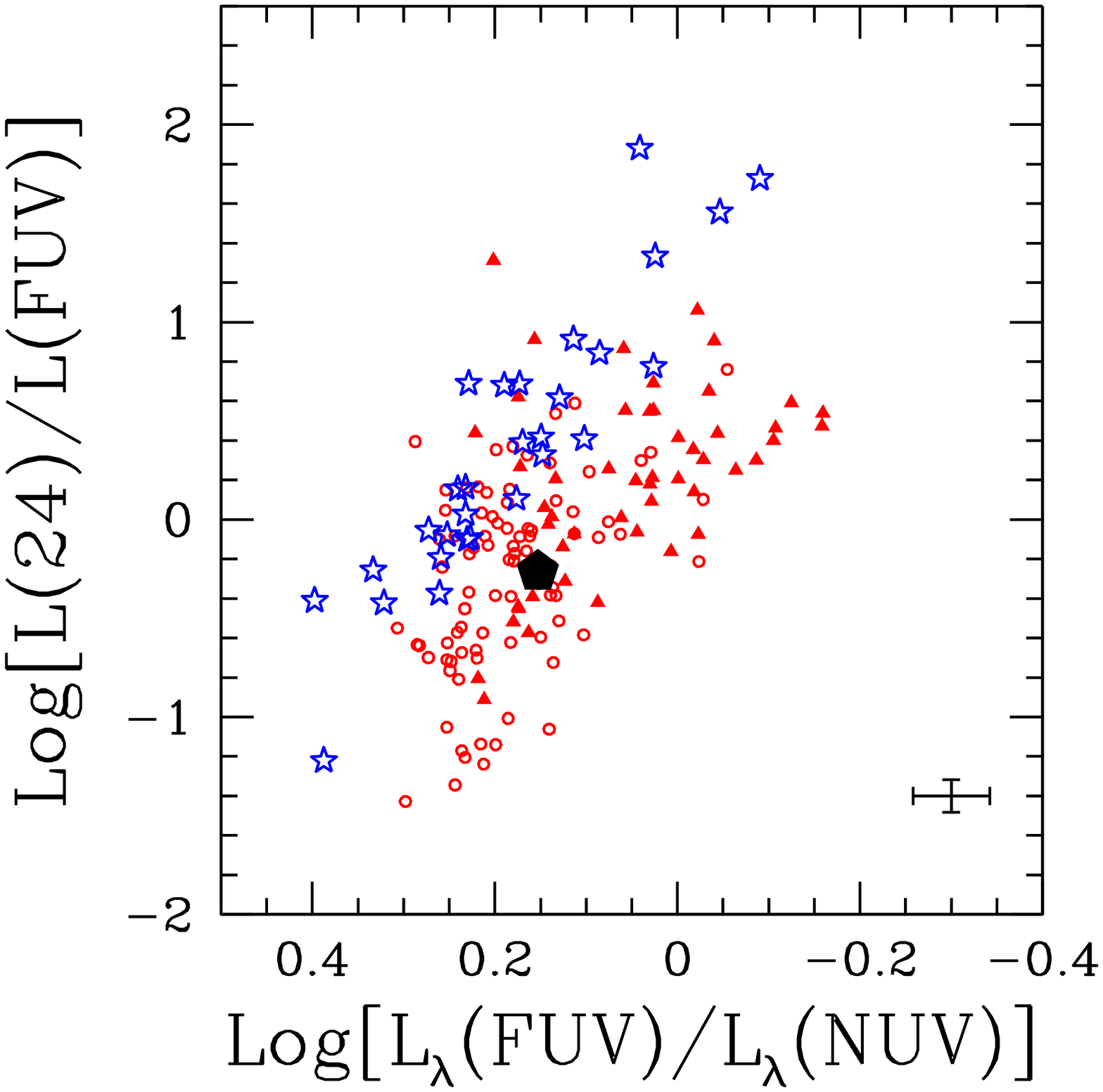}
\caption{{\bf (Left)} The total opacity, expressed as the infrared to
FUV luminosity ratio, as a function of the observed UV colors. L(IR)
and L(FUV) are total luminosities in the FIR and FUV, respectively
(L(FUV)=$\lambda$~L$_{\lambda}$(FUV)). Symbols are as in
Figure~\ref{fig6}. The filled pentagon is the position on the plot of
the emission from the entire NGC~5194 galaxy. Data for 29 starburst
galaxies from the sample of \citet{calz94} are plotted as star
symbols. Photometry for the starbursts has been performed on a
galaxy--wise basis (the entire starburst region at UV wavelengths,
from IUE, and the entire galaxy in the IR, from IRAS). Model lines are
obtained by convolving the spectral energy distribution of a constant
star formation stellar population \citep{leit99} with a range of dust
attenuation values and geometries: foreground, non--scattering dust
screens (MW extinction curve: short dashed line; SMC extinction curve:
dotted line), homogeneous mixtures of stars and dust (MW extinction
curve: dot--dashed line; SMC extinction curve: long--dashed line), and
the starburst dust distribution (continuous line, via the starburst
opacity curve, \citet{calz94,meur99,calz00}). {\bf (Right)} The same
plot with the abscissa given as the L(24)--to--L(FUV) ratio. For the
starbursts, the IRAS 25~$\mu$m flux is used as a close approximation
of the MIPS 24~$\mu$m. For both plots, median error bars are shown.
\label{fig9}}
\end{figure}

\clearpage 

\begin{figure}
\figurenum{10}
\plotone{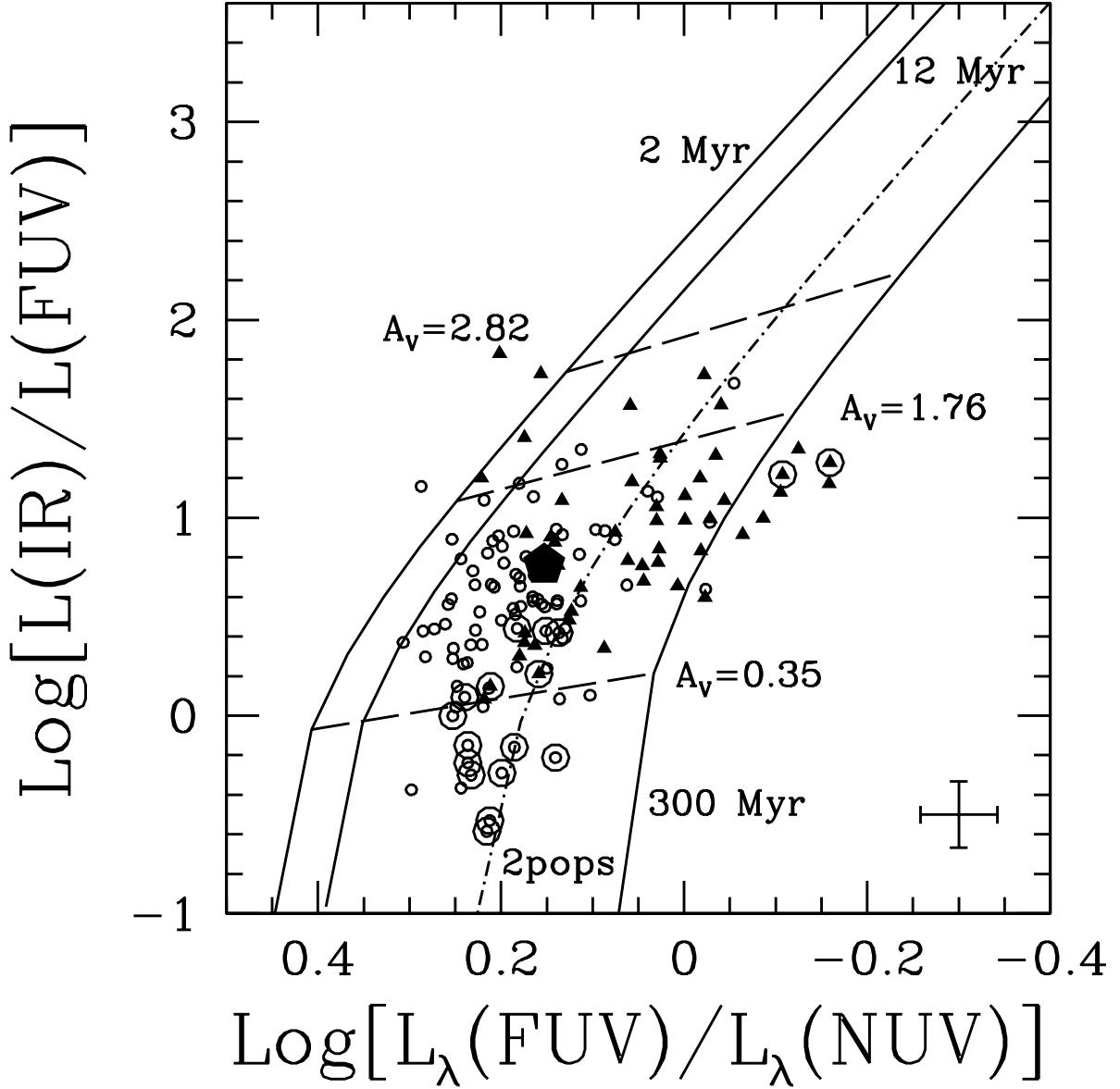}
\caption{The same plot and HII knots datapoints as Figure~\ref{fig9},
now compared to models of ageing burst populations convolved with the
starburst opacity curve for increasing amount of dust attenuation. The
continuous lines mark the locus of instantaneous burst populations,
2~Myr (left line), 12~Myr (center line), and 300~Myr old (right
hand--side line), respectively. The dot--dashed line marked `2pops'
shows the model track for the combination of a 5~Myr~old instantaneous
burst with a 300~Myr~old one. The mass of the 5~Myr~old burst is 300
times lower than than of the 300~Myr~old one, and its extinction
systematically higher by $\Delta$A$_V$=0.25~mag (except at A$_V$=0,
where both populations are extinction``free); both population models
are convolved with the starburst opacity curve. More details are given
in section~6.1.1. Loci of constant extinction A$_V$ are marked by
inclined dashed lines; these lines are not perfectly horizontal,
implying that fixed IR/FUV ratios do not exactly correpond to fixed
dust opacity A$_V$, in the presence of age variations. The vast
majority of the HII--emitting knots in NGC~5194 are bracketed by
models of stellar populations in the age range 2--300~Myr, with dust
extinction A$_V\lesssim$2.8~mag, for our adopted dust model.
\label{fig10}}
\end{figure}

\clearpage 
\figurenum{11}
\begin{figure}
\plottwo{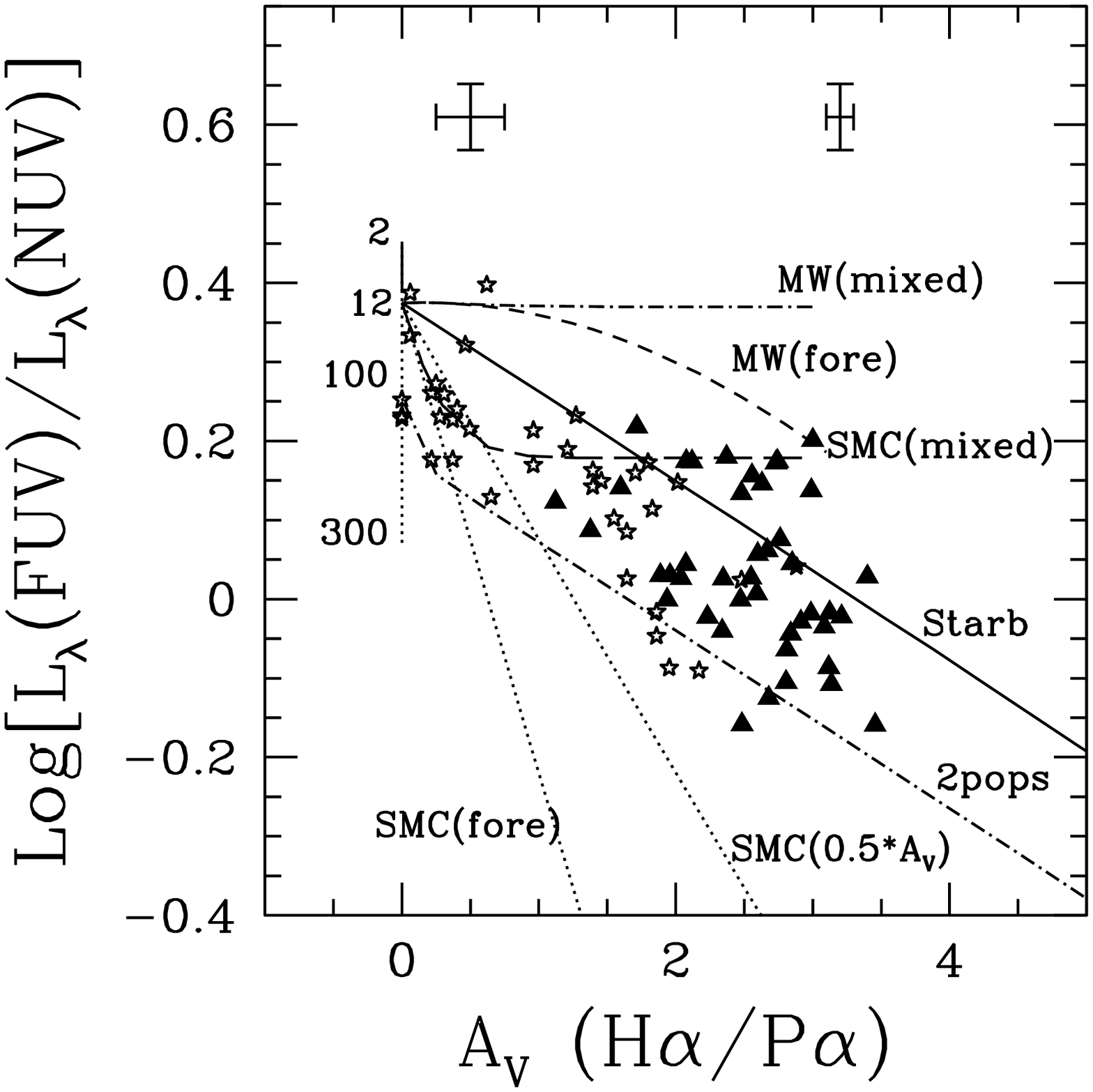}{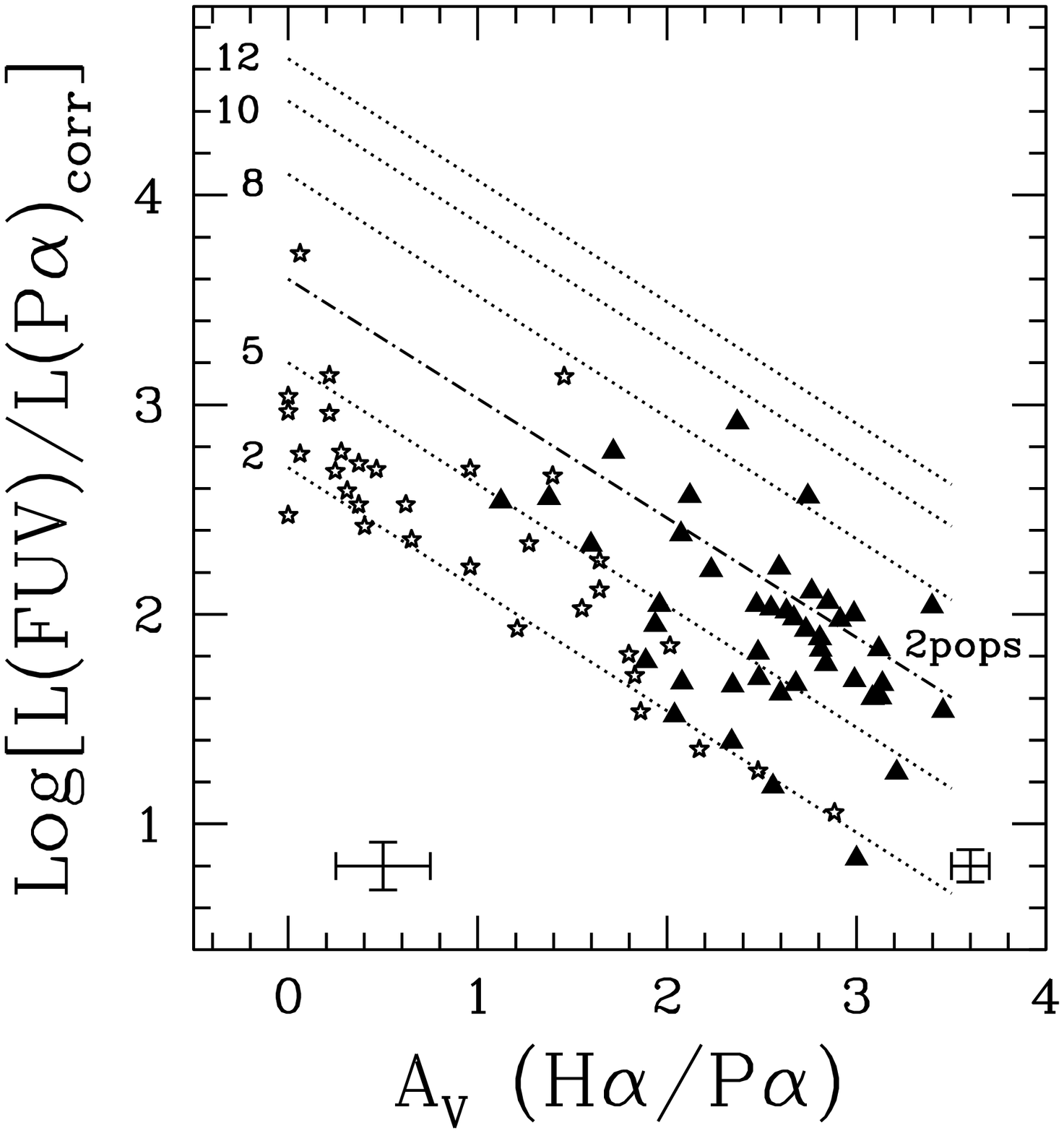}
\caption{{\bf (Left)} The UV flux ratio as a function of the optical
extinction A$_V$, in mag, for 42 knots in the Inner Region (filled
triangles). A$_V$ is calculated from the observed H$\alpha$/P$\alpha$
ratio, for a foreground dust screen.  Star symbols identify the
starburst galaxies. The vertical line at A$_V$=0 marks the range of
intrinsic colors for gas--ionizing instantaneous burst populations
(2--12~Myr, Leitherer et al. 1999, continuous line), and for
non--ionizing populations up to 300~Myr (dotted line). Dust models are
the same described in Figure~\ref{fig9} (left). One additional model
shown here is an SMC foreground, non--scattering, clumpy screen, with
the stellar continuum affected by half the reddening of the gas
(dotted line marked SMC(0.5 A$_V$)). The dot-dashed line marked
`2pops' represents the 2--populations model described in section~6.1.1
and Figure~\ref{fig10}. The discontinuity at A$_V$=0.25~mag is a model
artifact due to the different treatment of the two populations for
A$_V$=0~mag (both populations are extinction--free) and A$_V>$0~mag
(the younger population is systematically more extincted than the
older population). Median error bars at the two extremes of the
extinction range are shown on top of the plot; the uncertainties in
A$_V$ are mostly driven by the depth of the P$\alpha$ image. {\bf
(Right)} The ratio of the FUV luminosity to the extinction--corrected
P$\alpha$ luminosity as a function of the extinction A$_V$. The
datapoints, symbols, and the `2pops' model are as in the previous
panel. For the starbursts, L(P$\alpha$) is derived from the extinction
corrected H$\alpha$ luminosity, and A$_V$ is from the
H$\alpha$/H$\beta$ line ratio \citep{calz94}.  The intrinsic ratios
for unreddened instantaneous burst populations are marked by an
appropriate number (the age in units of Myr) at A$_V$=0. The dotted
lines mark the extinction trend in the FUV using the starburst opacity
curve.
\label{fig11}}
\end{figure}

\clearpage 
\begin{figure}
\figurenum{12}
\plotone{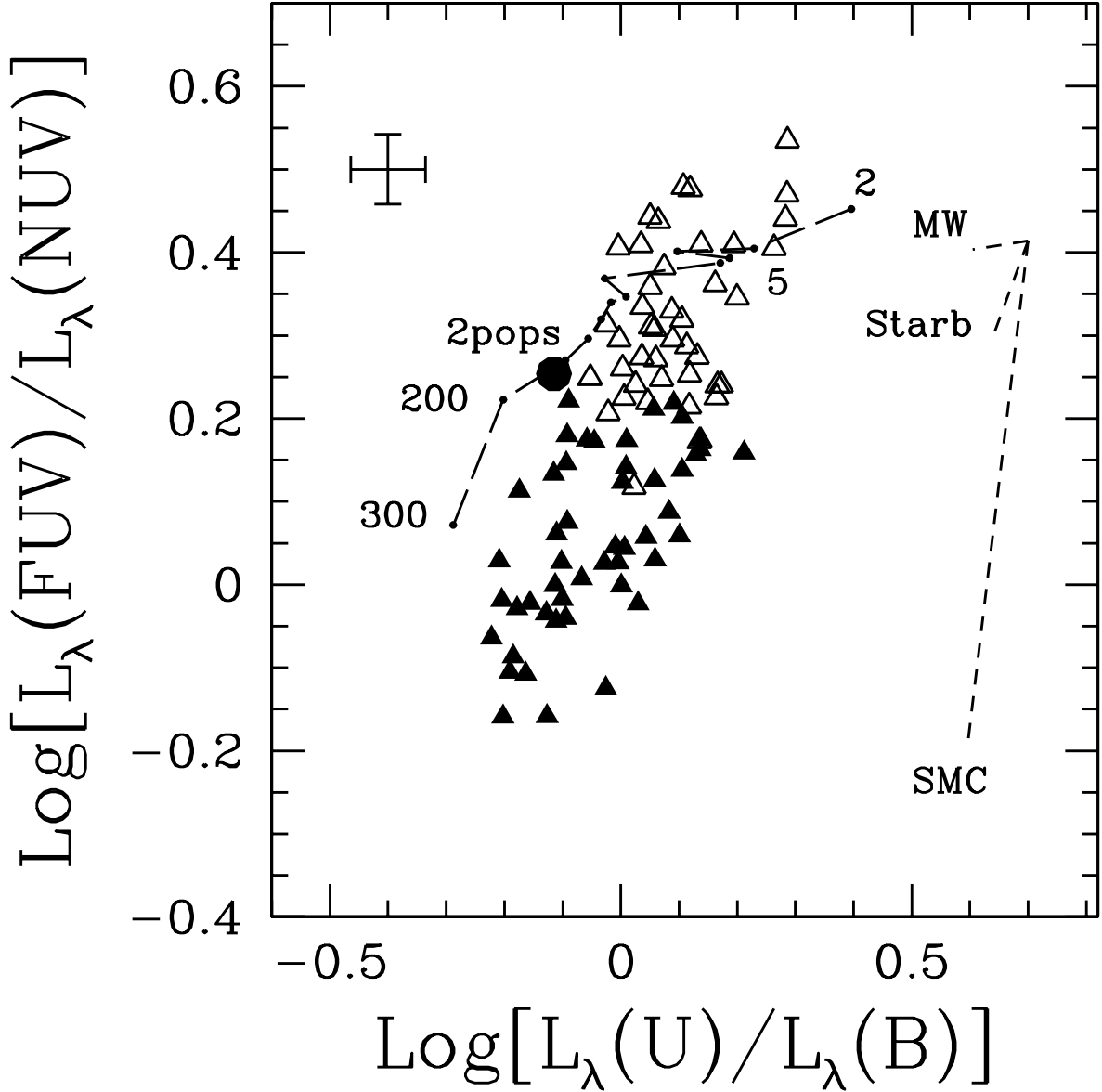}
\caption{The UV colors as a function of U$-$B for the 42 HII knots in
the Inner Region (filled triangles). U$-$B is expressed as the flux
ratios in the two bands: Log[L$_{\lambda}$(U)/L$_{\lambda}$(B)]. The
empty triangles are the extinction--corrected colors, using the
starburst opacity curve and the appropriate A$_V$ for each knot
(Figure~\ref{fig11}). The color evolution of ageing instantaneous
bursts between 2 and 300~Myr are shown as a long--dashed line, with a
few representative ages indicated in Myr. The large filled circle shows
the extinction--free colors of the 2--populations model described in
Figure~\ref{fig10} and section~6.1.1. Short--dashed straight lines show
the effect on the colors of 1~mag extinction at V for Galactic (MW),
starburst (Starb), and Small Magellanic Cloud (SMC) reddening curves.
\label{fig12}}
\end{figure}

\clearpage 
\begin{figure}
\figurenum{13}
\plotone{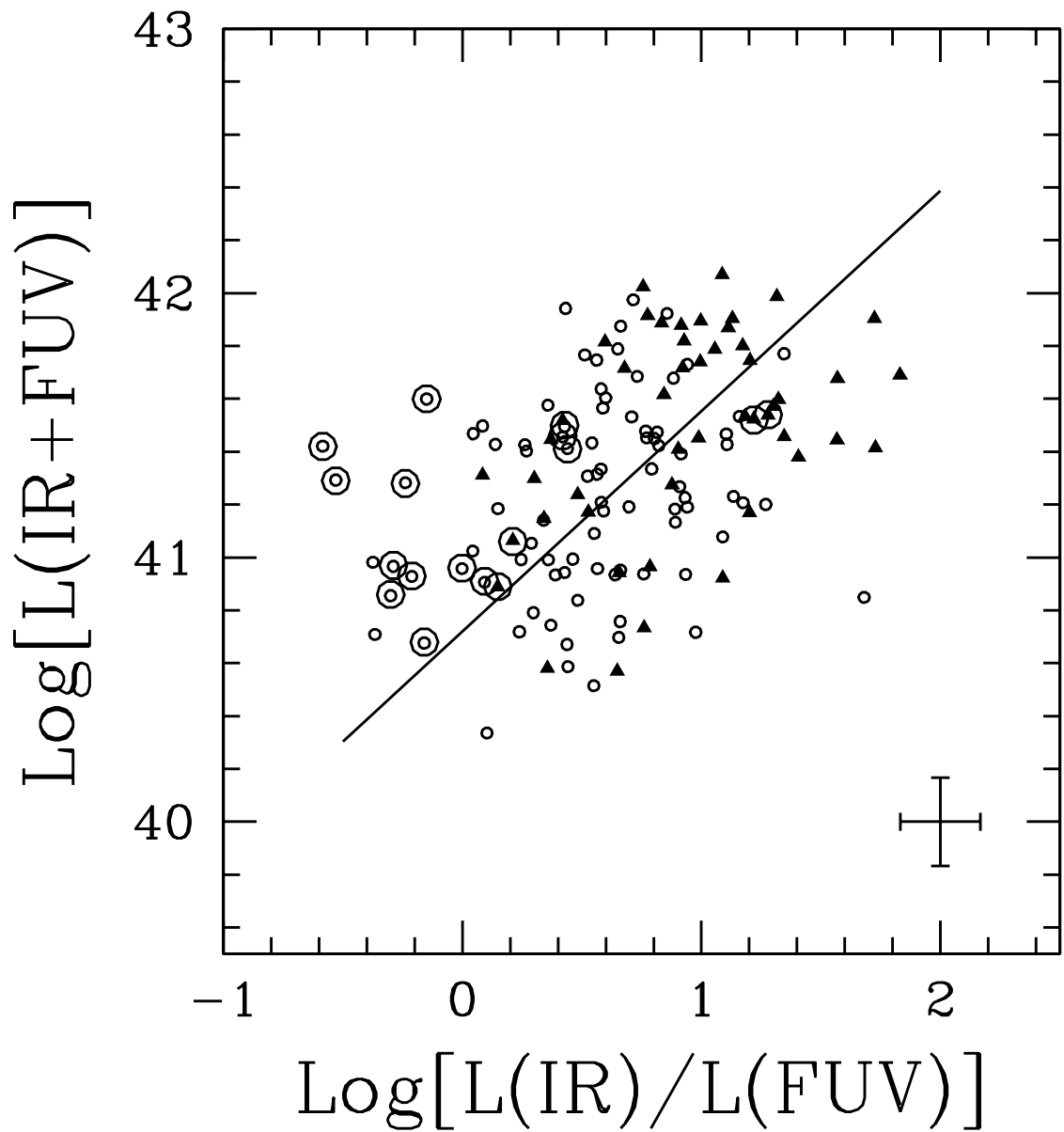}
\caption{The sum of the infrared and FUV luminosities as a function of
infrared-to-FUV ratio. Symbols are as in Figure~\ref{fig6}. The sum of
the IR and FUV emission is a proxy for total UV emission, and star
formation rate, in starburst galaxies \citep{heck98}. In NGC~5194 the
datapoints show some correlation (5.1~$\sigma$ significance). The
continuous line is from the best fit to starburst galaxies of
\citet{heck98}, shifted along the vertical axis by a factor
$\sim$40 to account for the lower luminosity of HII knots relative to
galaxies.
\label{fig13}}
\end{figure}

\clearpage 

\begin{figure}
\figurenum{14}
\plottwo{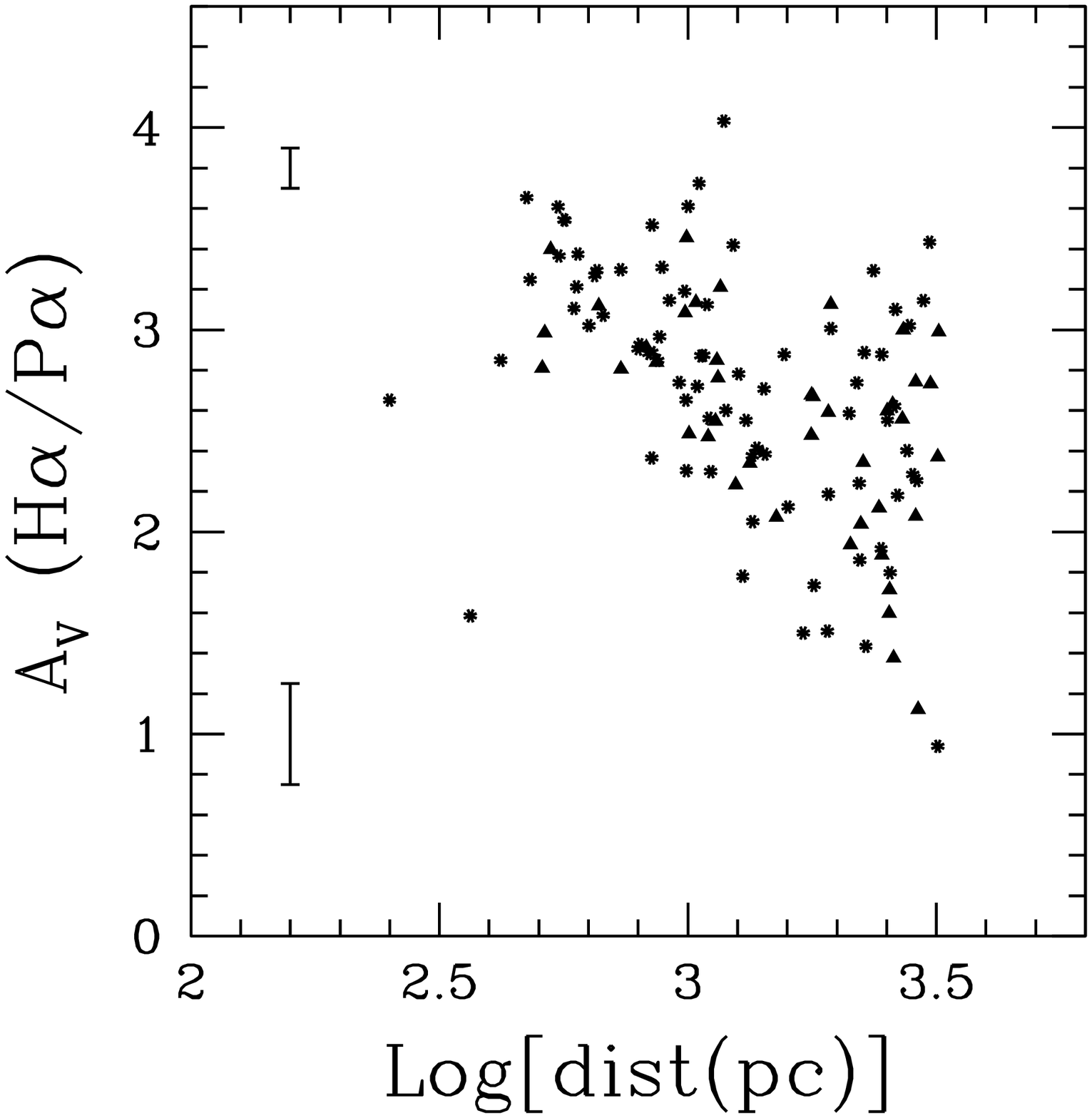}{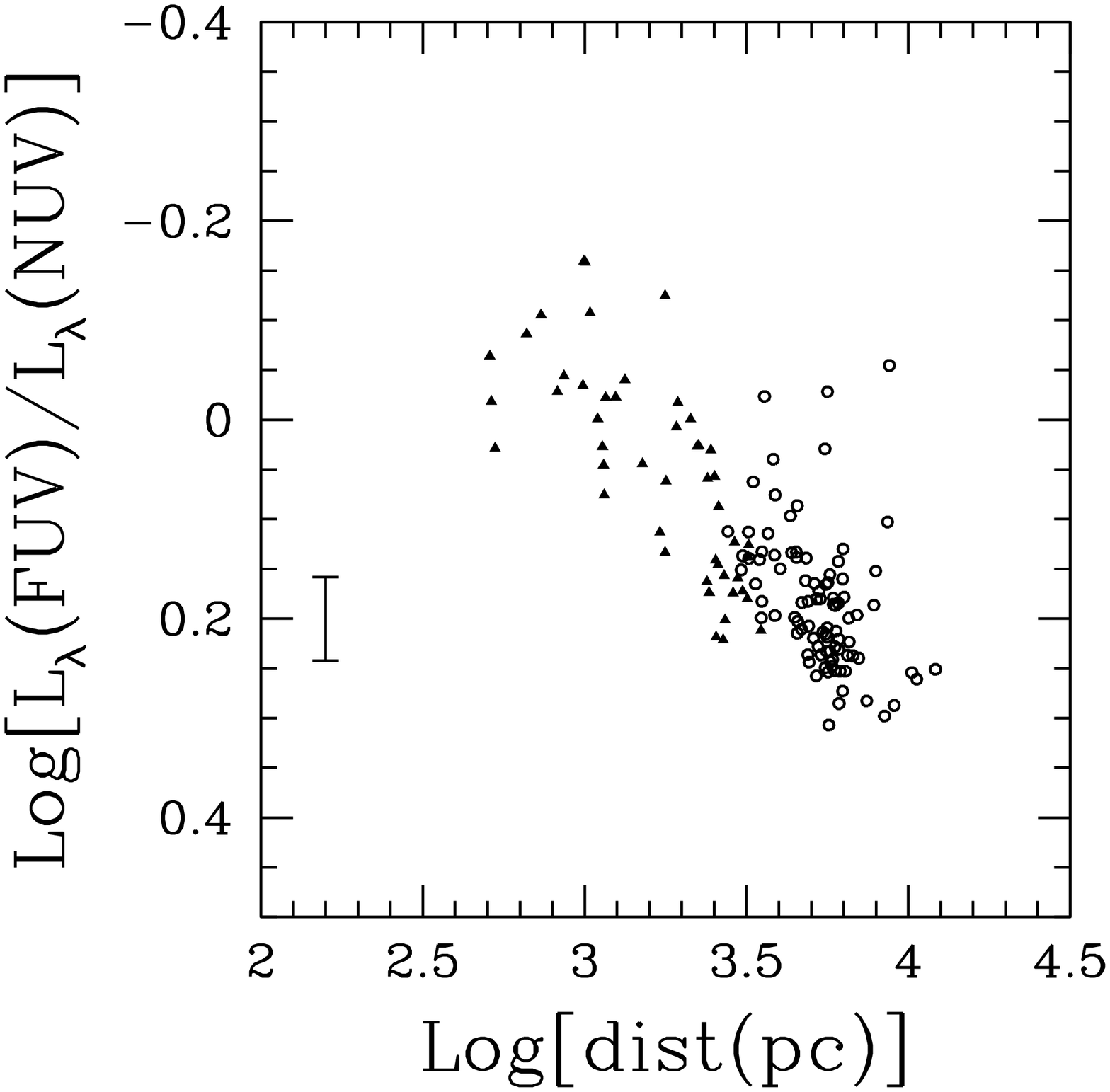}
\caption{{\bf (Left)}. The extinction at V, A$_V$, in magnitudes, as
a function of the distance from the galaxy's nucleus, from
$\sim$300~pc to 3.1~kpc. Data are shown for measurements performed in
the 13$^{\prime\prime}$--diameter apertures (filled triangles) and the
4$^{\prime\prime}$--diameter apertures (asterisks).  The two
asterisks closest to the nucleus show a strong
deviation (low A$_V$ values) from the general trend; the low A$_V$
values are consistent with those measured in the active nucleus
itself. The same `dip' is not measured in the larger aperture data, as
the closest datapoint to the nucleus (excluding the nucleus itself) is
over 500~pc away. {\bf (Right)}. The same plot for the observed UV
colors, up to $\sim$13~kpc distance from the nucleus.
\label{fig14}}
\end{figure}

\clearpage 

\begin{figure}
\figurenum{15}
\plottwo{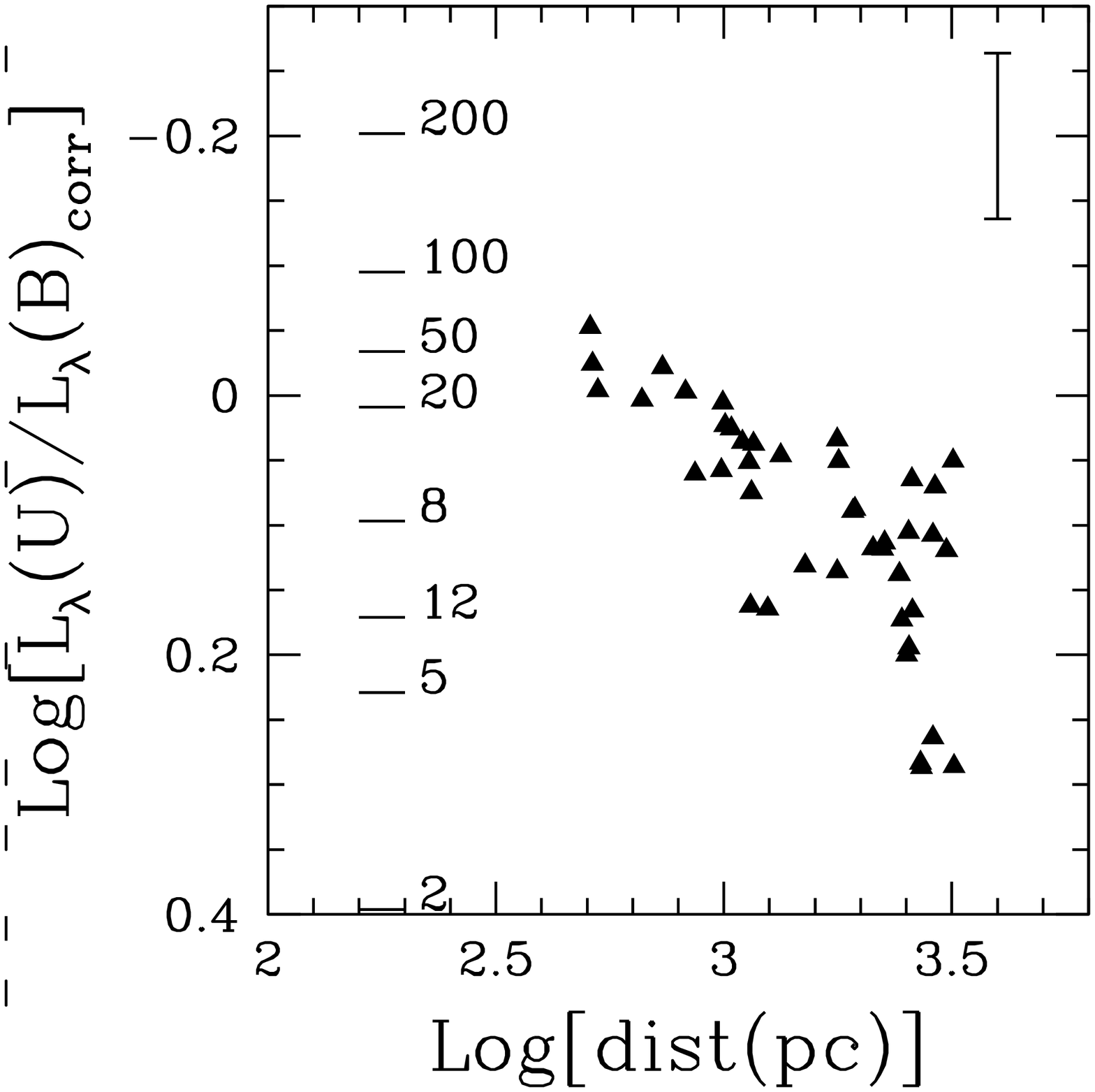}{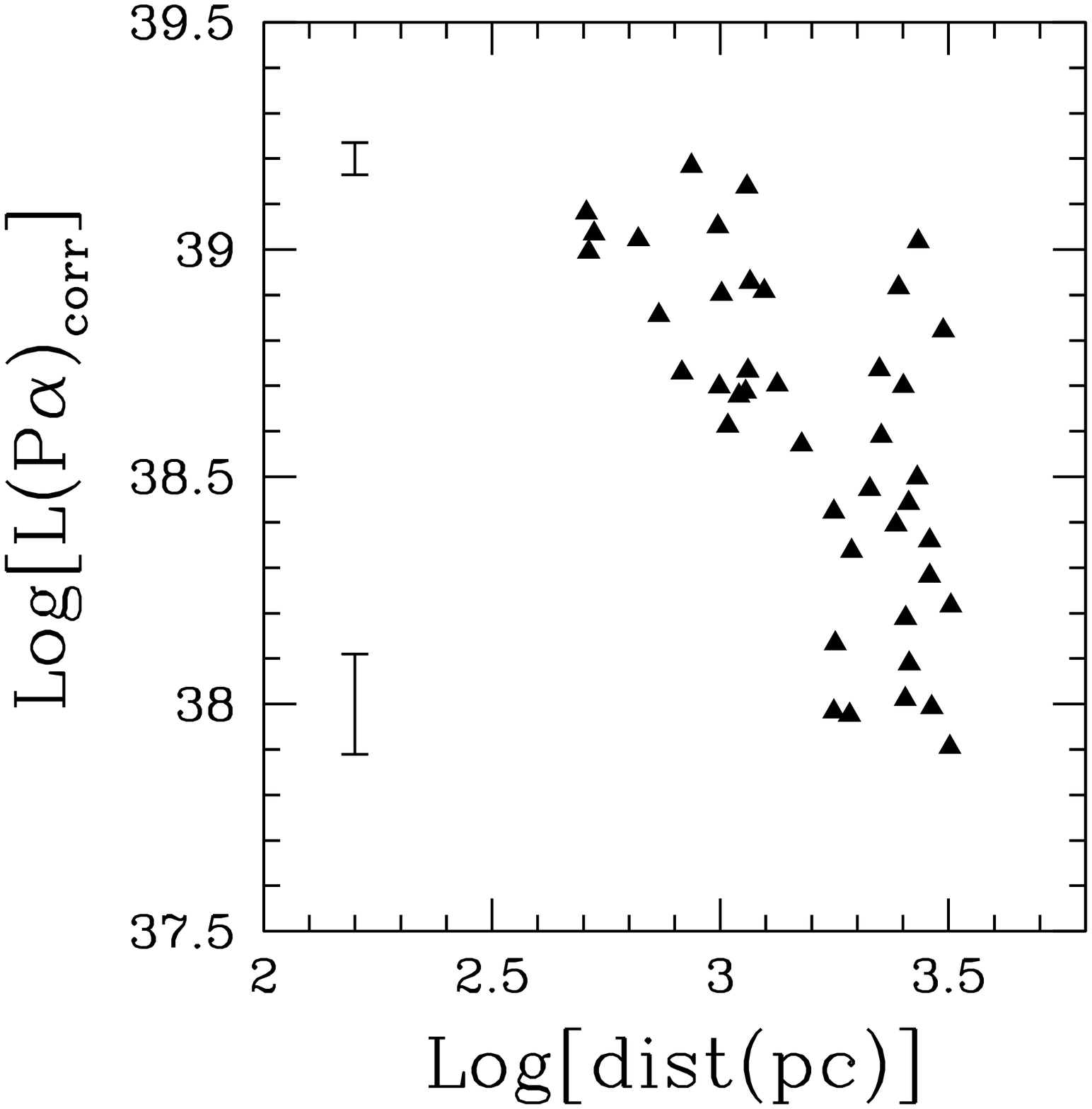}
\caption{The {\em extinction--corrected} L$_{\lambda}$(U)/L$_{\lambda}$(B)
color {\bf (left)} and the extinction--corrected P$\alpha$ luminosity
{\bf (right)} as a function of the distance from the galaxy's nucleus,
for the Inner Region. The expected colors of instantaneous burst
populations are shown as horizontal marks on the left diagram, marked
by their age (in Myr). The median uncertainties are shown in both
plots as vertical bars. The intrinsic colors of the UV--detected knots show a
trend for younger ages further away from the nucleus, while at the
same time the same regions show more massive young populations being
present closer to the nucleus.
\label{fig15}}
\end{figure}

\clearpage 
\begin{figure}
\figurenum{16}
\plotone{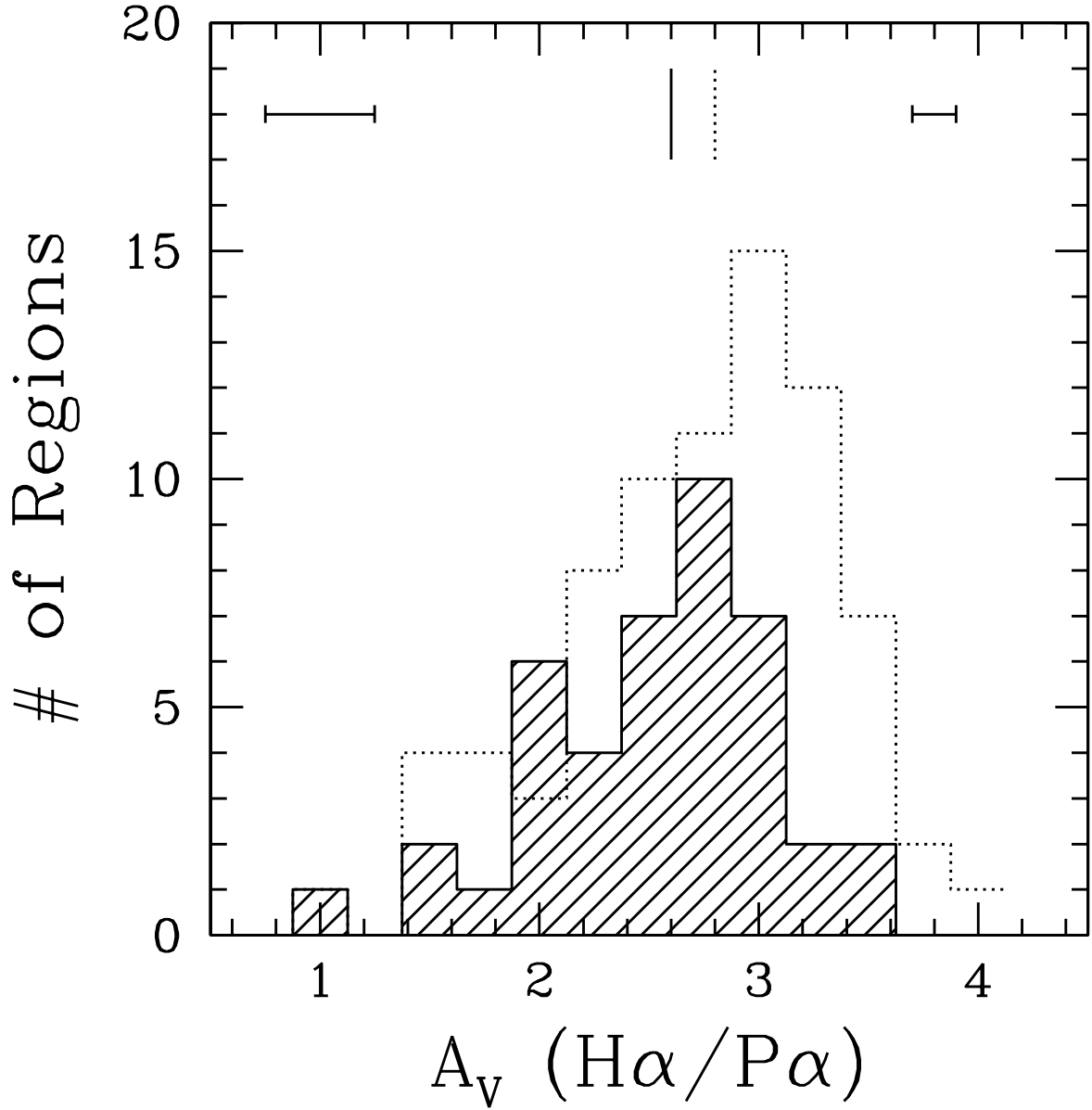}
\caption{Histogram of the extinction A$_V$ for the 42
13$^{\prime\prime}$--diameter regions (shaded histogram) and for the
78 4$^{\prime\prime}$--diameter regions (dotted histogram).  The
median values of the two distributions, A$_V\simeq$2.6~mag and 2.8~mag,
respectively, are shown as vertical bars at the top of the
diagram. The uncertainties in A$_V$ are shown as horizontal error
bars.
\label{fig16}}
\end{figure}

\clearpage 
\begin{figure}
\figurenum{17}
\plotone{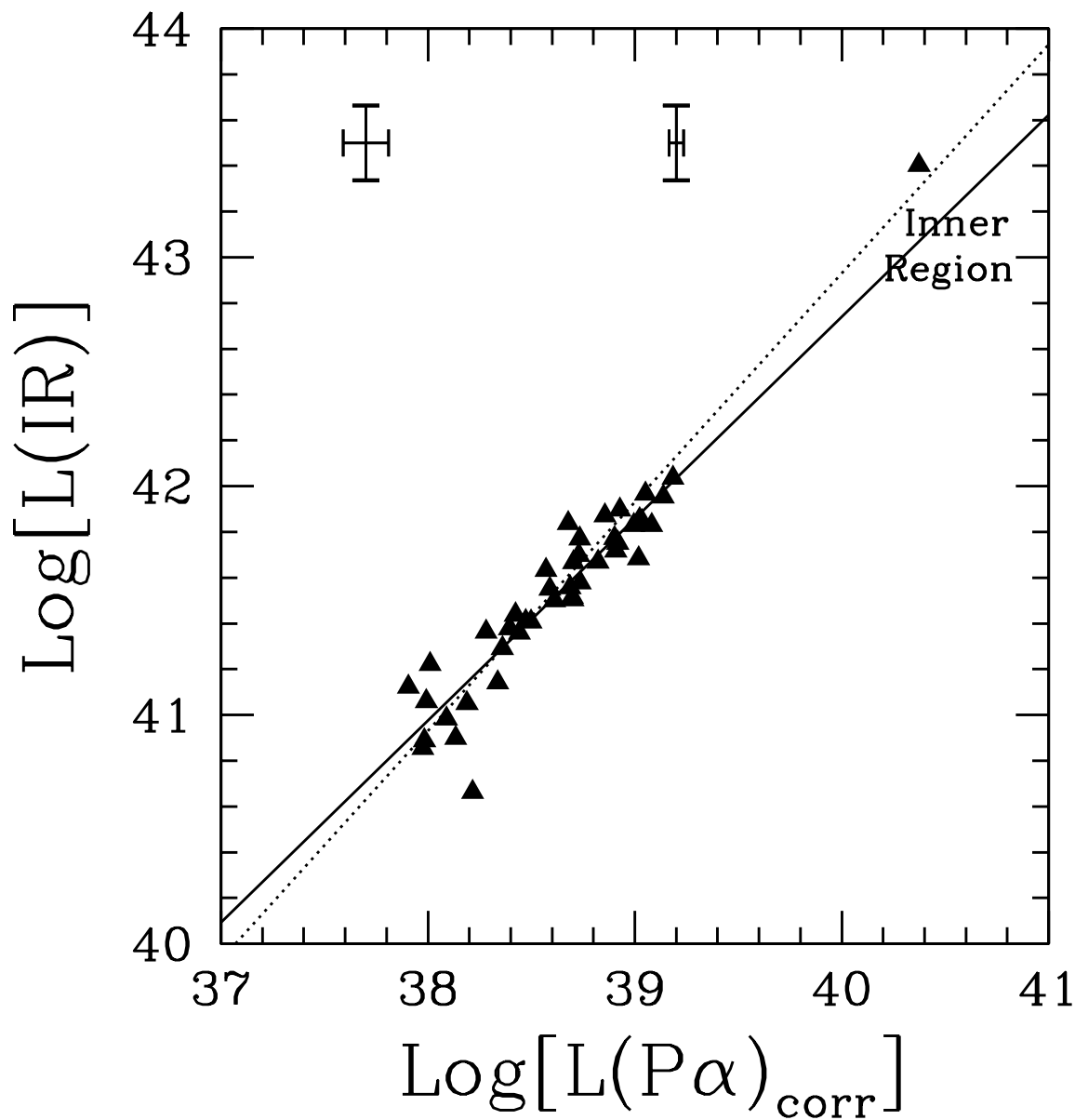}
\caption{The infrared luminosity as a function of the
extinction--corrected P$\alpha$ luminosity for the 42 knots in the
Inner Region.  The best fit line to the HII knots and the luminosity
relation with slope unity are shown as a continuous line and a dotted
horizontal line, respectively. The location on the plot of the
integrated light (background--subtracted) from the Inner Region is
also shown, and identified with its name. Median error bars are shown
at the two extremes of the P$\alpha$ luminosity range for the HII
knots.
\label{fig17}}
\end{figure}

\clearpage 
\begin{figure}
\figurenum{18}
\plotone{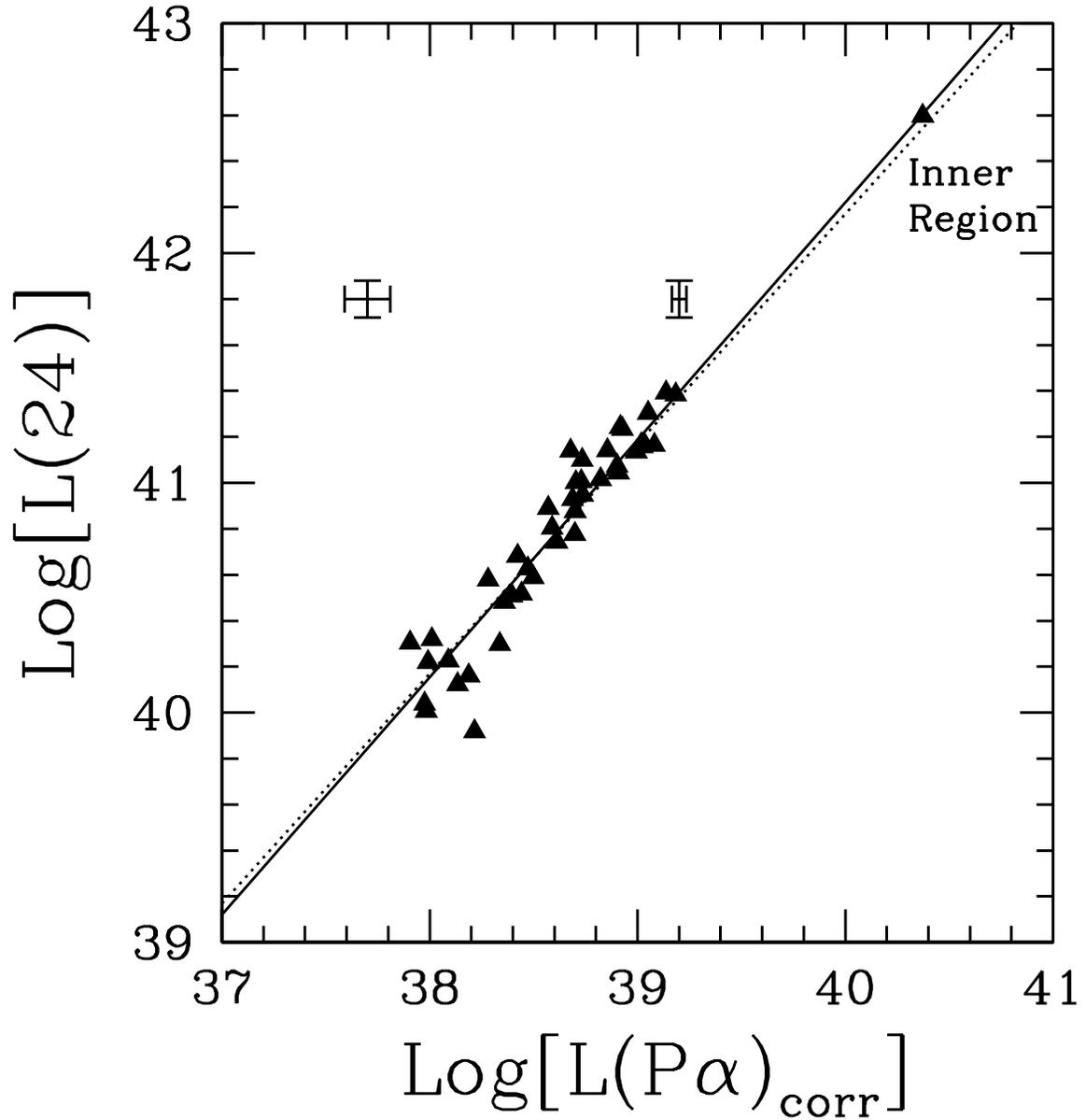}
\caption{The 24~$\mu$m luminosity as a function of the
extinction--corrected P$\alpha$ luminosity for the 42 knots in the
Inner Region. Points, lines, and error bars are as in Figure~\ref{fig17}.
\label{fig18}}
\end{figure}

\clearpage 
\begin{figure}
\figurenum{19}
\plotone{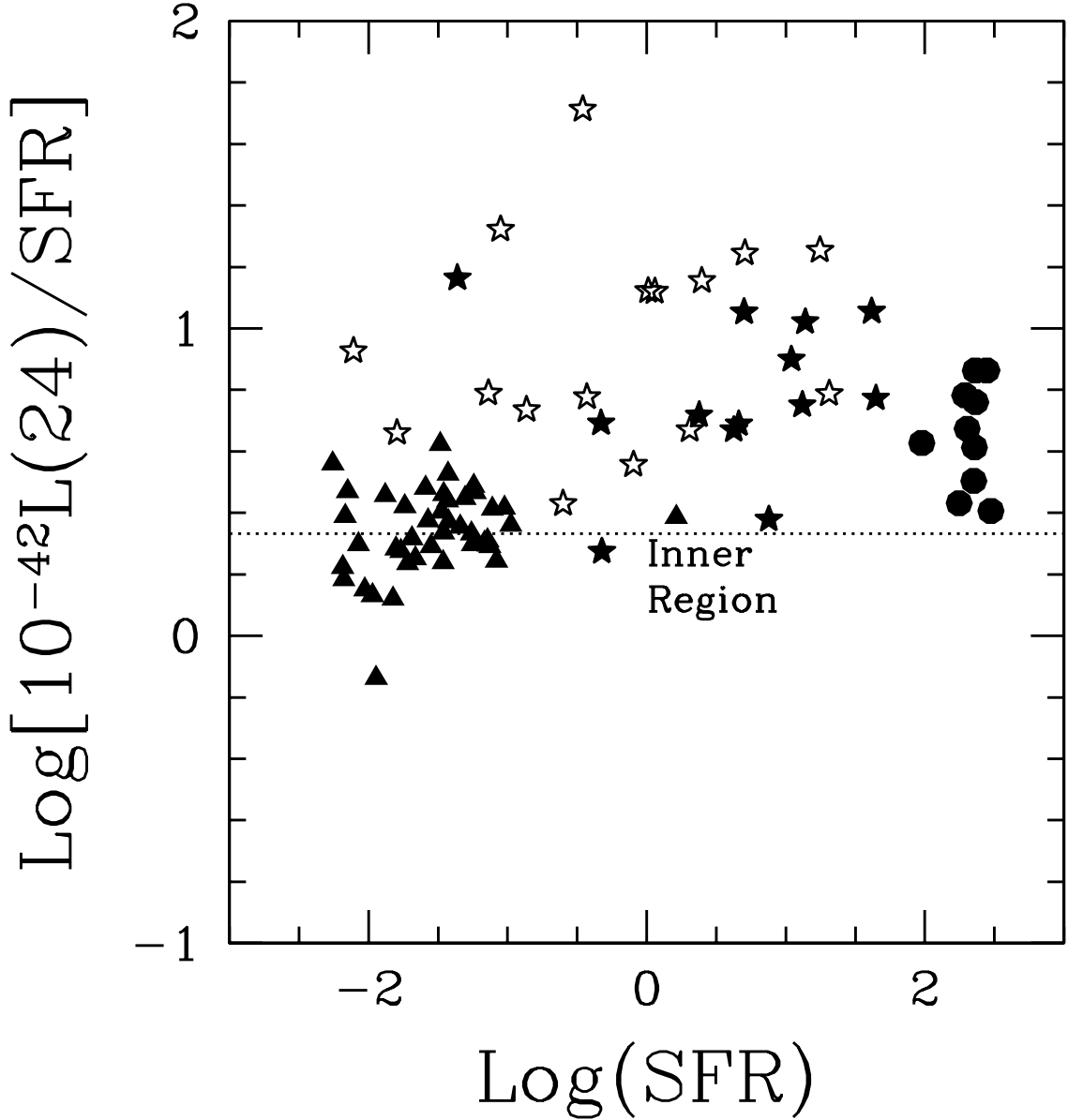}
\caption{The ratio of the 24~$\mu$m luminosity to the SFR as a
function of the SFR for the Inner Region's knots and for actively
star--forming galaxies. The units of the horizontal and vertical axes
are M$_{\odot}$~yr$^{-1}$ and
erg~s$^{-1}$~(M$_{\odot}$~yr$^{-1}$)$^{-1}$, respectively. The SFRs
for the 42 Inner Region's knots and the integrated Inner Region
(filled triangles) are from the extinction--corrected P$\alpha$
luminosity. The SFRs for the UV--selected starbursts are from the
extinction--corrected Br$\gamma$ line luminosity \citep[filled stars;
extinction correction from H$\beta$/Br$\gamma$][]{calz96} or from the
extinction--corrected H$\alpha$ luminosity (empty stars, extinction
correction from H$\alpha$/H$\beta$). Starburst--dominated ULIGs
(filled circles) are from \citet{gold02} and \citet{tren99}; for these
galaxies, SFRs are from the infrared luminosity using the formula of
\citet{kenn98}. For both sets of galaxies the IRAS 25~$\mu$m
luminosity is used as a good approximation of MIPS24. The horizontal
line is the best linear fit line through the Inner Region's knots
(from Figure~\ref{fig18}). This line is located along the lower
envelope of the locus occupied by galaxies (both starbursts and
ULIGs). The large spread of the UV--selected starbursts on this plane
is partially contributed by aperture mismatch between the IRAS and
emission line measurements, and, for the empty star symbols, also by
potentially insufficient extinction correction (lower inferred SFRs)
from the H$\alpha$/H$\beta$ line ratio.
\label{fig19}}
\end{figure}

\clearpage 
\begin{figure}
\figurenum{20}
\plottwo{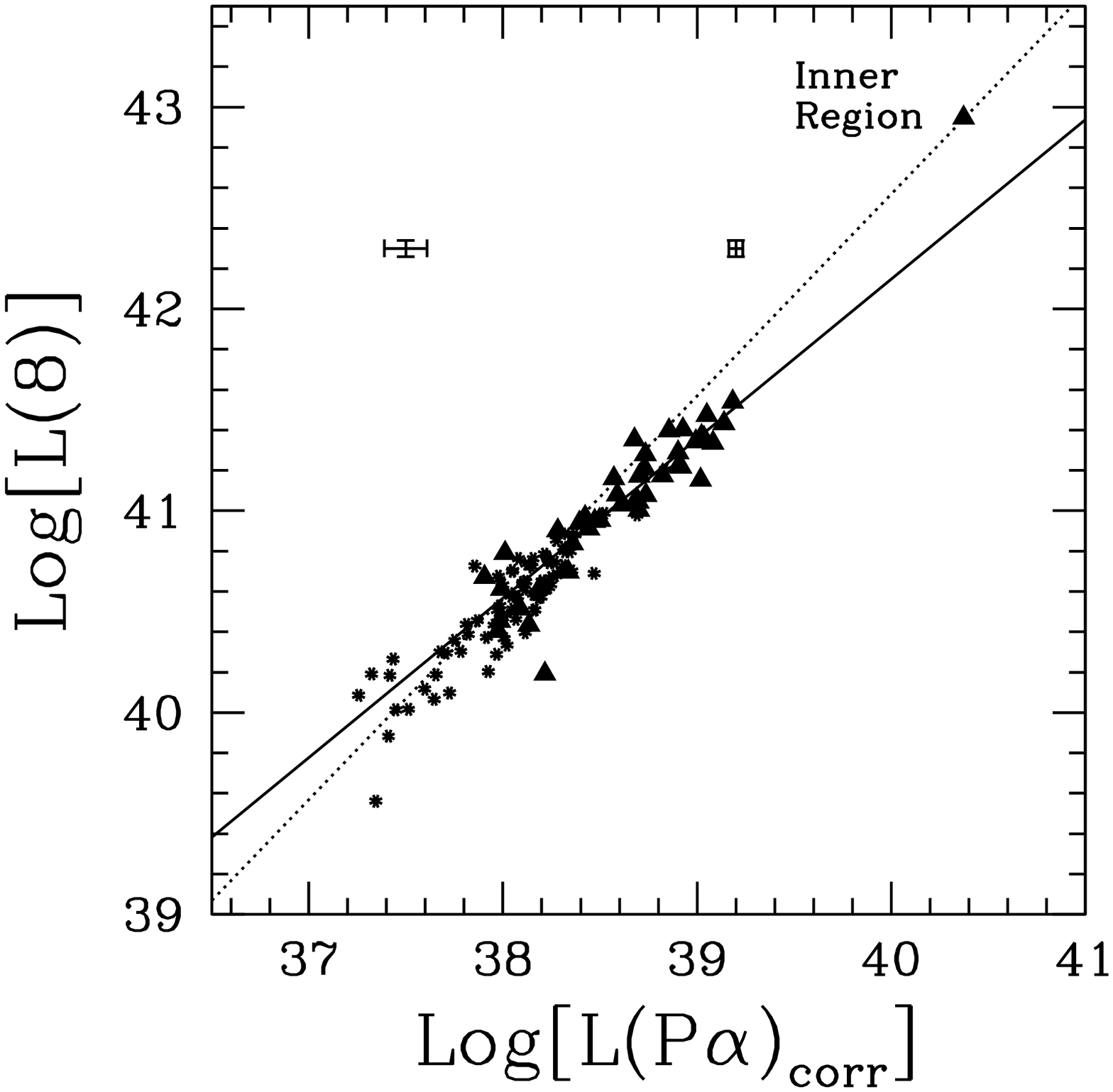}{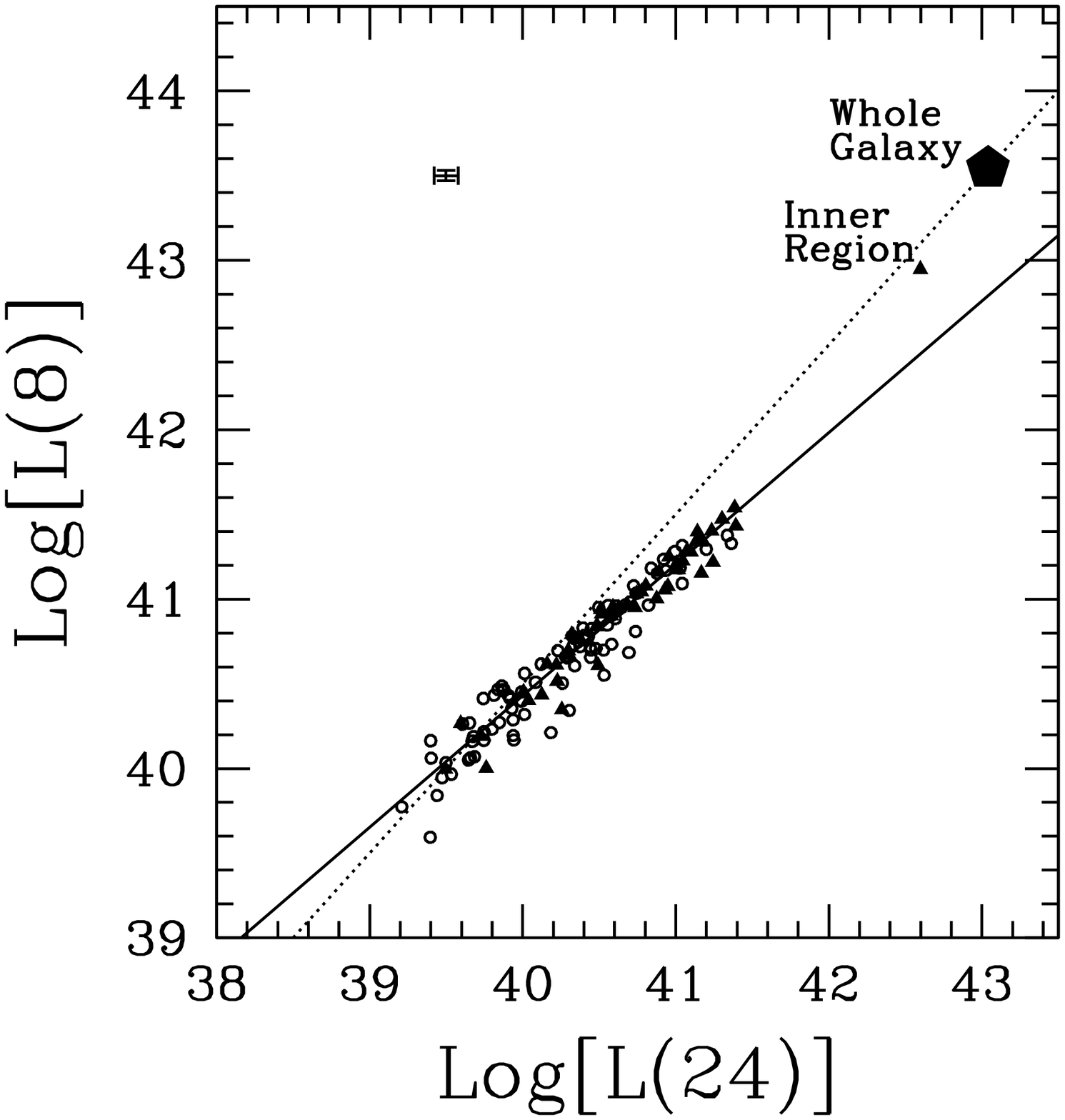}
\caption{{\bf (Left).} The 8~$\mu$m luminosity as a function of the
extinction--corrected P$\alpha$ luminosity in the Inner
Region. Triangles indicate photometry in the 42
13$^{\prime\prime}$--diameter apertures, while asterisks are for the
78 4$^{\prime\prime}$--diameter apertures. The triangle to the right
hand--side of the plot is the integrated value for the entire Inner
Region. Symbols, lines, and error bars are as in
Figure~\ref{fig17}. {\bf (Right).} The 8~$\mu$m luminosity as a
function of the 24~$\mu$m luminosity in the Inner$+$Outer
Regions. Data for the Outer Region are shown as empty
circles. Integrated values for both the Inner Region and the whole
galaxy (filled pentagon) are shown. The continuous line is the best
fit through the Inner$+$Outer Regions data, while the dotted line is
the slope unity relation through the whole galaxy's values.
\label{fig20}}
\end{figure}

%% If you are not including electonic art with your submission, you may
%% mark up your captions using the \figcaption command. See the 
%% No more than seven \figcaption commands are allowed per page, 
%% so if you have more than seven captions, insert a \clearpage 

%% Tables should be submitted one per page, so put a \clearpage before

%% Two options are available to the author for producing tables:  the
%% deluxetable environment provided by the AASTeX package or the LaTeX
%% table environment.  Use of deluxetable is preferred.
%%

%% Three table samples follow, two marked up in the deluxetable environment,
%% one marked up as a LaTeX table.

\clearpage

\begin{deluxetable}{lrrrrrrrrrr}
\rotate
\tabletypesize{\scriptsize}
\tablecaption{Positions and photometry of the HII knots.\label{tbl-1}}
\tablewidth{0pt}
\tablehead{
\colhead{ID\tablenotemark{a}} & \colhead{RA\tablenotemark{b}} 
& \colhead{DEC\tablenotemark{b}}  & \colhead{Log[L(FUV)]\tablenotemark{c}} 
& \colhead{Log[L(NUV)]\tablenotemark{c}} & \colhead{Log[L(U)]\tablenotemark{c}}
& \colhead{Log[L(B)]\tablenotemark{c}} 
& \colhead{Log[L(H$\alpha$)]\tablenotemark{c}} 
& \colhead{Log[L(P$\alpha$)]\tablenotemark{c}}
& \colhead{Log[L(8)]\tablenotemark{c}} 
& \colhead{Log[L(24)]\tablenotemark{c}}  \\
\colhead{} & \colhead{(J2000)} & \colhead{(J2000)} & \colhead{(erg~s$^{-1}$)} 
& \colhead{(erg~s$^{-1}$)} & \colhead{(erg~s$^{-1}$)} 
& \colhead{(erg~s$^{-1}$)} & \colhead{(erg~s$^{-1}$)} 
& \colhead{(erg~s$^{-1}$)} & \colhead{(erg~s$^{-1}$)} 
& \colhead{(erg~s$^{-1}$)}  
}
\startdata
 IR--02 & 13:29:53.1 & 47:11:55 &40.91 & 41.16 & 41.41 & 41.73 &39.10 &38.92 &41.34 &41.16\\
 IR--03 & 13:29:51.5 & 47:11:42 &41.07 & 41.22 & 41.45 & 41.75 &38.86 &38.83 &41.36 &41.17\\
 IR--04 & 13:29:52.6 & 47:11:30 &41.00 & 41.19 & 41.40 & 41.69 &38.96 &38.82 &41.34 &41.14\\
 IR--05 & 13:29:53.9 & 47:11:27 &40.74 & 41.03 & 41.28 & 41.56 &38.88 &38.69 &41.40 &41.14\\
 IR--06 & 13:29:54.5 & 47:11:40 &40.86 & 41.12 & 41.33 & 41.60 &38.94 &38.84 &41.37 &41.16\\
 IR--07 & 13:29:55.6 & 47:11:34 &40.84 & 40.95 & 41.03 & 41.21 &38.77 &38.57 &41.28 &41.10\\
 IR--08 & 13:29:55.7 & 47:11:47 &41.20 & 41.33 & 41.29 & 41.39 &39.15 &38.97 &41.43 &41.39\\
 IR--09 & 13:29:56.3 & 47:11:58 &40.95 & 41.09 & 41.07 & 41.15 &38.83 &38.45 &41.16 &40.89\\
 IR--10 & 13:29:54.9 & 47:11:58 &40.60 & 40.94 & 41.07 & 41.29 &39.03 &38.76 &41.29 &41.07\\
 IR--11 & 13:29:54.4 & 47:12:10 &41.12 & 41.32 & 41.33 & 41.39 &39.12 &38.78 &41.23 &41.04\\
 IR--12 & 13:29:53.2 & 47:12:11 &40.72 & 40.87 & 40.90 & 41.09 &38.79 &38.54 &41.06 &40.93\\
 IR--13 & 13:29:52.0 & 47:12:03 &40.95 & 41.17 & 41.28 & 41.48 &39.20 &39.02 &41.54 &41.38\\
 IR--14 & 13:29:50.9 & 47:11:57 &40.65 & 40.87 & 41.12 & 41.33 &38.98 &38.87 &41.47 &41.30\\
 IR--15 & 13:29:50.1 & 47:11:46 &40.72 & 40.91 & 41.08 & 41.28 &38.81 &38.53 &41.35 &41.14\\
 IR--16 & 13:29:50.1 & 47:11:33 &40.17 & 40.38 & 40.64 & 40.89 &38.82 &38.74 &41.40 &41.23\\
 IR--17 & 13:29:50.3 & 47:11:20 &40.10 & 40.32 & 40.60 & 40.79 &38.88 &38.57 &41.17 &41.00\\
 IR--18 & 13:29:51.3 & 47:11:29 &40.70 & 40.91 & 41.13 & 41.40 &38.72 &38.56 &41.21 &41.01\\
 IR--20 & 13:29:50.4 & 47:10:43 &40.97 & 40.93 & 40.85 & 40.84 &38.57 &38.09 &40.62 &40.16\\
 IR--21 & 13:29:51.6 & 47:10:39 &40.34 & 40.38 & 40.49 & 40.57 &38.43 &37.92 &40.79 &40.32\\
 IR--22 & 13:29:50.8 & 47:10:23 &39.90 & 39.95 & 40.24 & 40.23 &38.18 &38.04 &40.19 &39.92\\
 IR--24 & 13:29:53.1 & 47:10:40 &40.96 & 40.97 & 41.03 & 41.11 &38.64 &38.27 &40.94 &40.51\\
 IR--25 & 13:29:54.3 & 47:10:38 &40.46 & 40.49 & 40.67 & 40.86 &38.52 &38.29 &40.91 &40.52\\
 IR--28 & 13:30:00.6 & 47:11:24 &40.82 & 40.82 & 40.90 & 41.08 &38.07 &37.77 &40.67 &40.30\\
 IR--29 & 13:29:59.0 & 47:11:04 &40.92 & 40.93 & 40.93 & 41.08 &38.40 &38.20 &40.84 &40.48\\
 IR--30 & 13:30:00.0 & 47:11:12 &40.75 & 40.76 & 40.83 & 40.97 &38.87 &38.66 &41.17 &41.02\\
 IR--31 & 13:29:56.5 & 47:11:16 &39.80 & 39.85 & 40.21 & 40.42 &38.11 &37.84 &40.45 &40.01\\
 IR--32 & 13:29:57.3 & 47:11:35 &40.11 & 40.23 & 40.39 & 40.59 &38.20 &37.98 &40.43 &40.12\\
 IR--33 & 13:29:54.8 & 47:12:24 &40.09 & 40.40 & 40.55 & 40.67 &38.49 &38.27 &40.97 &40.68\\
 IR--34 & 13:29:56.0 & 47:12:27 &40.42 & 40.60 & 40.75 & 40.84 &38.78 &38.36 &40.95 &40.63\\
 IR--35 & 13:29:54.9 & 47:12:36 &40.25 & 40.40 & 40.66 & 40.78 &38.76 &38.45 &41.08 &40.80\\
 IR--36 & 13:29:53.3 & 47:12:39 &40.26 & 40.41 & 40.59 & 40.68 &39.01 &38.62 &41.08 &40.95\\
 IR--37 & 13:29:52.2 & 47:12:45 &40.69 & 40.84 & 40.86 & 40.89 &39.24 &38.81 &41.22 &41.24\\
 IR--38 & 13:29:50.9 & 47:12:44 &40.32 & 40.45 & 40.54 & 40.59 &38.79 &38.55 &41.00 &40.87\\
 IR--39 & 13:29:49.5 & 47:12:43 &39.85 & 39.83 & 40.29 & 40.28 &38.98 &38.84 &41.15 &41.17\\
 IR--41 & 13:29:47.3 & 47:12:22 &39.68 & 39.70 & 40.05 & 40.01 &38.60 &38.35 &40.95 &40.59\\
 IR--43 & 13:29:46.3 & 47:12:13 &39.96 & 39.96 & 40.33 & 40.28 &38.54 &38.16 &40.90 &40.58\\
 IR--47 & 13:29:57.6 & 47:11:52 &40.20 & 40.37 & 40.45 & 40.61 &38.07 &37.82 &40.40 &40.04\\
 IR--48 & 13:29:56.3 & 47:12:39 &40.64 & 40.74 & 40.73 & 40.73 &38.58 &38.01 &40.52 &40.23\\
 IR--49 & 13:29:47.8 & 47:12:36 &40.53 & 40.59 & 40.69 & 40.78 &38.57 &37.93 &40.61 &40.22\\
 IR--50 & 13:29:56.7 & 47:12:12 &39.94 & 40.14 & 40.46 & 40.65 &38.26 &38.15 &40.70 &40.30\\
 IR--53 & 13:29:51.9 & 47:11:18 &40.28 & 40.57 & 40.87 & 41.12 &38.53 &38.43 &41.03 &40.74\\
 IR--54 & 13:29:53.1 & 47:11:17 &40.24 & 40.58 & 40.87 & 41.16 &38.51 &38.50 &41.04 &40.78\\
 01--01 & 13:29:52.8 & 47:13:54 &40.42 & 40.42 & 40.37 & 40.57 &38.61 & ... &40.66 &40.28\\
 01--02 & 13:29:53.7 & 47:13:37 &39.95 & 40.05 & 39.88 & 40.03 &38.27 & ... &40.49 &39.86\\
 01--04 & 13:29:50.5 & 47:13:56 &40.00 & 40.00 & 40.05 & 40.06 &38.70 & ... &40.72 &40.38\\
 01--05 & 13:29:49.6 & 47:13:28 &39.91 & 39.95 & 39.77 & 39.91 &38.25 & ... &40.70 &40.45\\
 01--06 & 13:29:47.5 & 47:13:25 &40.31 & 40.29 & 39.99 & 40.08 &38.41 & ... &40.78 &40.32\\
 01--07 & 13:29:47.2 & 47:13:39 &41.14 & 41.10 & 40.90 & 41.05 &38.91 & ... &40.72 &40.44\\
 01--08 & 13:29:46.0 & 47:13:45 &40.70 & 40.63 & 40.41 & 40.45 &38.68 & ... &40.29 &39.94\\
 01--09 & 13:29:45.1 & 47:13:32 &40.58 & 40.59 & 40.46 & 40.55 &38.87 & ... &40.95 &40.50\\
 01--12 & 13:29:50.8 & 47:13:22 &40.28 & 40.31 & 40.22 & 40.44 &37.88 & ... &40.07 &39.69\\
 02--02 & 13:29:38.9 & 47:12:13 &40.22 & 40.09 & 39.96 & 40.08 &38.06 & ... &40.16 &39.67\\
 02--03 & 13:29:38.3 & 47:12:03 &39.96 & 39.96 & 39.83 & 39.93 &37.98 & ... &40.17 &39.75\\
 02--08 & 13:29:40.7 & 47:11:35 &40.29 & 40.33 & 40.18 & 40.43 &38.03 & ... &40.43 &39.91\\
 02--09 & 13:29:38.7 & 47:11:05 &40.45 & 40.44 & 40.31 & 40.54 &31.65 & ... &39.84 &39.44\\
 03--01 & 13:29:37.1 & 47:09:46 &40.25 & 40.24 & 40.24 & 40.44 &38.72 & ... &40.73 &40.33\\
 03--03 & 13:29:39.4 & 47:08:40 &40.35 & 40.24 & 40.32 & 40.43 &39.09 & ... &41.04 &40.74\\
 03--12 & 13:29:38.6 & 47:08:57 &39.16 & 39.39 & 38.98 & 39.42 &38.07 & ... &40.41 &39.92\\
 03--14 & 13:29:40.2 & 47:08:25 &38.13 & 39.23 & 38.89 & 39.07 &38.12 & ... &40.45 &39.99\\
 04--01 & 13:29:46.4 & 47:12:33 &40.65 & 40.72 & 40.59 & 40.81 &38.75 & ... &40.73 &40.58\\
 04--03 & 13:29:43.1 & 47:11:38 &41.15 & 41.20 & 41.01 & 41.21 &38.71 & ... &40.78 &40.43\\
 04--04 & 13:29:42.0 & 47:11:11 &40.43 & 40.47 & 40.36 & 40.54 &38.70 & ... &40.89 &40.52\\
 04--05 & 13:29:42.8 & 47:10:49 &40.53 & 40.57 & 40.47 & 40.74 &38.52 & ... &40.65 &40.29\\
 04--06 & 13:29:42.8 & 47:11:02 &40.20 & 40.29 & 40.32 & 40.69 &38.30 & ... &40.66 &40.44\\
 04--07 & 13:29:43.3 & 47:11:21 &40.24 & 40.35 & 40.29 & 40.63 &37.99 & ... &40.70 &40.23\\
 04--08 & 13:29:45.4 & 47:10:51 &40.21 & 40.41 & 40.94 & 41.39 &37.94 & ... &40.40 &39.99\\
 04--09 & 13:29:45.0 & 47:12:26 &40.55 & 40.55 & 40.39 & 40.64 &38.26 & ... &40.36 &39.93\\
 05--01 & 13:29:44.2 & 47:10:23 &41.18 & 41.18 & 41.01 & 41.17 &39.47 & ... &41.38 &41.34\\
 05--02 & 13:29:44.8 & 47:09:58 &41.37 & 41.33 & 41.12 & 41.25 &39.37 & ... &41.30 &41.20\\
 05--03 & 13:29:43.5 & 47:10:00 &40.33 & 40.48 & 40.40 & 40.61 &38.75 & ... &40.97 &40.67\\
 05--04 & 13:29:46.9 & 47:09:39 &41.05 & 41.02 & 40.84 & 41.01 &38.76 & ... &40.71 &40.48\\
 05--06 & 13:29:47.9 & 47:10:29 &40.40 & 40.44 & 40.30 & 40.58 &38.21 & ... &40.32 &40.01\\
 05--07 & 13:29:50.7 & 47:10:22 &40.01 & 40.13 & 39.97 & 40.33 &38.15 & ... &40.19 &39.94\\
 05--08 & 13:29:50.4 & 47:09:44 &40.64 & 40.66 & 40.55 & 40.71 &38.81 & ... &40.96 &40.56\\
 05--09 & 13:29:49.0 & 47:09:19 &40.21 & 40.16 & 40.14 & 40.38 &38.13 & ... &40.47 &39.84\\
 05--10 & 13:29:50.6 & 47:09:21 &40.65 & 40.67 & 40.52 & 40.69 &38.57 & ... &40.77 &40.34\\
 05--11 & 13:29:51.9 & 47:09:22 &39.70 & 39.91 & 39.34 & 39.79 &38.01 & ... &40.23 &39.80\\
 05--13 & 13:29:44.2 & 47:09:36 &40.38 & 40.27 & 40.01 & 40.24 &38.19 & ... &40.41 &39.74\\
 05--14 & 13:29:45.1 & 47:09:26 &40.10 & 40.00 & 39.63 & 39.82 &38.21 & ... &40.16 &39.40\\
 05--17 & 13:29:53.5 & 47:10:15 &40.79 & 40.77 & 40.52 & 40.74 &37.91 & ... &40.05 &39.65\\
 05--19 & 13:29:54.6 & 47:10:18 &40.72 & 40.76 & 40.56 & 40.79 &38.08 & ... &40.06 &39.66\\
 06--07 & 13:29:49.2 & 47:08:09 &39.98 & 40.06 & 39.88 & 40.07 &37.74 & ... &39.59 &39.40\\
 06--09 & 13:29:53.5 & 47:08:37 &40.32 & 40.21 & 39.98 & 40.18 &38.10 & ... &40.19 &39.68\\
 06--10 & 13:29:51.9 & 47:08:24 &39.86 & 39.89 & 39.44 & 39.45 &37.99 & ... &39.97 &39.53\\
 06--11 & 13:29:52.8 & 47:08:46 &40.56 & 40.50 & 40.26 & 40.47 &38.28 & ... &40.22 &39.75\\
 07--01 & 13:30:01.4 & 47:12:50 &41.01 & 40.99 & 40.79 & 40.96 &39.09 & ... &41.33 &41.36\\
 07--02 & 13:30:00.7 & 47:13:07 &41.13 & 41.10 & 40.93 & 41.09 &39.22 & ... &41.32 &41.04\\
 07--03 & 13:30:00.4 & 47:13:18 &41.05 & 41.02 & 40.85 & 41.03 &39.01 & ... &41.24 &40.92\\
 07--04 & 13:29:59.9 & 47:13:31 &41.08 & 41.00 & 40.83 & 41.03 &38.98 & ... &41.18 &40.84\\
 07--05 & 13:29:59.6 & 47:13:58 &41.14 & 41.13 & 40.95 & 41.02 &39.34 & ... &41.17 &40.94\\
 07--06 & 13:29:58.7 & 47:14:08 &40.88 & 40.90 & 40.74 & 40.79 &39.03 & ... &40.97 &40.82\\
 07--07 & 13:29:57.6 & 47:13:56 &40.75 & 40.76 & 40.65 & 40.81 &38.78 & ... &40.98 &40.70\\
 07--08 & 13:29:58.6 & 47:13:49 &40.91 & 40.92 & 40.80 & 40.96 &39.03 & ... &41.04 &40.75\\
 07--09 & 13:29:59.9 & 47:13:43 &41.06 & 41.01 & 40.81 & 40.97 &38.81 & ... &40.96 &40.61\\
 07--10 & 13:30:04.7 & 47:13:03 &40.98 & 40.91 & 40.70 & 40.80 &38.70 & ... &40.79 &40.40\\
 07--12 & 13:29:55.5 & 47:14:01 &40.74 & 40.71 & 40.56 & 40.63 &39.00 & ... &41.15 &40.88\\
 07--13 & 13:29:55.7 & 47:13:48 &40.29 & 40.30 & 40.28 & 40.45 &38.70 & ... &40.93 &40.61\\
 07--14 & 13:29:53.9 & 47:14:05 &40.19 & 40.12 & 40.08 & 40.15 &38.30 & ... &40.61 &40.34\\
 07--15 & 13:30:02.0 & 47:13:01 &41.37 & 41.31 & 41.09 & 41.27 &38.64 & ... &40.69 &40.69\\
 07--16 & 13:30:01.1 & 47:13:32 &41.32 & 41.29 & 40.99 & 41.15 &38.45 & ... &40.21 &40.18\\
 07--17 & 13:30:01.3 & 47:13:45 &41.18 & 41.15 & 40.92 & 41.04 &38.23 & ... &40.17 &39.94\\
 07--18 & 13:30:03.0 & 47:12:46 &40.84 & 40.84 & 40.62 & 40.77 &38.57 & ... &40.83 &40.45\\
 08--01 & 13:29:56.5 & 47:10:45 &40.41 & 40.47 & 40.53 & 40.75 &38.88 & ... &41.28 &41.00\\
 08--02 & 13:29:60.0 & 47:11:12 &40.74 & 40.78 & 40.68 & 40.90 &38.88 & ... &41.19 &41.03\\
 08--03 & 13:30:00.9 & 47:11:37 &40.96 & 40.97 & 40.83 & 41.04 &38.90 & ... &41.08 &40.73\\
 08--04 & 13:30:01.7 & 47:11:47 &40.60 & 40.66 & 40.61 & 40.86 &38.60 & ... &40.95 &40.64\\
 08--05 & 13:30:02.0 & 47:11:59 &40.61 & 40.60 & 40.49 & 40.70 &38.67 & ... &40.92 &40.59\\
 08--06 & 13:30:03.3 & 47:12:19 &40.54 & 40.51 & 40.20 & 39.97 &38.85 & ... &40.90 &40.57\\
 08--07 & 13:30:01.6 & 47:12:14 &40.07 & 40.21 & 40.26 & 40.58 &38.08 & ... &40.75 &40.37\\
 08--08 & 13:30:03.7 & 47:12:35 &40.48 & 40.41 & 39.78 & 40.14 &38.51 & ... &40.83 &40.39\\
 08--09 & 13:29:57.6 & 47:10:45 &40.90 & 40.94 & 40.84 & 41.10 &38.44 & ... &40.85 &40.55\\
 08--10 & 13:29:59.0 & 47:11:03 &40.93 & 40.96 & 40.82 & 41.06 &38.43 & ... &40.89 &40.61\\
 08--11 & 13:30:05.0 & 47:12:32 &41.08 & 41.03 & 40.76 & 40.88 &38.05 & ... &40.42 &39.91\\
 09--01 & 13:29:55.7 & 47:09:13 &40.11 & 40.15 & 39.99 & 40.10 &38.05 & ... &40.46 &39.87\\
 09--03 & 13:29:58.8 & 47:09:13 &40.95 & 40.89 & 40.68 & 40.77 &38.84 & ... &40.76 &40.40\\
 09--04 & 13:30:00.9 & 47:09:28 &40.43 & 40.43 & 40.38 & 40.24 &38.87 & ... &40.50 &40.26\\
 09--05 & 13:30:02.4 & 47:09:46 &40.88 & 40.83 & 40.73 & 40.82 &39.29 & ... &41.09 &41.04\\
 09--06 & 13:30:03.0 & 47:09:57 &40.78 & 40.78 & 40.59 & 40.70 &38.99 & ... &40.81 &40.74\\
 09--07 & 13:30:03.4 & 47:09:40 &40.67 & 40.63 & 40.48 & 40.59 &38.86 & ... &40.70 &40.53\\
 09--08 & 13:30:04.3 & 47:10:03 &40.58 & 40.51 & 40.35 & 40.51 &38.28 & ... &40.47 &39.88\\
 09--09 & 13:30:04.9 & 47:10:22 &40.64 & 40.57 & 40.34 & 40.44 &38.37 & ... &40.56 &40.01\\
 09--10 & 13:30:04.8 & 47:10:35 &39.95 & 39.92 & 39.81 & 39.77 &38.41 & ... &40.62 &40.12\\
 09--11 & 13:30:03.8 & 47:10:15 &40.68 & 40.63 & 40.42 & 40.57 &37.86 & ... &39.95 &39.47\\
 09--12 & 13:29:54.6 & 47:09:04 &40.66 & 40.59 & 40.40 & 40.56 &38.17 & ... &40.26 &39.61\\
 10--01 & 13:30:07.0 & 47:11:35 &40.80 & 40.73 & 40.56 & 40.67 &38.74 & ... &40.51 &40.08\\
 10--02 & 13:30:08.1 & 47:11:38 &40.01 & 40.06 & 39.60 & 39.76 &38.30 & ... &40.03 &39.50\\
 10--03 & 13:30:06.7 & 47:11:22 &40.55 & 40.49 & 40.25 & 40.28 &38.44 & ... &39.77 &39.21\\
 10--05 & 13:30:07.0 & 47:12:27 &40.47 & 40.43 & 40.19 & 40.26 &38.52 & ... &40.43 &39.81\\
 10--06 & 13:30:07.7 & 47:12:43 &40.23 & 40.21 & 40.06 & 40.15 &38.44 & ... &40.27 &39.85\\
 10--11 & 13:30:06.8 & 47:14:19 &40.83 & 40.71 & 40.48 & 40.47 &38.44 & ... &40.06 &39.40\\
 11--02 & 13:30:04.0 & 47:15:34 &40.49 & 40.41 & 40.25 & 40.34 &38.74 & ... &40.55 &40.53\\
 11--03 & 13:30:03.6 & 47:15:46 &40.40 & 40.32 & 40.25 & 40.37 &38.66 & ... &40.34 &40.30\\

 \enddata

%% Text for table notes should follow after the \enddata but before
%% the \end{deluxetable}. Make sure there is at least one \tablenotemark
%% in the table for each \tablenotetext.

\tablenotetext{a}{Identification of the HII knots for which photometry
has been measured in 13$^{\prime\prime}$ diameter apertures. The ID
format XX--YY identifies the background region (XX) where the aperture
is located and an internal progressive number (YY) for the aperture.}
\tablenotetext{b}{Position on the sky of the aperture.}
\tablenotetext{c}{Logarithm of the luminosities in the FUV and NUV
from GALEX, U, B, and H$\alpha$($\lambda$0.6563~$\mu$m) from
ground--based images, P$\alpha$($\lambda$1.8756~$\mu$m) from
HST/NICMOS, and 8~$\mu$m and 24~$\mu$m from Spitzer/IRAC and
MIPS. Stellar continuum luminosities are given as
$\lambda$L($\lambda$). The central wavelengths of the U and B filters
are 0.4312~$\mu$m and 0.3463~$\mu$m, respectively. Stellar and ionized
gas luminosities have been corrected for the Galactic foreground
extinction E(B$-$V)$_{MW}$=0.037.}

\end{deluxetable}

%% The following command ends your manuscript. LaTeX will ignore any text

\end{document}